\documentclass[fleqn,twoside]{article}
\usepackage{espcrc2}

\usepackage{graphicx}
\usepackage{epsfig}

\usepackage[figuresright]{rotating}

\newcommand{\AmS}{{\protect\the\textfont2
  A\kern-.1667em\lower.5ex\hbox{M}\kern-.125emS}}

\newcommand{\eq}[1]{eq.~(\ref{#1})}

\newcommand{\Eq}[1]{Eq.~(\ref{#1})}
\newcommand{\Eqs}[2]{Eqs.(\ref{#1}, \ref{#2})}

\newcommand{\ur}[1]{(\ref{#1})}

\newcommand{\nn}{\nonumber}
\newcommand{\n}{\nonumber}
\newcommand{\noi}{\noindent}
\def\nea{\nonumber\eea}

\newcommand{\beq}{\begin{equation}}
\newcommand{\eeq}{\end{equation}}

\newcommand{\la}[1]{\label{#1}}
\newcommand{\bea}{\begin{eqnarray}}
\newcommand{\eea}{\end{eqnarray}}
\newcommand{\beqa}{\begin{eqnarray}}
\newcommand{\eeqa}{\end{eqnarray}}
\newcommand{\ba}{\begin{array}}
\newcommand{\ea}{\end{array}}

\newcommand{\Nz}{$N_{CS}=0\;$}
\newcommand{\No}{$N_{CS}=1\;$}

\newcommand{\thb}[3]{\bar{\eta}^{#1}_{#2 #3}}
\newcommand{\tha}[3]{\eta^{#1}_{#2 #3}}

\newcommand{\half}{\mbox{$\frac{1}{2}$}}
\renewcommand{\vec}[1]{{\bf #1}}
\newcommand{\e}{\epsilon}

\newcommand{\Tr}{{\rm Tr}\,}

\renewcommand{\v}{{\rm v}}
\newcommand{\D}{r_{12}}
\newcommand{\x}{{\bf x}}

\newcommand{\bshd}{\overline{\mathrm{sh}}\,}
\newcommand{\bchd}{\overline{\mathrm{ch}}\,}
\newcommand{\shd}{\mathrm{sh}\,}
\newcommand{\chd}{\mathrm{ch}\,}
\newcommand{\Mphi}{\Phi}

\newcommand{\Det}{{\rm Det}}

\newcommand{\bv}{\overline{{\rm v}}}
\newcommand{\T}{T^{\rm\small o}}

\def\su2{{SU(2)}}

\def\det{\mbox{det\,}}
\def\Det{\mbox{Det\,}}

\title{{\bf TOPOLOGY AND CONFINEMENT}}

\author{{\bf Dmitri Diakonov}\address[THD]{Theory Division, Petersburg Nuclear Physics Institute,
188300, Gatchina, St. Petersburg, Russia}\thanks{
Lectures given at the ITEP Winter School (February 2009, Moscow) and Schladming
Winter School (March 2009, Schladming, Austria)}
}

\begin{document}

\begin{abstract}
These lectures contain an introduction to instantons, calorons and dyons
of the Yang--Mills gauge theory. Since we are interested in the mechanism of confinement and
of the deconfinement phase transition at some critical temperature, the Yang--Mills theory
is formulated and studied at nonzero temperatures. We introduce ``calorons with a nontrivial holonomy''
that are generalizations of instantons and can be viewed as ``made of'' constituent dyons.
The quantum weight with which these calorons contribute to the Yang--Mills partition function
is considered, and the ensuing statistical mechanics of the ensemble of interacting dyons is discussed.
We argue that a simple semiclassical picture based on dyons satisfies all known criteria of
confinement and explains the confinement-deconfinement phase transition. This refers not only
to the $SU(N)$ gauge groups where dyons lead to the expected behaviour of the observables with $N$,
but also to the exceptional $G(2)$ group whose group center, unlike $SU(N)$, is trivial.
Despite being centerless, the $G(2)$ gauge group possesses confinement at low temperatures,
and a $1^{\rm st}$ order deconfinement transition, according to several latest lattice simulations,
indicating that confinement-deconfinement is not related to the group center. Dyons, however,
reproduce this behaviour.

\vspace{1pc}
\end{abstract}

\maketitle
\vskip 0.7true cm

\tableofcontents
\vskip 1.5true cm

\section{WHAT IS `VACUUM'?}

Quantum field theory deals with fields fluctuating in space and time. If the fields are
free, {\it i.e.} non-interacting, their fluctuations are plane waves with any momenta,
called zero-point oscillations. The ground state of a system of free fields
with no sources (the `vacuum') is not a zero field but an infinite set of plane waves --
just as the ground state of a one-dimensional quantum-mechanical oscillator corresponds
not to a particle lying at the bottom of the potential but to a particle distributed
about the minima and having the energy $\frac{\hbar\omega}{2}$. The ground state of
a free quantum field theory has the energy $\sum\frac{\hbar\omega}{2}$ where one sums
over all eigenfrequencies of the free fields; in a $3\!+\!1$-dimensional theory this
sum diverges as the fourth power of the cutoff momentum.

However, if the fields are interacting, and the nonlinearity is strong enough such that
it is not tractable by perturbation theory, their can be large field fluctuations in the
vacuum, that cannot be reduced to weakly perturbed zero-point oscillations.
Those large vacuum fluctuations determine the content of the quantum field theory, but
in a strong interacting case it is difficult to describe them mathematically. One has
either to rely on some approximate methods, or study the dominant fluctuations numerically,
or hope to gain some insight from similar, {\it e.g.} supersymmetric field theories where,
because of additional symmetry, exact analytical methods can be developed. Confinement is
a very remarkable and still an unexplained phenomenon, therefore it may be helpful to use
all known methods to understand its microscopic mechanism.

There are two basic approaches to study the vacuum in quantum theory: one is the Hamiltonian
or Schr\"odinger approach, the other is the path integral or Feynman approach. We start from
simple quantum mechanics of a particle in a 1-dimensional potential $V(q)$ where $q$ is the
coordinate, to remind the connection between the two approaches.

Let the Lagrangian be $L=\frac{m\dot q^2}{2}-V(q)$ and the energy $H=\frac{m\dot q^2}{2}+V(q)$.
To find the (quantized) energy levels $E_n$ and the stationary wave function $\psi_n (q)$
one solves the Schr\"odinger equation:
\beq\n {\cal H}\psi_n(q) = E_n\,\psi_n(q),\quad
{\cal H}=-\frac{\hbar^2}{2m}\,\frac{d^2}{dq^2}+V(q).
\la{SchQM}\eeq

Another way to find all energy levels is to compute the partition function of a system,
as function of temperature $\T$. On the one hand, the partition function is defined as a sum
of Boltzmann exponents for all energy levels,
\beq
{\cal Z}=\sum_ne^{-\frac{E_n}{k\T}}.
\la{PF0}\eeq
On the other hand the partition function can be presented, according to Feynman, as
a path integral over all particle's trajectories periodic in time, with the period
$T=\frac{1}{k\T}$:
\bea\la{PT1}
{\cal Z}&=&\sum_n \int\!dq_0\,\psi_n^*(q_0)e^{-E_n T}\,\psi_n(q_0)\\
\n
&=&\int\!dq_0\int_{q(0)=q_0}^{q(T)=q_0}\!\!Dq(t)\\
\n
&\cdot &\exp\left(\!-\frac{1}{\hbar}
\int_0^T\!dt\left[\frac{m\dot q^2(t)}{2}\!+\!V\left(q(t)\right)\right]\right).
\eea
A generic path integral is illustrated in Fig.~1.

\begin{figure}[htb]
{\epsfig{figure=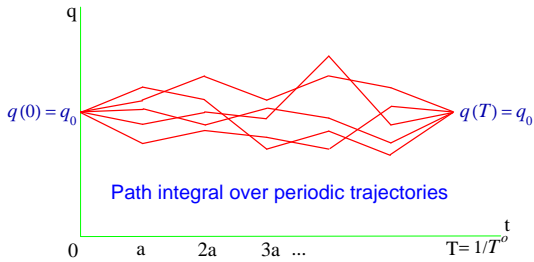,width=7cm}}
\caption{Feynman path integral in Quantum Mechanics, {\it i.e.} in $d=0+1$ quantum field theory.}
\end{figure}

\noindent
One can visualize typical trajectories contributing to the path integral \ur{PT1}
by discretizing it and computing a big number of ordinary integrals over coordinates
at discrete intermediate times $t_n=n a$:
\bea\la{PF2}
{\cal Z}&=&{\cal N}\;\lim_{a\to 0,N\to\infty}\prod_{n=1}^N\int\!dq_n\,e^{-\frac{S_{\rm discr}}{\hbar}},\\
\n
S_{\rm discr}\!\!\!&=&\!\!\!\sum_n a\!\left[\!\frac{m}{2}\!\left(\!\frac{q(t_n)\!-\!q(t_{n-1})}{a}\!\right)^2
\!+\!V\left(q(t_n)\!\right)\!\right].
\eea

In order to cut out the vacuum state with the lowest energy $E_0$ one has to take the limit
of large observation time $T\to\infty$, or small temperature $\T\to 0$. At first sight it may seem
that the leading contribution to the partition function \ur{PT1} will be from a trivial
trajectory corresponding to a particle lying all the time $T$ at the bottom of the potential well:
it minimizes both the potential and the kinetic energies both coming with the minus sign
in the exponent of \Eq{PT1}. However, such a trajectory is unique, and it comes with a vanishing weight
or entropy. If a trajectory oscillates somewhat about the minimum such that both the potential and kinetic
energies are not too large, one finds very many such trajectories, therefore they have a much larger statistical
weight. The outcome of this fight -- between minimizing the energy and maximizing the entropy --
is well known: typical trajectories contributing most to the path integral are of the type
shown in Fig.~2. This is why the ground-state energy is not zero but $E_0=\frac{\hbar\omega}{2}$,
plus corrections from the deviation of the potential from the quadratic one.

\begin{figure}[htb]
{\epsfig{figure=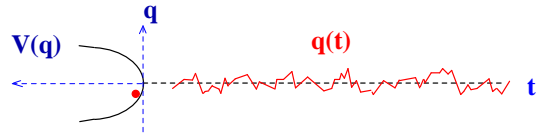,width=7cm}}
\caption{A typical trajectory in a potential well exhibits zero-point oscillations
giving rise to the vacuum energy $E_0=\frac{\hbar\omega}{2}$.}
\end{figure}

\noindent
For more complicated potentials, {\it e.g.} the double-well potential typical trajectories
are more complex, see Fig.~3.

\begin{figure}[htb]
{\epsfig{figure=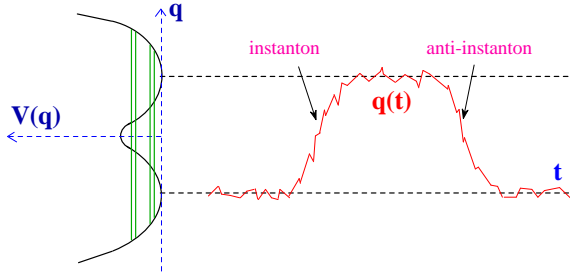,width=7.5cm}}
\caption{A typical trajectory in a double-well potential exhibits zero-point oscillations
about tunneling trajectories -- instantons and anti-instantons.}
\end{figure}

If the barrier between the potential wells is high and broad, the particle will
oscillate for a long time in one well but eventually will travel to the other one,
where it will oscillate again and eventually tunnel back to the first well, and so on.
The tunneling parts of the trajectory are called, in modern language, instantons
and anti-instantons. The definition is: Instantons are classical tunneling trajectories $q(t)$
with minimal action, satisfying the equation of motion (here: the Newton law)
\beq
\frac{\delta S}{\delta q(t)}=0\qquad {\rm or}\quad m\ddot q
= -\frac{\partial V}{\partial q}.
\la{EoM1}\eeq

As the particle transits from one well to another it still experiences zero-point oscillations
which, however, are now not around zero but about a nontrivial background trajectory.
The times of transition are random: in order to get physical quantities (like the level splitting
in the double-well potential) one has to integrate over all `positions' of instantons and anti-instantons
in time.

To visualize instantons in computer simulations, one can start computing the partition function
\ur{PF2} on a 1d Euclidean lattice by means of, say, the Metropolis algorithm. Stop it when
the system reaches equilibrium and draw the trajectory the computer is currently working with:
it will be something like presented in Fig.~3. While one will not see directly a smooth trajectory
interpolating between the two potential wells (because of the inevitable zero-point oscillations around it,
producing Gaussian noise), in Quantum Mechanics which is a (0+1)-dimensional quantum field theory it is not difficult to
smear out the quantum noise and reveal instantons. In (3+1) dimensions smearing out the normal quantum
fluctuations about smooth instanton configurations is not a straightforward task. A typical configuration
of the gluon field in a lattice simulation of the (3+1)-dimensional Yang--Mills theory is shown in
Fig.~4 borrowed from J.~Negele et al.~\cite{Negele}. It takes some effort to `cool' the configuration
down to something smooth. Eventually one sees instantons and anti-instantons with a naked eye.

\begin{figure}[htb]
{\epsfig{figure=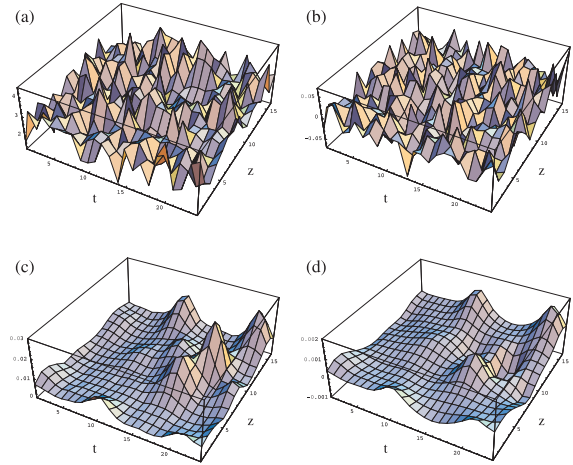,width=7.5cm}}
\caption{Instantons and anti-instantons in $4d$ Yang--Mills theory are revealed after zero-point oscillations
are smeared out. {\it Upper row}: a typical full configuration of the gluon field from lattice simulations~\cite{Negele}
in the $(z,t)$ hyperplane with $(x,y)$ fixed.
{\it Lower row}: the same configuration after smearing clearly shows that quantum noise of the upper row is around 3 instantons
and 2 anti-instantons. {\it Left column}: action density as function of Euclidean time and one spatial coordinate. {\it Right column}:
topological charge density of the same configuration. }
\end{figure}

Fig.~4, upper row, is a direct analogue of Fig.~3 but for a quantum field theory in (3+1) dimensions.

\section{INSTANTONS IN YANG--MILLS THEORY}

\subsection{Yang--Mills theory in Hamiltonian formulation}

A fundamental fact \cite{Fad,JR} is that the potential energy
of the gluon field is a periodic function in one particular direction
in the infinite-dimensional functional space; in all other directions the
potential energy is oscillator-like. This is illustrated in Fig.~5.
The nontrivial form of the potential energy implies that instantons exist in
the Yang--Mills (YM) theory, corresponding to tunneling through potential energy barriers.

\begin{figure}[htb]
{\epsfig{figure=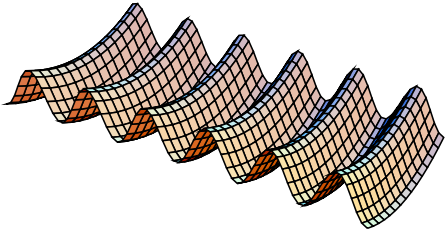,width=6cm}}
\caption{Potential energy of the gluon field is periodic in one direction
and oscillator-like in all other directions in functional space.}
\end{figure}

To observe the periodicity, we take the $A_0=0$ gauge, called Weyl
or Hamiltonian gauge. In Euclidean space it is more appropriate to
denote the time component of the YM field by $A_4$, so we use
the $A_4^a=0$ gauge. For simplicity we start from the $SU(2)$ gauge group,
$a=1,2,3$.

The spatial YM potentials $A_i^a(\x, t)$ can be considered
as an infinite set of the coordinates of the system, where $i=1,2,3,\;\;
a=1,2,3$ and $\x$ are ``labels" denoting various coordinates. The
YM action is
\bea\la{YMA}
S&=&\frac{1}{4g^2}\int\! d^4x\; F_{\mu\nu}^aF_{\mu\nu}^a\\
\n
&=&\int dt\left(\frac{1}{2g^2}\int\! d^3\x\;{\bf E}^2 - \frac{1}{2g^2}\int\!d^3\x\;{\bf B}^2 \right)
\eea
where $E_i^a={\dot A}_i^a$ is the electric field strength and $
B_i^a=\frac{1}{2}\epsilon_{ijk}\left(\partial_jA_k^a-
\partial_kA_j^a+\epsilon^{abc}A_j^bA_k^c\right)$ is the magnetic
field strength.

Apparently, the first term in \Eq{YMA} is the kinetic energy of the
system of coordinates $\{A_i^a(\x,t)\}$ while the second term is minus
the potential energy being just the magnetic energy of the field.
Upon quantization the electric field is replaced by the
variational derivative,
$E_i^a(x) \rightarrow -ig^2\delta/\delta A_i^a(x)$,
if one uses the `coordinate' representation for the wave
functional. The functional Schr\"odinger equation for the wave
functional $\Psi[A_i^a(x)]$ takes the form
\bea\n
{\cal H}\Psi[A_i]&=&
\int \!d^3x\left\{-\frac{g^2}{2}\frac{\delta^2}{(\delta A_i^a(x))^2}\right.\\
&+&\left.\frac{1}{2g^2}(B_i^a(x))^2\right\}\Psi[A_i]
={\cal E}\Psi[A_i]
\la{schr1}\eea
where ${\cal E}$ is the eigenenergy of the state in question. The YM
vacuum is the ground state of the Hamiltonian \ur{schr1},
corresponding to the lowest energy ${\cal E}$.

Not all $\Psi[A_i]$ solving formally the Schr\"odinger equation describe physical states,
however, but only those that are gauge invariant:
\bea\n
&&\Psi[A_i^\Omega]=\Psi[A_i],\qquad A_i^\Omega=\Omega^\dagger A_i\Omega+i\Omega^\dagger\partial_i \Omega,\\
&&\Omega(\x)\in SU(2).
\la{gauge invariance}\eea

\noindent
\underline{Example: perturbative wave functional}\\

At small gauge coupling constant $g^2\to 0$ one can neglect the commutator term
in the magnetic field, $B_i^a\approx\epsilon_{ijk}\,\partial_jA_k^a$, the Schr\"odinger
equation \ur{schr1} becomes that for an infinite set of coupled harmonic oscillators, and
can be solved exactly. The ground state wave functional is given by a Gaussian that can be
best written in terms of the coordinates $A_i^a({\bf p})$ labeled by the 3-momenta ${\bf p}$
of the fields:
\bea\la{psi0}
\Psi_0^{(0)}[A]\!\!\!&=&\!\!\!\exp\left[-\frac{1}{2}\!\int\!\frac{d^3{\bf p}}{(2\pi)^3}\,
A^a_i(-{\bf p})\right.\\
\n
\!\!\!&\cdot &\!\!\!\left.\left(|{\bf p}|\delta_{ij}-\frac{p_ip_j}{|{\bf p}|}\right)A^a_j({\bf p})\right]\\
\nn
\!\!\!&=&\!\!\!\exp\left[-\frac{1}{4\pi^2}\!\int\!\int\!d^3{\bf x}d^3{\bf y}\,\frac{B^a_i({\bf x})B^a_i({\bf y})}
{|{\bf x}-{\bf y}|^2}\right],
\eea
describing $2\cdot(N^2-1)$ gluon waves (for the $SU(N)$ gauge group) with transverse polarizations
and energy $|{\bf p}|$. The vacuum energy of these zero-point oscillations is the sum of
the harmonic oscillators' energies $\sum\frac{\hbar\omega}{2}$ or, more precisely,
\beq
{\cal E}_0^{(0)}=
\frac{2(N^2-1)}{2}V\!\int\!\frac{d^3{\bf p}}{(2\pi)^3}\,|{\bf p}|
\la{E00}\eeq
where $V$ is the $3d$ volume. We see that the energy of the zero-point oscillations is
quartically divergent.

\subsection{Topology}

Let us introduce an important quantity called the Pontryagin index or
the four-dimensional topological charge of the YM fields:
\bea\la{FFd}
Q_T&=&\frac{1}{32\pi^2}\int d^4x\; F_{\mu\nu}^a {\tilde F}_{\mu\nu}^a,\\
\n
&&{\tilde F}_{\mu\nu}^a \equiv \frac{1}{2}\epsilon_{\mu\nu\alpha\beta}F_{\alpha\beta}^a.
\eea
The integrand in \eq{FFd} happens to be a full derivative of the four-vector $K_\mu$:
\bea\la{K}
&&\frac{1}{32\pi^2} F_{\mu\nu}^a {\tilde F}_{\mu\nu}^a=\partial_\mu K_\mu,\\
\n
&&K_\mu\!=\!\frac{1}{16\pi^2}\epsilon_{\mu\alpha\beta\gamma}
\left(\!A_\alpha^a\partial_\beta A_\gamma^a\!\!+\!\!\frac{1}{3}\epsilon^{abc}
A_\alpha^a A_\beta^b A_\gamma^c\!\right)\!.
\eea
Therefore, assuming the fields $A_i$ are decreasing rapidly enough at
spatial infinity, one can rewrite the 4-dimensional topological charge
\ur{FFd} as
\beq
Q_T=\!\int\!\! d^4x (\partial_t K_4\!+\!\partial_i K_i)=\!\int\!\! dt \frac{d}{dt}
\int\!\! d^3\x K_4.
\la{Gauss}\eeq
Introducing the {\it Chern--Simons number}
\bea\la{NCS}
N_{CS}\!\!\!&=&\!\!\!\int d^3\x\;K_4\\
\n
\!\!\!&=&\!\!\!\frac{1}{16\pi^2}\!\int\!\! d^3\x\epsilon^{ijk}\!
\left(\!A_i^a\partial_j A_k^a\!+\!\frac{1}{3}\epsilon^{abc}
A_i^a A_j^b A_k^c\!\right)
\eea
we see from \Eq{Gauss} that $Q_T$ can be rewritten as the difference
of the Chern--Simons numbers characterizing the fields at $t=\pm\infty$:
\beq
Q_T=N_{CS}(+\infty)-N_{CS}(-\infty).
\la{dif}\eeq

The Chern--Simons number of the field has an important property
that it can change by integers under large gauge transformations. Indeed,
under a general time-independent gauge transformation,
\beq
A_i \rightarrow U^\dagger A_iU + iU^\dagger\partial_iU, \;\;\;
A_i\equiv A_i^a\frac{\tau^a}{2},
\la{gt}\eeq
the Chern--Simons number transforms as follows:
\bea\la{CSt}
N_{CS}&\rightarrow &N_{CS}+N_W\\
\n
&+&\frac{i}{8\pi^2}\int\!d^3x\,\epsilon^{ijk}\partial_j\,\Tr(\partial_i UU^\dagger A_k).
\eea
The last term is a full derivative and can be omitted if $A_i$ decreases
sufficiently  fast  at spatial infinity. $N_W$ is the winding number of the gauge
transformation \ur{gt}:
\bea\la{wn}
N_W&=&\frac{1}{24\pi^2}\int d^3\x\,\epsilon^{ijk}\\
\n
&\cdot &\Tr\left[(U^\dagger\partial_iU)(U^\dagger\partial_jU) (U^\dagger\partial_kU)\right].
\eea

The $SU(2)$ unitary matrix $U$ of the gauge transformation \ur{gt}
can be viewed as a mapping from the 3-dimensional space onto the
3-dimensional sphere of parameters $S^3$. If at spatial infinity we
wish to have the same matrix $U$ independently of the way we approach
the infinity (and this is what is usually assumed), then the spatial
infinity is in fact one point, so the mapping is topologically
equivalent to that from $S^3$ to $S^3$. This mapping is known to be
nontrivial, meaning that mappings with different winding numbers
are irreducible by smooth transformations to one another. The winding
number of the gauge transformation is, analytically, given by \eq{wn}.
As it is common for topological characteristics, the integrand in \ur{wn}
is in fact a full derivative. For example, if we take the matrix $U(\x)$
in a ``hedgehog" form, $U=\exp[i(r\cdot \tau)/r\, P(r)]$, \eq{wn}
can be rewritten as
\bea\n
N_W&=&\frac{2}{\pi}\int\!dr  \frac{dP}{dr}\sin^2 P = \frac{1}{\pi}
\left[P-\frac{\sin 2P}{2}\right]_0^\infty \\
&=& \mbox{integer}
\la{wnh}\eea
since $P(r)$ both at zero and at infinity needs to be multiples of $\pi$
if we wish $U({\bf r})$ to be unambiguously defined at the origin and
at the infinity.

Let us return now to the potential energy of the YM fields,
\beq
{\cal V}=\frac{1}{2g^2} \int d^3\x \left(B_i^a\right)^2.
\la{potene}\eeq

One can imagine plotting the potential energy surfaces over the
Hilbert space of the coordinates $A_i^a(\x)$. It will be some complicated
mountain country. If the field happens to be a pure gauge, $A_i=
iU^\dagger\partial_i U$, the potential energy at such points of the
Hilbert space is naturally zero. Imagine that we move along the
``generalized coordinate'' being the Chern--Simons number \ur{NCS},
fixing all other coordinates whatever they are. Let us take some
point $A_i^a(\x)$ with the potential energy ${\cal V}$. If we move to
another point which is a gauge transformation of $A_i^a(\x)$ with a
winding number $N_W$, its potential energy will be exactly the same as
it is strictly gauge invariant.  However the Chern--Simons
``coordinate'' of the new point will be shifted by an integer $N_W$
from the original one. We arrive to the conclusion first pointed out
by Faddeev \cite{Fad} and Jackiw and Rebbi \cite{JR} in 1976, that
the potential energy of the YM fields is {\em periodic} in the
particular coordinate called the Chern--Simons number. This is
illustrated in Fig.~5.

\subsection{YM instantons in simple terms}

In perturbation theory one deals with zero-point quantum-mechanical
fluctuations of the YM fields near one of the minima, say, at $N_{\rm CS}=0$.
The non-linearity of the YM theory is taken into account as a
perturbation, and results in series in $g^2$ where $g$ is the gauge
coupling. In such approach one is apparently missing a possibility for
the system to tunnel to another minimum, say, at $N_{\rm CS}=1$.
Tunneling is a typical nonperturbative effect in the coupling constant.

Instanton is a large fluctuation of the gluon field in imaginary
(or Euclidean) time corresponding to quantum tunneling from one minimum
of the potential energy to the neighbour one. Mathematically, it was
discovered by Belavin, Polyakov, Schwarz and Tyupkin~\cite{BPST};
the tunneling interpretation was given by Gribov~\footnote{Gribov made this comment
at a seminar given by A.~Polyakov in 1976 at the Petersburg Nuclear Physics Institute,
that I attended. Polyakov acknowledges Gribov's tunneling interpretation
of instantons in his famous paper~\cite{Pol}.}.
The name `instanton' has been introduced by 't~Hooft~\cite{tH} who
studied many of the key properties of those fluctuations. Anti-instantons
are similar fluctuations but tunneling in the opposite direction.
Physically, one can think of instantons in two ways: on the one hand it is
a tunneling {\em process} occurring in time, on the other hand it is a
localized {\em pseudoparticle} in the Euclidean space.

Following the WKB approximation, the tunneling amplitude can be estimated as $\exp(-S)$,
where $S$ is the action along the classical trajectory in imaginary time, leading from
the minimum at $N_{\rm CS}=0$ at $t=-\infty$ to that at $N_{\rm CS}=1$ at $t=+\infty$.

According to \eq{dif} the $4d$ topological charge
of such trajectory is $Q_T=1$. To find the best tunneling trajectory
having the largest amplitude one has thus to minimize the YM action
\ur{YMA} provided the topological charge \ur{FFd} is fixed to be unity.
This can be done using the following trick. Consider
the inequality
\[
0\leq \int d^4x \left(F_{\mu\nu}^a-{\tilde F}_{\mu\nu}^a\right)^2
\]
\beq
=\int d^4 x \left(2 F^2-2F{\tilde F}\right) =
8g^2S-64\pi^2Q_T,
\la{inequ}\eeq
hence the action is restricted from below:
\beq
S\geq \frac{8\pi^2}{g^2}Q_T =  \frac{8\pi^2}{g^2}.
\la{inequa}\eeq

Therefore, the minimal action for a trajectory with a unity topological
charge is equal to $8\pi^2/g^2$, which is achieved if the trajectory
satisfies the {\em self-duality} equation~\cite{BPST}:
\beq
F_{\mu\nu}^a={\tilde F}_{\mu\nu}^a.
\la{selfdual}\eeq

Notice that any solution of \Eq{selfdual} is simultaneously a solution of
the general YM equation of motion $D_\mu^{ab}F_{\mu\nu}^b=0$: that is
because the ``second pair" of the Maxwell equations,
$D_\mu^{ab}{\tilde F}_{\mu\nu}^b=0$, is satisfied identically.

Thus, the tunneling amplitude can be estimated as
\bea\n
{\cal A} &\sim& \exp(-{\rm Action})=\exp\left(-\frac{1}{4g^2}\int\!d^4x\,F_{\mu\nu}^2\right)\\
\la{ta}
&=&\exp\left(-\frac{8\pi^2}{g^2}\right)=\exp\left(-\frac{2\pi}{\alpha_s}\right).
\eea
It is non-analytic in the gauge coupling constant and hence instantons are
missed in all orders of the perturbation theory. However, it is not
a reason to ignore tunneling. For example, tunneling of electrons from one
atom to another in a metal is also a nonperturbative effect but we would get
nowhere in understanding metals had we ignored it.

\subsection{Instanton configuration}

To solve \Eq{selfdual} let us recall a few facts about the Lorentz group
$SO(3,1)$. Since we are talking about the tunneling fields which can only
develop in imaginary time, it means that we have to consider the fields
in Euclidean space-time, so that the Lorentz group is just $SO(4)$
locally isomorphic to $SU(2)\times SU(2)$. The gauge potentials $A_\mu$ belong to the
$(\frac{1}{2},\frac{1}{2})$ representation of the $SU(2)\times SU(2)$
group, while the field strength $F_{\mu\nu}$ belongs to the reducible $(1,0)+(0,1)$
representation. In other words it means that one linear combination
of $F_{\mu\nu}$ transforms as a vector of the left $SU(2)$, and another
combination transforms as a vector of the right $SU(2)$. These
combinations are
\bea\la{oneone}
F_L^A&=&\eta_{\mu\nu}^A(F_{\mu\nu}+\tilde{F}_{\mu\nu}),\\
\n
F_R^A&=&\bar{\eta}_{\mu\nu}^A(F_{\mu\nu}-\tilde{F}_{\mu\nu}),
\eea
where $\eta, \bar{\eta}$ are the so-called 't Hooft symbols, see below.
We see therefore that a self-dual
field strength is a vector of the left $SU(2)$ while its right part is
zero.  Keeping this experience in mind we look for the solution of the
self-dual equation in the form
\bea\la{tHanz}
A_\mu^a&=&\thb{a}{\mu}{\nu}\, x_\nu\,\frac{1+\Phi(x^2)}{x^2};\\
\nn
&&\tha{a}{i}{j}=\epsilon_{aij},\quad \tha{a}{4}{j}=-\tha{a}{j}{4}=-\delta_{a\,j},\\
\nn
&&\thb{a}{i}{j}=\epsilon_{aij},\quad \thb{a}{4}{j}=-\thb{a}{j}{4}=+\delta_{a\,j}.
\eea
Using the formulae for the 't Hooft's $\eta$ symbols one can
easily check that the YM action can be rewritten as
\bea\n
S&=&\frac{8\pi^2}{g^2}\frac{3}{2} \int d\tau
\left[\frac{1}{2}\left(\frac{d\Phi}{d\tau}\right)^2+
\frac{1}{8}(\Phi^2-1)^2\right], \\
\la{doublewell}
&&\tau = \ln\left(\frac{x^2}{\rho^2}\right).
\eea
This can be recognized as the action of the double-well potential whose
minima lie at $\Phi=\pm 1$, and $\tau$ plays the role of ``time"; $\rho$
is an arbitrary scale. The trajectory which tunnels from $1$ at
$\tau=-\infty$ to $-1$ at $\tau = +\infty$ is
\beq
\Phi=-\tanh\left(\frac{\tau}{2}\right),
\la{QMinst}\eeq
and its action \ur{doublewell} is $S=8\pi^2/g^2$, as needed. Substituting
the solution \ur{QMinst} into \ur{tHanz} we get
\bea\n
A_\mu^a(x)&=&\thb{a}{\mu}{\nu}\, x_\nu\,
\frac{1+\tanh\left(\frac{\ln\left(\frac{x^2}{\rho^2}\right)}{2}\right)}{x^2}\\
&=&\frac{2\thb{\mu}{\nu}{a}\rho^2}{x^2(x^2+\rho^2)}.
\la{YMinst}\eea
The correspondent field strength is
\bea\la{fstrength}
F_{\mu\nu}^a\!\!\!&=&\!\!\!-\frac{4\rho^2}{(x^2+\rho^2)^2}\left(\thb{\mu}{\nu}{a}-
2\thb{\mu}{\alpha}{a}\frac{x_\alpha x_\nu}{x^2}\right.\\
\n
\!\!\!&&\!\!\!-\left.2\thb{\beta}{\nu}{a}\frac{x_\mu x_\beta}{x^2}\right),\quad
F_{\mu\nu}^aF_{\mu\nu}^a=\frac{192\rho^4}{(x^2+\rho^2)^4},
\eea
and satisfies the self-duality condition \ur{selfdual}.

The {\em anti-instanton} corresponding to tunneling in the opposite
direct\-ion, from \No to \Nz, satis\-fies the {\em anti}-self-dual
equation, with $\tilde{F}\to -\tilde{F}$; its concrete form is
given by \Eqs{YMinst}{fstrength} with the replacement
$\bar{\eta}\to \eta$.

\begin{figure}[htb]
{\epsfig{figure=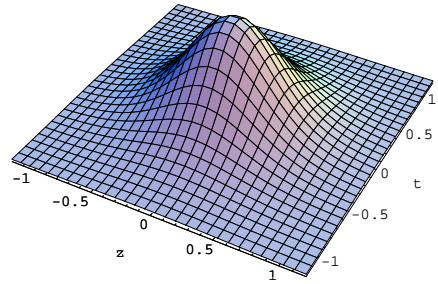,width=6cm}}
\caption{Action density of the YM instanton as function of $z,\;t$ at fixed
$x=y=0$. Instanton is a `ball' in $4d$ possessing $O(4)$ symmetry.}
\end{figure}

\Eqs{YMinst}{fstrength} describe the field of the instanton in the
singular Lorenz gauge; the singularity of $A_\mu$ at $x^2=0$ is a gauge
artifact: the gauge-invariant field strength squared is smooth at the
origin. Formulae for instantons are more compact in the Lorenz gauge
but an instructive homework is to pass on to the Hamiltonian $A_4=0$ gauge
and to check explicitly that instantons indeed correspond to tunneling
from \No to \Nz.

\subsection{Instanton collective coordinates}

The instanton field, \Eq{YMinst}, depends on an arbitrary scale
parameter $\rho$ which we shall call the instanton size, while the
action, being scale invariant, is independent of $\rho$. One can
obviously shift the position of the instanton to an arbitrary 4-point
$z_\mu$ -- the action will not change either. Finally, one can rotate
the instanton field in colour space by constant unitary matrices $U$.
For the $SU(2)$ gauge group this rotation is characterized by 3
parameters, {\it e.g.} by Euler angles. For a general $SU(N)$ group
the number of parameters is $N^2-1$ (the total number of the
$SU(N)$ generators) {\em minus} $(N-2)^2$ (the number of generators
which do not affect the left upper $2\times 2$ corner where the
standard $SU(2)$ instanton \ur{YMinst} is residing), that is $4N-5$.
These degrees of freedom are called instanton orientation in colour
space.  All in all there are
$$
4\; ({\rm centre})\;+\;1\; ({\rm size})\;+\;
(4N-5)\; ({\rm orientations})
$$
\beq\la{collcoo}
\:\:\:=\;\;4N
\eeq
so-called collective coordinates describing the field of the instanton,
of which the action is independent.

It is convenient to introduce $2\times 2$ matrices
\beq
\sigma^{\pm}_\mu = (\pm i \overrightarrow{\sigma}, 1),\qquad
x^{\pm}=x_\mu\sigma^{\pm}_\mu,
\la{sigma}\eeq
such that
\bea\n
2i\tau^a
\tha{\mu}{\nu}{a}&=&\sigma^+_\mu\sigma^-_\nu
-\sigma^+_\nu\sigma^-_\mu, \\
\n
2i\tau^a\thb{\mu}{\nu}{a}&=&\sigma^-_\mu\sigma^+_\nu
-\sigma^-_\nu\sigma^+_\mu,
\la{thsig}\eea
then the instanton field with arbitrary center $z_\mu$, size $\rho$ and
colour orientation $U$ in the $SU(N)$ gauge group can be written as
$$
A_\mu=A_\mu^at^a
=\frac{-i\rho^2U[\sigma^-_\mu(x-z)^+-(x-z)_\mu]U^\dagger}
{(x-z)^2[\rho^2+(x-z)^2]}$$
\beq
\Tr(t^at^b)=\frac{1}{2}\delta^{ab},
\la{instgen}\eeq
or as
\beq
A_\mu^a=\frac{2\rho^2O^{ab}\thb{\mu}{\nu}{b}(x-z)_\mu}
{(x-z)^2[\rho^2+(x-z)^2]},
\la{instgena}\eeq
$$
O^{ab}=\Tr(U^\dagger t^aU\sigma^b),\qquad O^{ab}O^{ac}=\delta^{bc}.
$$
This is the explicit expression for the $4N$-parameter instanton field
in the $SU(N)$ gauge theory, written down in the singular Lorenz gauge.

\subsection{One-instanton weight}

The contribution of a saddle point -- one isolated instanton -- to the partition function is called
the one-instanton weight. We have already estimated the tunneling amplitude in \Eq{ta} but
it is not sufficient: the pre-factor is very important. To the 1-loop accuracy, it has been
first computed by 't Hooft~\cite{tH} for the $SU(2)$ colour group, and generalized to
arbitrary $SU(N)$ by C.~Bernard~\cite{Bernard}.

The general field can be decomposed as a sum of a classical field of an instanton
$A_\mu^I(x,\xi)$ where $\xi$ is a set of $4N$ collective coordinates characterizing
a given instanton (see \Eq{instgen}), and of a presumably small quantum field $a_\mu(x)$:
\beq
A_\mu(x)=A_\mu^I(x,\xi) + a_\mu(x).
\la{genfi}\eeq
There is a subtlety in this decomposition due to the gauge freedom. The action is
\bea\la{quaform}
{\rm Action}\!\!\!&=&\!\!\!\frac{1}{4g^2}\int d^4x\, F_{\mu\nu}^2\\
\n
\!\!\!&=&\!\!\!\frac{8\pi^2}{g^2}+\frac{1}{g^2}\int d^4x\, D_\mu F_{\mu\nu}a_\nu\\
\n
\!\!\!&+&\!\!\!\frac{1}{2g^2}\int d^4x\, a_\mu W_{\mu\nu}a_\nu\! + \!O(a^3).
\eea
Here the term linear in $a_\mu$ drops out because the instanton field satisfies the equation
of motion. The quadratic form $W_{\mu\nu}$ being a second-order differential operator,
has $4N$ zero modes related to the fact that
the action does not depend on $4N$ collective coordinates. This brings in a divergence
in the functional integral over the quantum field $a_\mu$ which, however, can and should
be qualified as integrals over the collective coordinates: center, size and orientations.
Formally the functional integral over $a_\mu$ gives
\beq
\frac{1}{\sqrt{\det\;W_{\mu\nu}}},
\la{funcdet}\eeq
which must be {\it i}) normalized (to the determinant of the free quadratic form, {\it i.e.} with no
background field), {\it ii}) regularized (for example by using the Pauli--Villars method),
and {\it iii}) accounted for the zero modes. Actually one has to compute a ``quadrupole"
combination,
\beq
\left[\frac{\det^\prime W \; \det (W_0+\mu^2)}{\det W_0 \; \det (W+\mu^2)}\right]^{-\frac{1}{2}},
\la{quadrup}\eeq
where $W_0$ is the quadratic form with no background field and $\mu^2$ is the
Pauli--Villars mass playing the role of the ultraviolet cutoff; the prime reminds that the zero
modes should be removed and treated separately. The resulting one-instanton contribution to
the partition function (normalized to the free one) is~\cite{tH,Bernard}:
\bea
\la{1instw}
\frac{{\cal Z}_{\rm 1-inst}}{{\cal Z}_{\rm P.T.}} &=&\int\! d^4z_\mu\int\! d\rho\int\!
d^{4N-5}U\, d_0(\rho), \\
\n
d_0(\rho)&=&\frac{C(N)}{\rho^5}\left[\frac{2\pi}{\alpha_s(\mu)}\right]^{2N}
(\mu\rho)^{\frac{11}{3}N}\\
&\cdot &\exp\left(-\frac{2\pi}{\alpha_s(\mu)}\right),\quad \alpha_s=\frac{g^2}{4\pi}.
\la{1instw1}\eea
The fact that there are all in all $4N$ integrations over the collective coordinates
$z_\mu,\rho,U$ reflects $4N$ zero modes in the instanton background.
The combination of the UV cutoff $\mu$ and the coupling constant given at that cutoff
$\alpha_s(\mu)$ is the YM scale parameter
\beq
\Lambda=\mu\,\exp\left(-\frac{3}{11\,N}\,\frac{2\pi}{\alpha_s(\mu)}\right).
\la{Lambda}\eeq
The numerical coefficient $C(N)$ depends implicitly on the regularization scheme used.
In the Pauli--Villars scheme exploited above
\beq
C(N)=\frac{4.60\exp(-1.68N)}{\pi^2(N-1)!(N-2)!}.
\la{CNc}\eeq
This is how the cutoff-independent combination $\Lambda$ \ur{Lambda} comes into being.
The mechanism is called the `transmutation of dimensions'. Henceforth all dimensional quantities
will be expressed through $\Lambda$, the only scale parameter of the massless YM theory,
which is very much welcome.

\begin{figure}[htb]
{\epsfig{figure=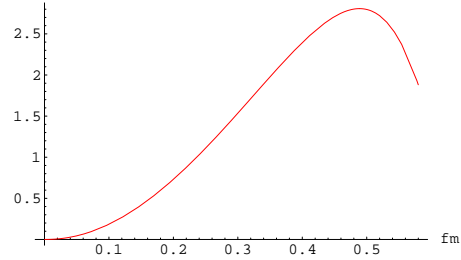,width=6cm}}
\caption{Distribution in instanton sizes computed in two-loop order~\cite{DP-84}.
The fall-off at large $\rho$ cannot be trusted as the two-loop approximation breaks down there.}
\end{figure}

\noi
Notice that  the integral over the instanton sizes in \Eq{1instw} diverges as a high power of
$\rho$ at large $\rho$: this is of course the consequence of asymptotic freedom. It means that
individual instantons tend to swell! This circumstance plagued the instanton calculus for
many years.

\section{YANG--MILLS THEORY AT NONZERO TEMPERATURES}

In the previous section we have introduced instantons at zero temperature, that is
in the Euclidean $R^4$ space. We attempt, however, not only to understand confinement at zero temperature
but also deconfinement at a high temperature. Therefore, we have to introduce the formalism
for studying YM theory at a nonzero temperature~\cite{GPY}.

We recall Feynman's representation for the partition function (see section 1):
\bea\nn
{\cal Z}\!\!\! &=&\!\!\!  \sum_n\langle n\left| e^{-\beta {\cal H}}\right| n\rangle
\qquad\quad{\left[ \beta=\frac{1}{T} \right]}\\
\nn
\!\!\!&=&\!\!\! \sum_n \int\!dq\, \psi^*_n(q)\; e^{-\beta {\cal E}_n}\; \psi_n(q)\\
\n
\!\!\!&=&\!\!\!\int\!\!dq\int_{q(0)=q}^{q(\beta)=q}\!\!\!Dq(t)\,\exp\!\left(\!-\!\!\int_0^\beta\!\!\!dt{\cal L_{\rm Euclid}}[q,\dot q]
\!\right).
\eea
In YM theory the role of coordinates $q$ is played by the spatial components of the gluon field
$A_i^a(x)$:
\bea\nn
&&{\cal Z}\!=\!\!\int\! DA_i(x) \int_{A_i(0,x)=A_i(x)}^{A_i(\beta,x)=A_i(x)}\!\!DA_i(t,x)\\
\nn
&&\cdot\exp\left\{\!-\frac{1}{2g^2}\,\int_0^\beta\!dt\, \int\!d^3x\left[\!\left(\!\dot A_i^a\right)^2
\!+\!\left(B_i^a\right)^2\!\right]\!\right\}.
\eea
However, in a gauge theory one sums not over all possible but only over
physical states, {\it i.e.} satisfying Gauss' law. In the absence of external
sources it means that only those states  need to be taken into account
that are invariant under gauge transformations (see subsection 2.1):
\bea\n
A_i(x)\!\!\!&\to &\!\!\!\left[A_i(x)\right]^{\Omega(x)},\qquad\left[\Omega(x)\!=\!e^{i\omega_a(x)t^a}\right]\\
\n
\!\!\!&=&\!\!\!\Omega(x)^{\dagger}\,A_i(x)\,\Omega(x)\! +\!
i\Omega(x)^{\dagger}\,\partial_i \Omega(x).
\eea
To restrict the summation to physical states, one projects to the physical {\it i.e.} gauge invariant
states by averaging the initial and final configurations over gauge rotations. [This is as if
in a quantum mechanics problem with a central-symmetric potential we would like to restrict
the summation to spherically-symmetric states only.] We have
\bea\nn
{\cal Z_{\rm phys}}\!\!\!\!&=&\!\!\sum_{\rm phys\; states}
\langle n\left| e^{-\beta {\cal H}}\right|n\rangle\\
\n
\!\!\!\!&=&\!\!\!\! \sum_n \int\!d\Omega_{1,2}\!\!\int\!\!dq\, \psi^*_n(\Omega_1q)\; e^{-\beta {\cal E}_n}\; \psi_n(\Omega_2q)\\
\nn
\!\!\!\!\!\!&=&\!\!\!\!\!\!\! \int \!\!\! D\Omega_{1,2}(x)D\!A_i(x) \!\!\!
\int_{A_i(0,x)=A_i(x)^{\Omega_1(x)}}
^{A_i(\beta,x)=A_i(x)^{\Omega_2(x)}} \!\!\!\!\!\!\!\!\!\!\!D\!A_i(t,x)\\
\n
\!\!\!\!&\cdot &\!\!\!
\exp\!\left\{\!\!-\frac{1}{2g^2}\!\! \int_0^\beta\!\!\!dt\!\!
\int \!\!\!d^3x \!\!\left[\!\left(\!\dot A_i^a\!\right)^2
\!\!\! + \!\!\left(B_i^a\right)^2\right]\!\right\}\!.
\eea
Renaming the initial field $A_i^{\Omega_1(x)} \to A_i$ and introducing the
relative gauge transformation $\Omega(x)=\Omega_2(x)\,\Omega_1^{\dagger}(x)$
  one can rewrite this as
\beq
{\cal Z_{\rm{phys}}}=\!\!\! \int\!\!\! D\Omega(x)DA_i(x)\!\!\!
\int_{A_i(x)}^{A_i(x)^{\Omega(x)}}\!\!\!\!\!\!\!\!\!\!\!\!\!\!\!\!
DA_i(t,x)\;e^{-S[A_i]}.
\la{Zph1}\eeq
A more customary form of the YM partition function is obtained
if one introduces, instead of $\Omega(x)$, a new variable $A_4(t,x)$,  {\it e.g.} as
$A_4=i\exp(-it\omega^at^a)\partial_t\exp(it\omega^at^a)$. Then the partition function
can be presented as a path integral over strictly periodic gauge fields:
\bea\la{Zph2}
{\cal Z}_{\rm phys}\!\!\!&=&\!\!\!\!\int\!\!DA_\mu\,
\exp\left\{-\frac{1}{4g^2}\int\!d^4x\,F_{\mu\nu}^aF_{\mu\nu}^a\right\},\\
\n
&&A_{\mu}(t, x) = A_{\mu}(t+\beta, x),\quad \beta = \frac{1}{T}.
\eea

An important variable is the Polyakov line or holonomy; it is the path-ordered exponent
in the time direction, hence it can depend only on the space point ${\bf x}$:
\beq
L({\bf x})={\cal P}\,\exp\left(i\int_0^{\frac{1}{T}}\!dt\,A_4(t,{\bf x})\right).
\la{L}\eeq
In the first formulation \ur{Zph1} it is nothing but the group element
$\Omega(x)$ by which the final $A_i(x,\beta)$ can differ from the initial $A_i(x,0)$:
$$
L(x)=\Omega(x).
$$
Under space-dependent gauge transformations it transforms
as $L\to U^{-1}LU$. The eigenvalues of $L({\bf x})$
are gauge invariant; for the $SU(N)$ gauge group we parameterize them as
\beq
L={\rm diag}\left(e^{2\pi i \mu_1},e^{2\pi i \mu_2},\ldots ,e^{2\pi i \mu_N}\right),
\la{mu}\eeq
$\mu_1\!+\!\ldots\!+\mu_N=0$, and assume that the phases of these eigenvalues are ordered:
$\mu_1\leq\mu_2\leq\ldots\leq\mu_N\leq\mu_{N+1}\equiv\mu_1\!+\!1$.
We shall call the set of $N$ phases $\{\mu_m\}$ the ``holonomy'' for short.
Apparently, shifting $\mu$'s by integers does not change the eigenvalues, hence all
quantities have to be periodic in all $\mu$'s with a period equal to unity.

The holonomy is said to be ``trivial'' if $L$ belongs to one of the $N$ elements
of the group center $Z_N$. For example, in $SU(3)$ the three trivial holonomies
are
\bea\n
\mu_1=\mu_2=\mu_3=0\quad &\Longrightarrow &\quad
L={\bf 1}_3,\\
\n
\mu_1=-\frac{2}{3},\mu_2=\frac{1}{3},\mu_3=\frac{1}{3} &\Longrightarrow &
L=e^{\frac{2\pi i}{3}}\,{\bf 1}_3,\\
\n
\mu_1=-\frac{1}{3},\mu_2=-\frac{1}{3},\mu_3=\frac{2}{3} &\Longrightarrow &
L=e^{-\frac{2\pi i}{3}}\,{\bf 1}_3.
\eea
Trivial holonomy corresponds to equal $\mu$'s, {\it modulo} unity. Out of all possible
combinations of $\mu$'s a distinguished role is played by {\em equidistant} $\mu$'s
corresponding to $\Tr L=0$:
\beq
\mu_m^{\rm conf}=-\frac{1}{2}-\frac{1}{2N}+\frac{m}{N}.
\la{muconf}\eeq
For example, in $SU(3)$ it is
\bea\n
&&\mu_1=-\frac{1}{3},\mu_2=0,\mu_3=\frac{1}{3}\\
&&\Longrightarrow
L={\rm diag}\left(e^{-\frac{2\pi i}{3}},1,e^{\frac{2\pi i}{3}}\right).
\la{muconf3}\eea
We shall call it ``most non-trivial'' or ``confining'' holonomy.

Immediately, an interesting question arises: Imagine we take the YM partition
function \ur{Zph1} and integrate out all degrees of freedom except
the eigenvalues $\{\mu_m\}$ of the Polyakov loop $L({\bf x})$
which, in addition, we take slowly varying in space. What is the effective action
for $\mu$'s? What set of $\mu$'s is preferred dynamically by the YM system of fields?

\begin{figure}[h]
\includegraphics[width=0.40\textwidth]{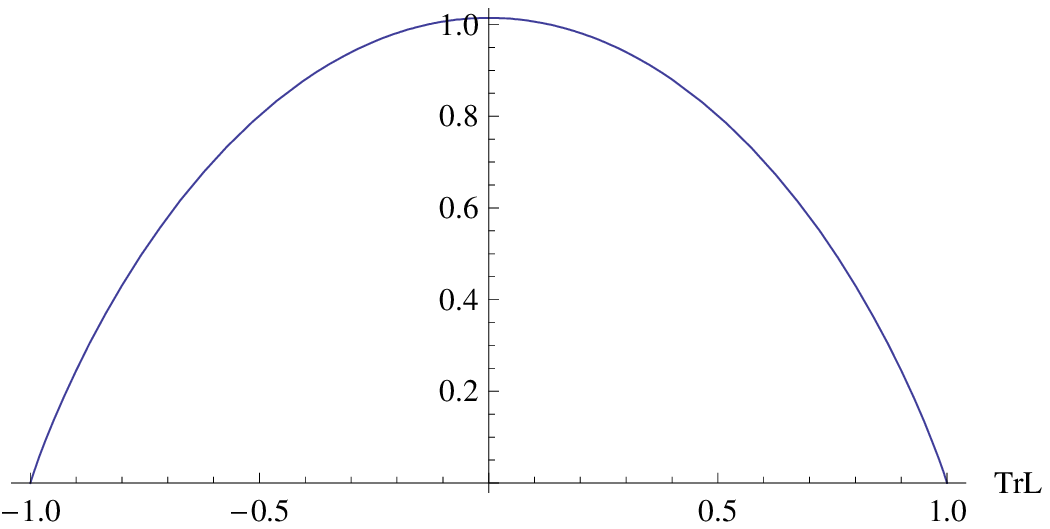} 
\includegraphics[width=0.40\textwidth]{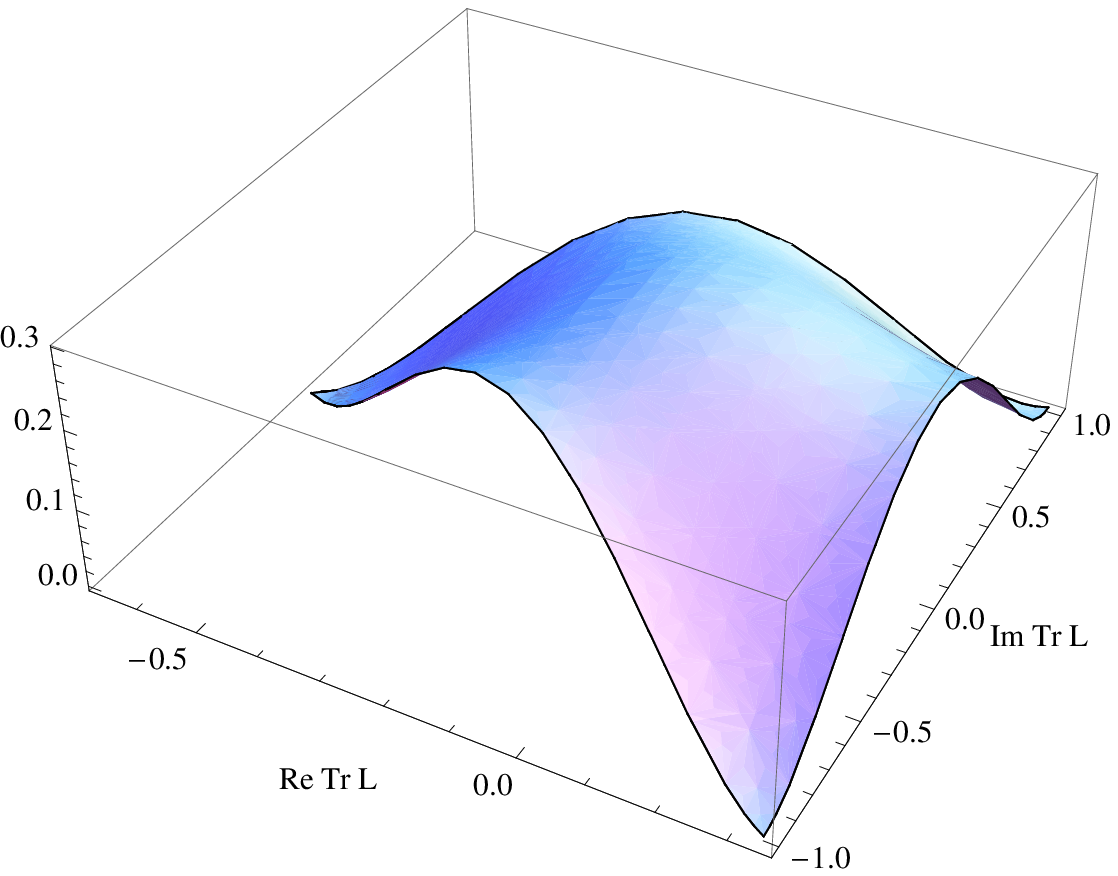}  
\caption{The perturbative potential energy
as function of the Polyakov line for the $SU(2)$ ({\it top}) and $SU(3)$ ({\it bottom}) groups.
It has minima where the Polyakov loop is one of the $N$ elements
of the center $Z_N$ and is maximal at the ``confining'' holonomy.}
\end{figure}

First of all, one can address this question in perturbation theory: the result
for the potential energy as function of $\mu$'s is~\cite{GPY,NW}
\beq
P^{\rm pert}\!=\!
\frac{(2\pi)^2T^3}{3}\!\sum_{m>n}^N\!
(\mu_m\!-\!\mu_n)^2[1\!-\!(\mu_m\!-\!\mu_n)]^2\!\!.
\la{Ppert}\eeq
It is proportional to $T^3$ (by dimensions) and has exactly $N$ zero minima
when all $\mu$'s are equal {\it modulo} unity, see Fig.~8. Hence, $P^{\rm pert}$
says that at least at high temperatures the system prefers one of the
$N$ trivial holonomies corresponding to the Polyakov loop being one of the
$N$ elements of the center $Z_N$. However, terms with gradients of $\mu$'s in
the effective action become negative near ``trivial'' holonomy, signalling
its instability even in perturbation theory~\cite{DO}.

\begin{figure}[h]
\includegraphics[width=0.40\textwidth]{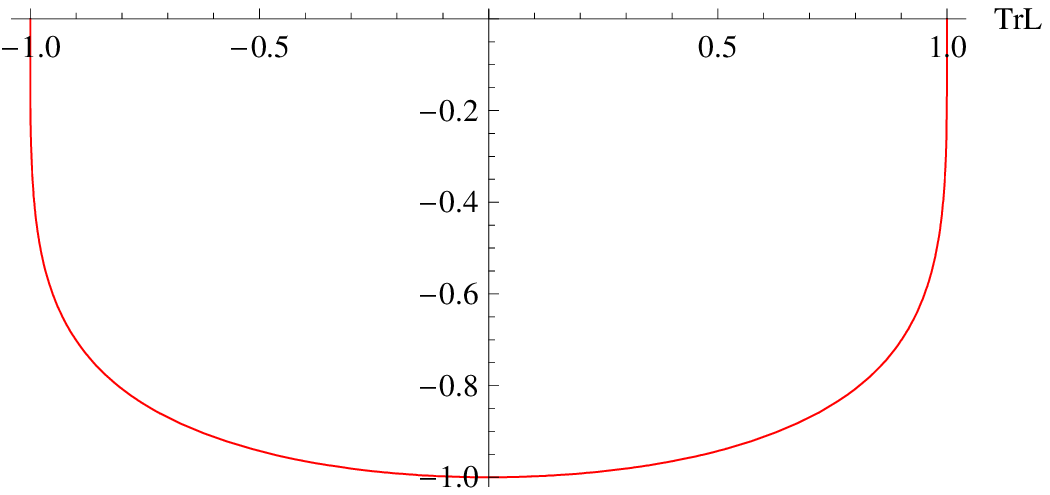} 
\includegraphics[width=0.40\textwidth]{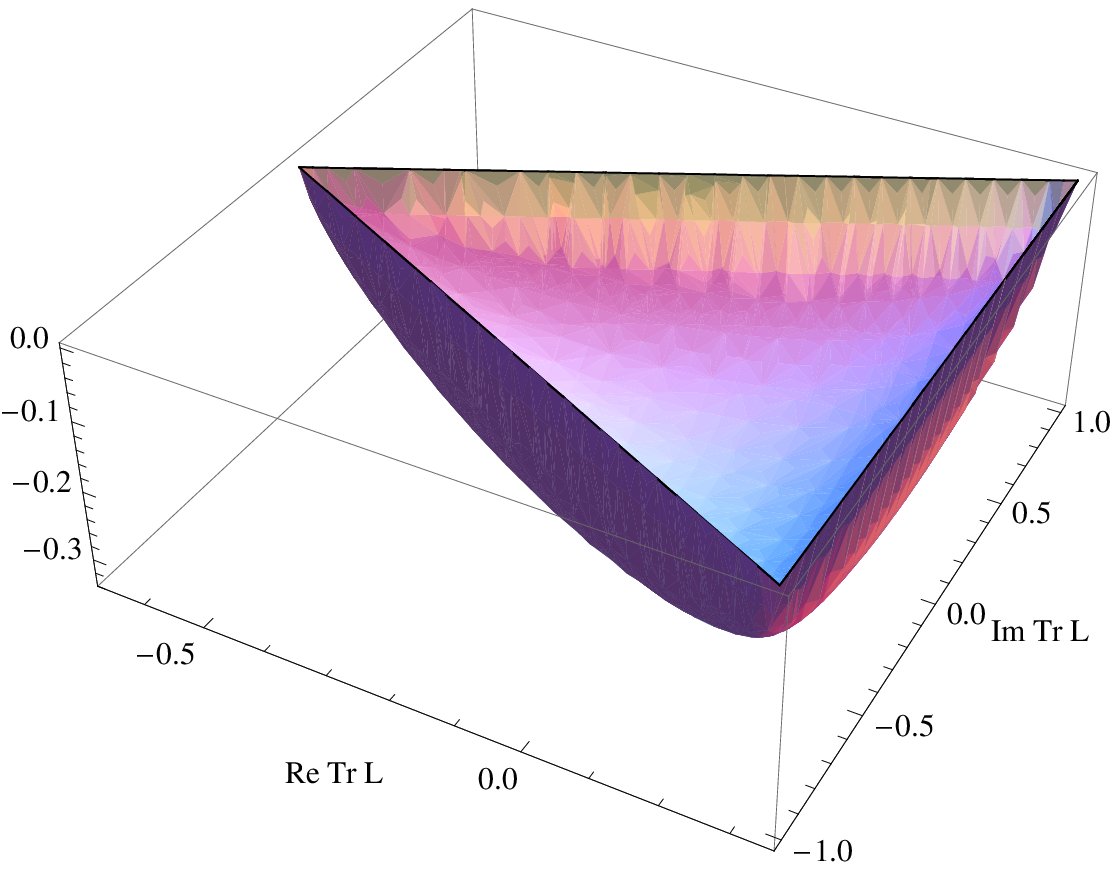}  
\caption{Dyon-induced nonperturbative potential energy
as function of the Polyakov line for the $SU(2)$ ({\it top}) and $SU(3)$ ({\it bottom}) groups.
Contrary to the perturbative potential energy, it has a single and non-degenerate
minimum at the confining holonomy corresponding to $\Tr L=0$.}
\end{figure}

It is interesting that in the supersymmetric ${\cal N}\!=\!1$ version
of the YM theory (where in addition to gluons there are spin-$\half$ gluinos
in the adjoint representation) the perturbative potential energy \ur{Ppert}
is absent in all orders owing to fermion-boson cancelation, but the
nonperturbative potential energy is nonzero. Moreover, it is known exactly
as function of $\mu$'s~\cite{DHKM-99}: it has a single minimum at precisely
the ``most non-trivial'' or ``confining'' holonomy \ur{muconf}.
The result can be traced to the semiclassical contribution of
dyons, which turns out to be exact owing to supersymmetry.

In the non-supersymmetric pure YM theory, the dyon-induced contribution
cannot be computed exactly but only in the semiclassical approximation (this is
what the rest of the paper is about), and the perturbative contribution \ur{Ppert} is present, too.
We shall show below that a semiclassical configuration -- an ensemble of dyons with quantum
fluctuations about it -- generates a nonperturbative free energy shown in Fig.~9.
It has the opposite behavior of the perturbative one, having
the minimum at the equidistant (confining) values of the $\mu$'s. There is a fight between
the perturbative and nonperturbative contributions to the free energy~\cite{D-02}.
Since the perturbative contribution to the free energy is $\!\sim\!T^4$ with respect to the nonperturbative one,
it certainly wins when temperatures are high enough, and the system is then forced into one
of the $N$ vacua thus breaking spontaneously the $Z_N$ symmetry. At low temperatures the
nonperturbative contribution prevails forcing the system into the confining vacuum.
At a critical $T_c$ there is a confinement-deconfinement phase transition. It turns out to
be of the second order for $N\!=\!2$ but first order for $N\!=\!3$ and higher, in agreement
with lattice findings.

For the centerless $G(2)$ group, there is a $1^{\rm st}$ order deconfinement transition
(also in agreement with the lattice) although it is not associated with any symmetry.

\section{CONFINEMENT CRITERIA}

Confinement, as we understand it today and learn from lattice experiments with
a pure glue theory, has in fact many facets, and all have to be explained.
For example, in a general $SU(N)$ group one can consider ``quarks'' in
various irreducible representations. From the confinement viewpoint all
representations are characterized by the phase it acquires under the
gauge transformation from the group center. The representation is said to
have ``$N$-ality $=k$'' if the phase is $\frac{2\pi k}{N}$.
Let us formulate mathematically the main confinement requirements that
need to be satisfied:
\begin{itemize}
\item the average Polyakov line in any $N$-ality nonzero representation
of the $SU(N)$ group is zero below $T_c$ and nonzero above it
\item the potential energy of two static color sources (defined through
the correlation function of two Polyakov lines) asymptotically rises linearly
with the separation; the slope called the string tension depends only on
the $N$-ality of the sources
\item the average of the spatial Wilson loop decays exponentially with
the area spanning the contour; at vanishing temperatures the spatial (``magnetic'') string
tension has to coincide with the ``electric'' one, for all representations
\item the mass gap: no massless gluons should be left in the spectrum; the massive glueballs
must exhibit the Hagedorn spectrum, {\it i.e.} an exponentially rising density of states
as function of mass
\item no Stefan--Boltzmann law for the free energy, $F\neq N^2T^4$; instead
the free energy should be ${\cal O}(N^0)$ and very small below $T_c$.
\end{itemize}

\section{SADDLE-POINT GLUON FIELDS}

The dynamics of the quantum YM system is governed by the YM partition function \ur{Zph1} or
\ur{Zph2}. A reasonable question to ask is whether there are field configurations
$A_\mu^a({\bf x},t)$ that give relatively large contributions to the partition function.
Naturally, when we speak about certain field configurations we always imply that there
are zero-point quantum fluctuations around, as one cannot stop them.

The privileged field configurations are apparently those that are local maxima
of the weight for a given configuration, $e^{-S[A]}$ or local minima of the
YM action $S[A]$. They are called saddle-point fields and satisfy the non-linear Maxwell equation,
$$
\frac{\delta S}{\delta A_\mu^a(x)}=0\qquad {\rm or}\quad D_\mu^{ab}F^b_{\mu\nu}=0,
$$
and have finite action $S=\frac{1}{4g^2}\int\!d^4x\,F^a_{\mu\nu}F^a_{\mu\nu}<\infty$.

\subsection{Topological classification of classical configurations}

According to Gross, Pisarski and Yaffe~\cite{GPY}, classical configurations with
a finite action can be classified by the following numbers:\\

\noi
1. Topological charge
$$
Q_{\rm T}=\frac{1}{16\pi^2}\int\!d^4x\,\epsilon^{\kappa\lambda\mu\nu}\,
F^a_{\kappa\lambda}(x)\,F^a_{\mu\nu}(x).
$$
2. Holonomy, more precisely the gauge-invariant eigenvalues
of the Polyakov loop at spatial infinity:
$$
{\rm eigenvalues\;of}\quad L={\cal P}\exp\left(i\int_0^{1/T^{\rm o}}\!dt\,A_4({\bf x},t)\right).
$$
3. Magnetic charge, more precisely the gauge-invariant eigenvalues
of the chromo-magnetic flux at spatial infinity:
$$
{\bf B}=\frac{{\bf r}}{|{\bf r}|^3}\times\left(\begin{array}{ccc}
1&0&0\\0&-1&0\\0&0&0\end{array}\right),\quad {\rm etc.}
$$

\subsection{Generalization of standard instantons to $T\neq 0$}

In section 2 we have introduced the standard Belavin--Polyakov--Schwarz--Tyupkin (BPST)
instanton that is a saddle-point YM field configuration at zero temperature.
It is easy to generalize it to nonzero temperatures: to that end one has to make
the fields periodic in the Euclidean time direction, with the period $\beta=\frac{1}{T}$.
This has been done by Harrington and Shepard~\cite{HS}. According to the classification of the
previous subsection one obtains a configuration that has $Q_T=1$ and trivial holonomy,
as the instanton field falls off fast enough at spatial infinity, see \Eq{instgena}.
For the same reason the magnetic charge of such configuration is zero.
It has not $O(4)$ but only the $O(3)$ symmetry: it is a `ball' in $3d$ but squeezed by the
periodic conditions in the time direction, see Fig.~10.

\begin{figure}[htb]
{\epsfig{figure=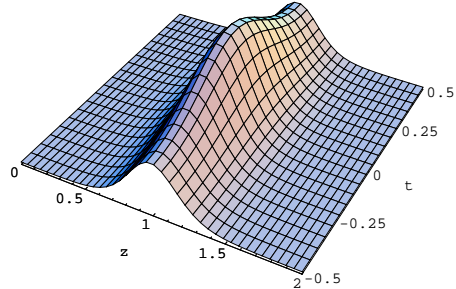,width=6cm}}
\caption{A typical action density of the periodic instanton with trivial holonomy as function of
$z,\;t$ at fixed $x=y=0$. It is $O(3)$ symmetric in space.}
\end{figure}

The 1-loop weight of the periodic instanton (also called the caloron) has been computed
by Gross, Pisarski and Yaffe~\cite{GPY} and has the form similar to \Eq{1instw1}:
\bea\la{caloron_weight}
&&\int_0^{\frac{1}{T}} \!dz_4\int\!d^3z\!\int\!d^3U\!\int\!\frac{d\rho}{\rho^5}\,
\frac{C(2)}{4\pi^2}\left(\frac{8\pi^2}{g^2(\mu)}\right)^4\\
\n
&&\cdot(\Lambda\rho)^{\frac{22}{3}}(\pi\rho T)^{\frac{8}{3}}\,e^{-\frac{4}{3}(\pi\rho T)^2}\quad {\rm at}\; \pi\rho T\gg 1.
\nea
Important: large-size calorons are suppressed by the Debye screening of the chromoelectric field,
represented by the last factor in \Eq{caloron_weight}.

If the temperatures are high enough, the caloron sizes are, according to \Eq{caloron_weight},
strongly suppressed. Therefore, a `dilute gas' of many calorons and anti-calorons is an
approximate saddle point of the partition function, and one can easily evaluate it in that
approximation. At lower temperatures the approximation fails, however one can still evaluate
the partition function using the variational principle developed in Ref.~\cite{DP-84} for
zero-temperature instantons and applied to calorons in Ref.~\cite{DMir}.

Do Harrington--Shepard calorons with trivial holonomy lead to confinement?
In Fig.~11 we show the average $<\!\!\Tr L\!\!>$ as function of temperature.

\begin{figure}[htb]
{\epsfig{figure=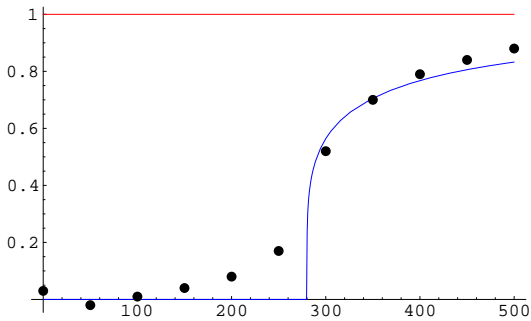,width=7cm}}
\caption{The average Polyakov line (big dots) as function of temperature in MeV, for the vacuum built of
periodic Harrington--Shepard calorons with trivial holonomy (D.~Diakonov and V.~Petrov, 1998, unpublished).
The solid line is drawn approximately through lattice data for the $SU(2)$ gauge group (the error bars are not shown).}
\end{figure}

It can be seen that it qualitatively follows the lattice data rising fast at $T\approx\Lambda$,
however $<\!\Tr L\!>$ at low temperatures is close to but not strictly zero;
there is no abrupt deconfinement phase transition either. Therefore calorons with trivial
holonomy do not and can not explain confinement. The reason is that periodic instantons have
$A_4=0$ at spatial infinity, hence a strict zero for $<\!\Tr L\!>$ needed for confinement
can be achieved only through a miraculous cancelation, but it does not happen.

The conclusion is that a modification of periodic instantons such that the field has
$A_4=\pi T\,\frac{\tau_3}{2}$ at infinity, leading to $\Tr L=0$, would be helpful to obtain confinement.

A conceptual difficulty, however, is that a nonzero $A_4$ is forbidden by the perturbative potential
\ur{Ppert}. The philosophy is then~\cite{D-02} that one has to admit that $A_4$ is nonzero at low temperatures,
look for saddle-point configurations with a nontrivial holonomy, and check that they induce a nonperturbative potential
that indeed has a minimum at a needed `most nontrivial' value of $A_4$.

\section{BOGOMOLNY--PRASAD--SOMMER\-FIELD (BPS) MONOPOLE}

BPS monopole or dyon is a static, time-independent self-dual solution of the YM equation of
motion $D_\mu F_{\mu\nu}=0$ having a finite action and topological charge, possessing both chromoelectric
and chromomagnetic charges, and corresponding to nontrivial eigenvalues of the Polyakov line (or holonomy)
at spatial infinity.

These are gauge invariant characterizations of the solution, however to find it explicitly one
needs to fix the gauge. We first use the so-called `hedgehog gauge' where the solution is $O(3)$ symmetric
and time-independent. Later we discuss BPS monopoles in other gauges.

The key point is the nontrivial holonomy, {\it i.e.} nontrivial eigenvalues of the
Polyakov line far away from the center of the BPS monopole. To ensure it, we choose the
gauge where $A_4$ is static and its modulus is constant at spatial infinity. BPS monopole is basically
an $SU(2)$ construction, therefore we first build it for the $SU(2)$ gauge group and then
generalize it to higher-rank groups by embedding.

We look for the monopole solution in the form
\bea\n
A_4^a(x)&=&n_a\frac{E(r)}{r},\\
\n\\
\n
A_i^a(x)&=&\e_{aij}n_j\frac{1-A(r)}{r}+(\delta_{ai}-n_an_i)\frac{B(r)}{r}\\
&+&n_an_i\frac{C(r)}{r},
\la{BPS1}\eea
where $n_i$  is a unit vector in 3-space and $A,B,C,E$  are functions of
the distance $r$ from the origin only. They are all dimensionless.

There is a time-independent gauge transformation which leaves the form
of the above Ansatz unchanged. Indeed, if we perform a gauge
transformation
$$
A_\mu \rightarrow UA_\mu U^\dagger + i U\partial_\mu U^\dagger,\;\;\;\;
U= \exp\left(i(n\cdot\tau)P(r)\right),
$$
\noi
we remain inside the Ansatz but with the transformed functions:
\bea\n
A&\rightarrow& A\cos 2P -B \sin 2P,\\
\n
B&\rightarrow& A\sin 2P + B\cos 2P,\\
C&\rightarrow& C+2rP^\prime, \;\;\;\;E\rightarrow E.
\nea

Let us introduce the electric ($E$) and magnetic ($B$) field strengths:
$$
F_{\mu\nu}^a=\partial_\mu A_\nu^a-\partial_\nu A_\mu^a
+\e^{abc}A_\mu^bA_\nu^c,
$$
$$
E_i^a=F_{i4}^a=D_i^{ab}A_4^b,\;\;\;\;B_i^a=\frac{1}{2}\e_{ijk}F_{jk}^a.
$$

\noi
Substituting the potentials we get for the electric and magnetic fields:
\bea\n
E_i^a\!\!\!\!&=&\!\!\!\!\e_{aij}n_j\frac{B}{r}\frac{E}{r}
\!+\!(\delta_{ai}\!-\!n_an_i)\frac{A}{r}\frac{E}{r}\!+\!
n_an_i\frac{d}{dr}\frac{E}{r},\\
\n
B_i^a\!\!\!\!&=&\!\!\!\!\e_{aij}n_j\!\left(\!\frac{B^\prime}{r}\!-\!\frac{AC}{r}\!\right)\!
+\!(\delta_{ai}\!-\!n_an_i)\!\left(\!\frac{A^\prime}{r}\!-\!\frac{BC}{r}\!\right)\\
\n
\!\!\!&+&\!\!\!n_an_i\frac{A^2+B^2-1}{r^2}.
\nea
\noi
The correspondent field strengths squared are
\bea\n
\left(E_i^a\right)^2&=&\left(\frac{d}{dr}\frac{E}{r}\right)^2
+2\frac{E^2R^2}{r^4},\\
\n
\left(B_i^a\right)^2&=&2\frac{R^{\prime 2}}{r^2}+2\frac{R^2D^2}{r^4}+\frac{(R^2-1)^2}{r^4},
\nea
where we have parameterized the functions $A,B,C$  as follows:
$$
A=R\cos\alpha,\;\;\;B=R\sin\alpha,\;\;\;C=D+r\alpha^\prime.
$$
\noi
Under this gauge transformation $\alpha(r)\rightarrow \alpha(r)+2P(r)$
while the functions $R(r)$ and $D(r)$ rest invariant,
hence only they appear in the gauge invariant expressions.
\noi
The Euclidean energy functional is
$$
{\cal E}=\frac{1}{2}\int d^3 {\bf r}\left[\left(E_i^a\right)^2+\left(B_i^a\right)^2\right] .
$$

Varying this functional with respect to the gauge invariant functions $D,E$ and $R$ one finds the equations of motion:
\bea\n
E^{\prime\prime}&=&E\frac{2R^2}{r^2},\qquad D(r)=0,\\
R^{\prime\prime}&=&R\frac{E^2+R^2-1}{r^2}.
\nea
Since $D(r)$ occurs to be zero one can choose the gauge transformation
such that $\alpha(r)=0,\;B(r)=0,\;C(r)=0$, and $A(r)=R(r)$.
A solution of these equations was first found (numerically) by Prasad and Sommerfield.
Later Bogomolny showed that the solutions can be easily found analytically:
\bea\n
A(r)\!\!\!&=&\!\!\!R(r)=\frac{v r}{\sinh v r} =
\left\{\!\! \begin{array}{ll}
1\!-\!\frac{(v r)^2}{6},\;\;r\rightarrow 0,\\
O\left( e^{-v r}\right),\;\;r\rightarrow \infty \end{array}\right.\\
\n
E(r)\!\!\!&=&\!\!\! \pm \left(1\!-\!v r \coth v r\right)\\
\n
&=&
\left\{ \begin{array}{ll}
\mp\frac{v^2 r}{3},\;\;r\rightarrow 0,\\
\mp\left(v r\!-\!1\!+\!O\left(\!e^{-v r}\!\right)\!\right),\;\;r\rightarrow \infty \end{array}\right. .
\nea
\noi
This analytical solution can be found in a simple way, by solving
a {\em first} order equation. To see that, we notice that the
energy functional satisfies the so-called  Bogomolny inequality:
\bea\n
{\cal E}\!\!\!&=&\!\!\! \left\{\!\begin{array}{ll}
\frac{1}{2g^2}\int d^3r\left(E_i^a\!-\!B_i^a\right)^2\!+\!
\frac{1}{g^2}\int d^3r E_i^aB_i^a \\
\frac{1}{2g^2}\int d^3r\left(E_i^a\!+\!B_i^a\right)^2\!-\!
\frac{1}{g^2}\int d^3r E_i^aB_i^a
\end{array}\right.\\
\n
\!\!\!&\geq &\!\!\!\frac{1}{g^2}\left|\int d^3 r E_i^aB_i^a \right|.
\nea
The equality is achieved when the fields satisfy the Euclidean
self-duality (or anti-self-duality) equations,
$$
F_{\mu\nu}^a=\pm \frac{1}{2}\e_{\mu\nu\alpha\beta}F_{\alpha\beta}^a
\;\;\;\;\mbox{ or}\;\;\;\; E_i^a=\pm B_i^a.
$$
It is easy to check that the two solutions satisfy
exactly these equations, so that the upper sign  is in
correspondence with the upper sign in the previous equation and {\em vice versa}.
In both cases the energy of the solution can be computed as
\bea\n
&&{\cal E}_{\rm mon}=\frac{1}{g^2}\left|\int d^3 r E_i^aB_i^a \right|\\
\n
&&=\frac{4\pi}{g^2}\left|\int dr \frac{d}{dr}\frac{(1-A^2)E}{r}\right|
=\frac{4\pi v}{g^2},\;\;\;\;v>0 .
\nea

Let us write down explicitly the electric and magnetic field strengths:
\bea\n
B_i^a&=&(\delta_{ai}-n_an_i)\frac{v}{\sinh v r}
\left(\frac{1}{r}-v \coth v r\right)\\
\n
&&+n_an_i
\left(-\frac{1}{r^2}+\frac{v^2}{\sinh^2 v r}\right),\\
E_i^a&=&\pm B_i^a.
\nea

We see that both electric and magnetic field strengths fall off at
spatial infinity as $1/r^2$; this behaviour is characteristic of the
fields of electric and magnetic charges. That is why the BPS monopole
is often called `dyon'.

\subsection{Four dyons of the $SU(2)$ gauge group ($M,\bar M,L,\bar L$)}

\noi
In $SU(2)$ there are four dyons with all four possible signs
of the electric and magnetic charges. The two usual
BPS dyons in the regular (`hedgehog') gauge have the form:
\bea\n
A_4^a &=& \mp n_a\,v\Phi(vr),\\
\n
\Phi(z)&=&\coth z\! -\!\frac{1}{z}\quad
\stackrel{z\to\infty}{\longrightarrow}\quad 1\!-\!\frac{1}{z}\!+\!O(e^{-z}),\\
\nonumber \\
\n
A_i^a &=& \epsilon_{aij}n_j\,\frac{1-R(vr)}{r},\\
\n
R(z)&=&\frac{z}{\sinh z}\quad \stackrel{z\to\infty}{\longrightarrow}\quad O(ze^{-z}).
\nea
The upper sign in $A_4$ corresponds to the self-dual ($E^a_i=B^a_i$) and the lower sign
to the anti-self-dual ($E^a_i=-B^a_i$) solution.
We shall call them $M$- and $\bar M$-monopoles, respectively.
\noi
The magnetic field strength in the hedgehog gauge is given by two structures:
$$
B^a_i = (\delta_{ai}-n_an_i)\,F_1(r)+n_an_i\,F_2(r)
$$
where
\bea\la{F1}
F_1(r)\!\!\! &=&\!\!\! \frac{1}{r}\frac{d}{dr}R(vr)=-v\frac{R(vr)\Phi(vr)}{r}\\
\n
\!\!\!&=&\!\!\!\frac{v^2}{\sinh(vr)}\!\left(\!\frac{1}{vr}\!-\!\coth(vr)\!\right)=v^2O\left(e^{-vr}\right),\\
\la{F2}
F_2(r)\!\!\! &=&\!\!\! -\frac{d}{dr}v\Phi(vr)=\frac{R^2(vr)\!-\!1}{r^2}\\
\n
\!\!\!&=&\!\!\!\frac{v^2}{\sinh(vr)}-\frac{1}{r^2}
\!=\!-\frac{1}{r^2}\!+\!v^2O\left(e^{-2vr}\right).
\nea

If there is more than one monopole in the vacuum it is impossible to
add them up in the hedgehog gauge: one has to ``gauge-comb'' them to
a gauge where $A_4^a$ has the same asymptotic value at spatial
infinity for all monopoles involved, -- say, along the third colour
axis. It is achieved with the help of two unitary matrices dependent
on the spherical angles $\theta,\phi$:
\bea\n
S_+(\theta,\phi) &=& e^{-i\frac{\phi}{2}\tau^3}
e^{i\frac{\theta}{2}\tau^2}e^{i\frac{\phi}{2}\tau^3},\\
\n
&&S_+(n\cdot \tau)S_+^\dagger=\tau^3,\\
\n
S_-(\theta,\phi) &=& e^{i\frac{\phi}{2}\tau^3}
e^{i\frac{\pi-\theta}{2}\tau^2}e^{i\frac{\phi}{2}\tau^3},\\
\n
&&S_-(n\cdot \tau)S_-^\dagger=-\tau^3.
\nea

\noi
We shall gauge-transform the $M$-monopole field with $S_-$ and the $\bar M$-monopole with $S_+$.
As the result their $A_4$ components become equal:
$$
A_4^{M,\bar M}=v\Phi(vr)\frac{\tau^3}{2}=\left[v-\frac{1}{r}
+O\left(e^{-vr}\right)\right]\frac{\tau^3}{2}.
$$
On the contrary, the spatial components differ in sign. We write
them in spherical components:
$$
\pm A_i^{M,\bar M}=\left\{\begin{array}{ccc}
A_r &=&\!\!\!\!\!\!\!\!\!\!\!\!\!\!\!\!\!\!\!\!\!\!\!\!\!\!\!\!\!\!\!\!\!\!\!\!\!\!\! 0, \\
\nonumber\\
A_\theta &=& \frac{R(vr)}{2r}(\tau^1\,\sin\phi+\tau^2\,\cos\phi),\\
\nonumber\\
A_\phi &=& \frac{R(vr)}{2r}(\tau^1\,\cos\phi-\tau^2\,\sin\phi)\\
\n\\
&+&\!\!\!\!\!\!\!\!\!\!\!\!\!\!\!\!\!\!\!\!\!\!\!\!\!\!\!\!\!\!\!\!\!\!\!\!\!\!\!\!\! \frac{1}{2r}\tan\frac{\theta}{2}\tau^3.
\end{array}\right.
$$

\noi
The azimuthal component of the gauge field has a singularity along
the negative $z$ axis, therefore we shall call it the {\em ``stringy gauge''}.
The field strength, however, has no singularities. The electric field
both of $M$ and $\bar M$ monopoles in the stringy gauge is
$$
E_i^{M,\bar M}=\left\{\begin{array}{ccc}
E_r &=&\!\!\!\!\!\!\!\!\!\!\!\!\! - \frac{F_2(r)}{2}\tau^3
\; \stackrel{r\to\infty}{\longrightarrow}
\; \frac{1}{r^2}\frac{\tau^3}{2} \\
\nonumber \\
\n
E_\theta &=& \frac{F_1(r)}{2}(-\tau^1\,\cos\phi+\tau^2\,\sin\phi)\\
\n\\
&=&\!\!\!\!\!\!\!\!\!\!\!\!\!\!\!\!\!\!\!\!\!\!\!\!\!\! v^2 O\left(e^{-vr}\right)\\
\nonumber \\
\n
E_\phi &=& \frac{F_1(r)}{2}(\tau^1\,\sin\phi+\tau^2\,\cos\phi)\\
\n\\
&=&\!\!\!\!\!\!\!\!\!\!\!\!\!\!\!\!\!\!\!\!\!\!\!\!\!\! v^2 O\left(e^{-vr}\right)
\end{array}\right.
$$
while the magnetic field is $B_i=\pm E_i$. We see that the $\theta,\phi$ components
are non-Abelian but exist only inside the dyon core, whereas the radial component
is Coulomb-like and Abelian as it has only the $3^{\rm d}$ colour component.
Therefore, $M$ monopole has Abelian (electric, magnetic) charges $(++)$ whereas the
$\bar M$ one has $(+-)$. \\

There is a second pair of dyons: a self-dual one with the
charges $(--)$ which we shall name $L$-monopole, and an
anti-self-dual one with charges $(-+)$ which we shall name $\bar L$
monopole. They are obtained from the above equations by replacing $v\to 2\pi T-v$.
One first transforms them from the hedgehog
to the stringy gauge with the help of the unitary matrices $S_+$ and $S_-$, respectively.
As a result, they get the same asymptotics $A_4(\infty)=\left(-2\pi T+v
+\frac{1}{r}\right)\frac{\tau^3}{2}$. To put the asymptotics in the
same form as for $M,\bar M$-monopoles one makes an
additional gauge transformation with the help of the time-dependent
matrix
$$
U=\exp\left(-i\pi T x^4\tau^3\right).
$$
This gives the following fields of $L,\bar L$ monopoles in the
stringy gauge:
\bea\n
A_4^{L,\bar L}\!\!\!&=&\!\!\!\left[\left(2\pi T-v\right)
\Phi\left(\left|2\pi T-v\right|r\right)-2\pi T\right]
\frac{\tau^3}{2}\\
\n
& \stackrel{r\to\infty}{\longrightarrow} &
=\left(v+\frac{1}{r}\right)\frac{\tau^3}{2},
\nea

\bea\n
E_{r,\theta,\phi}^{L,\bar L} &=& \left\{\begin{array}{ccc}
\frac{F_2(r)}{2}\tau^3
\quad \stackrel{r\to\infty}{\longrightarrow}
\quad -\frac{1}{r^2}\frac{\tau^3}{2} \\
\nonumber \\
\n
\!\!-\frac{F_1(r)}{2}\,U(-\tau^1\cos\phi+\tau^2\sin\phi)U^\dagger\\
\nonumber \\
\n
\!\!-\frac{F_1(r)}{2}\,U(\tau^1\sin\phi+\tau^2\cos\phi)U^\dagger,
\end{array}\right. \\
\nonumber \\
\n
B_i^{L,\bar L} &=& \pm E_i^{L,\bar L}.
\nea

\noi
The `profile' functions $F_{1,2}$ are given by \Eqs{F1}{F2}, with
the replacement $v\to 2\pi T-v$. We notice that ``the
interior'' of the $L,\bar L$ dyons, represented by the $\theta,\phi$
field components are {\it time dependent}. This is why in the true $3d$
case these objects do not exist!

Properties of the four fundamental
dyons of the $SU(2)$ group are summarized in the Table:
\begin{table}[h]
\begin{center}
\begin{tabular}{|c|c|c|c|c|}
\hline
&&&& \\
                & $M$ & $\bar M$ & $L$ & $\bar L$  \\
&&&& \\
\hline
&&&& \\
 e-charge      &  +  &  +     &  $-$   &   $-$      \\
&&&& \\
 m-charge    & + &  $-$       &   $-$  &   +      \\
&&&& \\
 action, $\frac{1}{\alpha_s\,T}$ & $v$ & $v$ & $2\pi T\!-\!v$ &
$2\pi T\!-\!v$  \\
&&&& \\
top-charge & $\frac{v}{2\pi T}$ & $\!-\frac{v}{2\pi T}$ & $1\!-\!\frac{v}{2\pi T}$ & $\frac{v}{2\pi T}\!-\!1$  \\
&&&& \\
\hline
\end{tabular}
\end{center}
\end{table}

\subsection{Dyons of higher-rank gauge groups}

Dyons or BPS monopoles are basically $SU(2)$ constructions, therefore if one wants to construct dyon
solutions for higher-rank gauge groups one has to embed the $SU(2)$ construction of the previous subsection
into a higher-rank group. This, of course, can be done by infinitely many ways; the problem is to
construct a set of `fundamental' dyons from which all other types can be built as `bound states'.
The problem has been addressed in Ref.~\cite{DHK} for all Lie groups. Here we shall concentrate on
the $SU(N)$ series, and later add some remarks on the exceptional $G(2)$ group which is interesting both
from the mathematical and physical points of view since, contrary to $SU(N)$, this group has only
a trivial center.

In $SU(N)$ there are exactly $N$ kinds of fundamental self-dual dyons (and $N$ anti-self-dual anti-dyons)
which can be described as follows. Let us introduce a basis of Cartan generators given by diagonal matrices
\bea\la{Cartan}
C_m&=&{\rm diag}(0,...,0,1,-1,0,...,0),\\
\n
&&m=1,...,N-1,
\eea
where $+1$ is in the $m^{\rm th}$ place, and supplement this set by the matrix $C_N={\rm diag}(-1,0,...,0,1)$,
such that $\sum_{m=1}^N C_m=0$. All $N$ types of dyon solutions are labeled by the holonomy or the
set of eigenphases $\mu$'s of the Polyakov line at spatial infinity. In the gauge where $A_4$
is time-independent it means the asymptotic field
\bea\la{A4}
\!\!\!&&\!\!\!A_4(\infty)= 2\pi T{\rm diag}(\mu_1,\mu_2,\ldots ,\mu_N),\\
\n
\!\!\!&&\!\!\!\mu_1\leq\mu_2\leq\ldots \mu_N\leq\mu_1+1,\quad \sum_{m=1}^N\mu_m=0.
\eea
We adopt cyclic-symmetric notations such that $\mu_{N+1}\equiv\mu_1+1$ and
introduce the differences
\beq
\nu_m\equiv\mu_{m+1}-\mu_m,\quad \sum_{m=1}^N\nu_m=1.
\la{nu}\eeq

The dyon of the $m^{\rm th}$ kind is an object of the $SU(2)$ whose generators are $C_m$
playing the role of the Pauli matrix $\tau^3$, and the two corresponding off-diagonal matrices
playing the role of the Pauli $\tau^\pm$ matrices. At the origin its $A_4$ component is
\bea\n
&&A_4^{(m)}(0)=2\pi T \,{\rm diag}\left(\mu_1,...\right.\\
\n
&&\left. ...\mu_{m-1},\frac{\mu_m\!+\!\mu_{m+1}}{2},\frac{\mu_m\!+\!\mu_{m+1}}{2},\mu_{m+2},...\right),
\eea
and its core (where the fields are strong and non-Abelian) has the size $\sim 1/(2\pi T\nu_m)$.
Beyond the core the fields are Abelian as they have only a diagonal component, and are Coulomb-like:
\bea\n
A_4^{(m)}&\stackrel{|{\bf x}|\!\to\!\infty}{=}&-C_m\,\frac{1}{2{|\bf x}|}+A_4(\infty),\\
\la{m-asympt}
\pm{\bf B}^{(m)}={\bf E}^{(m)}&\stackrel{|{\bf x}|\!\to\!\infty}{=}&C_m\,\frac{{\bf x}}{2|{\bf x}|^3}.
\eea
The upper sign for the magnetic field is for the self-dual dyons, and the lower sign is for the
anti-self-dual anti-dyons. In the gauge where $A_4$ is static, the last $N^{\rm th}$ dyon is obtained
through a time-dependent gauge transformation, like the `L' dyon of the previous subsection. It is
sometimes called the Kaluza--Klein monopole; in a true $3d$ theory it does not exist.

The action density is, however, time-independent for all $N$ monopoles. The full action and the
topological charge are temperature-independent:
\beq
S^{(m)}=\frac{2\pi}{\alpha_s}\nu_m,\qquad Q^{(m)}=\nu_m.
\la{dyon-action}\eeq
The full action of all $N$ kinds of well-separated dyons together are that of one standard instanton:
$S_{\rm inst}=2\pi/\alpha_s$. We also note that the total electric and magnetic
charge of a set of $N$ dyons is zero, and that the topological charge is unity.
Therefore, a set of $N$ kinds of well-separated dyons have the quantum numbers (and the action)
of one instanton. However, contrary to the standard instanton or its periodic generalization,
the holonomy is not trivial but arbitrary. Remarkably, there exists an exact classical solution
that preserves these properties even as one moves the $N$ dyons close to each other, see the next
section.

For the $G(2)$ group, there are three kinds of fundamental dyons (as in $SU(3)$) but a neutral combination
(`the instanton') is obtained from four dyons: one of the three kinds has to be taken twice.

In the semiclassical approach, one has first of all to find the statistical weight with which
a given classical configuration enters the partition function. It is given by $\exp(-{\rm Action})$,
times the (determinant)$^{-1/2}$ from small quantum oscillations about the saddle point. For
an isolated dyon as a saddle-point configuration, this factor diverges linearly in the infrared
region owing to the slow Coulomb decrease of the dyon field. It means that isolated
dyons are not acceptable as saddle points: they have zero weight, despite finite classical
action. However, one may look for classical solutions that are superpositions of $N$ fundamental
dyons, with zero net magnetic and electric charges. The small-oscillation determinant must be infrared-finite
for such classical solutions, if they exist.\\

\section{INSTANTONS WITH NONTRIVIAL HOLONOMY}

The needed classical solution has been found a decade ago by Kraan
and van Baal~\cite{KvB} and independently by Lee and Lu~\cite{LL},
see also~\cite{LeeYi}. We shall call them for short the ``KvBLL instantons''; an
alternative name is ``calorons with nontrivial holonomy''. The solution was first found
for the $SU(2)$ group but soon generalized to an arbitrary $SU(N)$~\cite{KvBSUN},
see~\cite{vBZak} for a review.

The general solution $A_\mu^{\rm KvBLL}$ depends on Euclidean time $t$ and
space ${\bf x}$ and is parameterized by $3N$ positions of $N$ kinds of
`constituent' dyons in space ${\bf x}_1,\ldots, {\bf x}_N$
and their $U(1)$ phases $\psi_1,\ldots,\psi_N$. All in all, there are $4N$ collective
coordinates characterizing the solution (called the moduli space), of which the action
$S_{\rm inst}=2\pi/\alpha_s$ is in fact independent, as it should be for a general
solution with a unity topological charge. The solution also depends explicitly on
temperature $T$

\begin{figure}[t]
\hspace{1.5cm}{\epsfig{figure=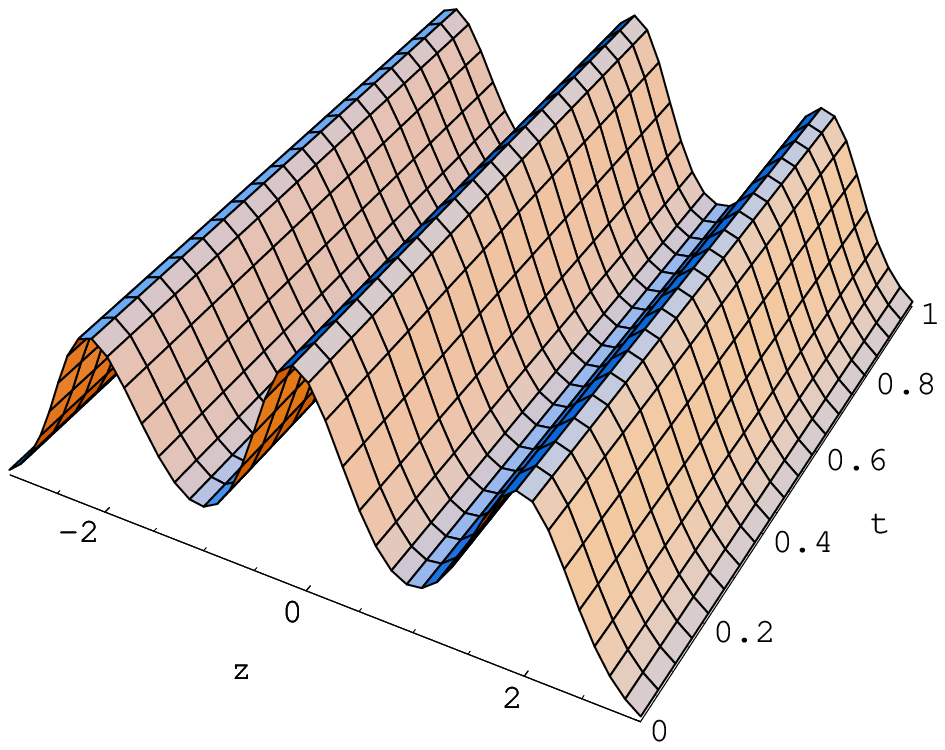,width=4cm}}

\hspace{1.5cm}{\epsfig{figure=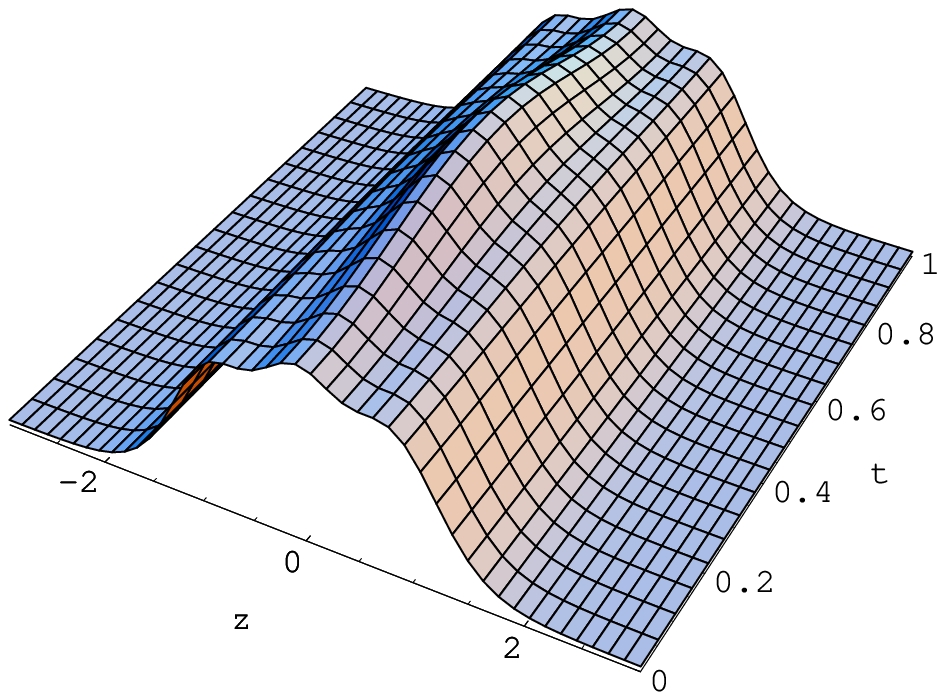,width=4cm}}

\hspace{1.5cm}{\epsfig{figure=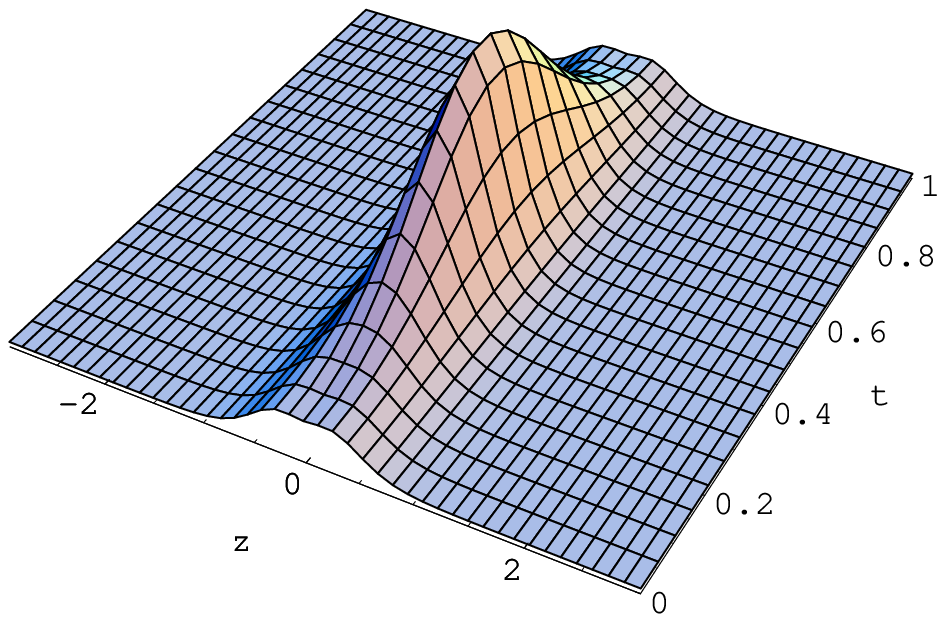,width=4cm}}
\caption{Action density inside the $SU(3)$ KvBLL instanton as function of time and
one space coordinate, for large ({\it top}), intermediate ({\it middle}) and small
({\it bottom}) separations between the three constituent dyons.}
\end{figure}

\noi
and on the holonomy $\mu_1,\ldots,\mu_N$:
\bea\n
&&A_\mu^{\rm KvBLL}=\bar A_\mu^a\left(t, {\bf x};\, {\bf x}_1,\ldots, {\bf x}_N,\psi_1,\ldots,\psi_N;\right.\\
\la{KvBLLi}
&&\qquad\qquad\qquad\;\;
\left. T,\mu_1,\ldots,\mu_N\right).
\eea
The solution is a relatively simple expression given by elementary functions.
If the holonomy is trivial (all $\mu$'s are equal {\it modulo} unity) the expression
takes the form of the strictly periodic $O(3)$ symmetric caloron~\cite{HS} reducing further
to the standard $O(4)$ symmetric BPST instanton~\cite{BPST} in the $T\to 0$ limit.
At small temperatures but arbitrary holonomy, the KvBLL instanton also has only a small
${\cal O}(T)$ difference with the standard instanton.

One can plot the action density of the KvBLL instanton in various corners of the
parameter (moduli) space, see Fig.~12.

When all dyons are far apart one observes $N$ static ({\it i.e.} time-independent)
objects, the isolated dyons. As they merge, the configuration is not static anymore,
it becomes a {\it process} in time. The reason why time dependence creeps in is
that the field of one of the $N$ kinds of dyons is time-dependent from the start
(in the notations of subsection 6.1 it is the L dyon). The time dependence of that
particular dyon is pure gauge rotation for isolated objects, therefore the action density
and other gauge-invariant quantities are time-independent. However, as we merge together
dyons and wish to support the property that the configuration remains to be a solution
of the equation of motion, time dependence ceases to be a gauge rotation but enters
the field in an essential way, see the next subsection.

In the limiting case of a complete merger, the configuration becomes a $4d$ lump resembling
the standard instanton. The full (integrated) action is exactly the same $S_{\rm inst}=2\pi/\alpha_s$
for any choice of dyon separations. It means that classically dyons do not interact. However,
they do experience a peculiar interaction at the quantum level, which we discuss in Section 8.

\subsection{Explicit form of the KvBLL instanton for the $SU(2)$ gauge group}

We follow in this subsection Ref.~\cite{KvB}. We consider a classical solution whose
asymptotic $A_4$ component is $A_4\rightarrow \v\frac{\tau^3}{2}$. In the notations of the
previous subsection $\v=4\pi T \mu_1$. We introduce also a notation $\bv=2\pi\T-\v$.

The KvBLL instanton is constructed as a combination of $L$ and $M$ monopoles, see
subsection 6.1. Let the $L$ dyon be centered at a $3d$ point ${\bf x}_1$ and the $M$ dyon is
at a point ${\bf x}_2$. The dyon separation is $r_{12}=|{\bf x}_1-{\bf x}_2|=\pi T\rho^2.$
Here $\rho$ is the size of the standard instanton (see subsection 2.4) to which the KvBLL instanton
reduces at $\v\to 0$ or at $T\to 0$.

We introduce the distances from the `observation point' ${\bf x}$ to the dyon centers,
${\bf r}={\bf x}-{\bf x}_1,\quad{\bf s}={\bf x}-{\bf x}_2$. Correspondingly, $r=|{\bf r}|,\;s=|{\bf s}|$.
We choose the separation between dyons to be along the third spatial direction, ${\bf r}_{12}={\bf e}_3 r_{12}$.

The KvBLL instanton field is
\bea\la{KvBLL_field}
A_\mu\!\!\! &=&\!\!\! \delta_{\mu,4}\,\v\frac{\tau^3}{2}
+\frac{1}{2}\bar\eta^3_{\mu\nu}\tau_3\partial_\nu\ln\Phi\\
\n
\!\!\!&+&\!\!\!\Mphi\;{\rm  Re}\left[\!(\!\bar\eta^1_{\mu\nu}\!-\!i\bar\eta^2_{\mu\nu}\!)
(\tau^1\!+\!i\tau^2)(\partial_\nu\!+\!i\v\delta_{\nu,4})\tilde\chi\!\right],
\nea
where $\tau^a$ are Pauli matrices, $\bar\eta^a_{\mu\nu}$ is the 't~Hooft
symbol (see subsection 2.4). The functions used are
\bea
\nn
&&\hat\psi= -\cos(2\pi T x^4)+\bchd\chd+\frac{{\bf r}\cdot{\bf s}}{2rs}\,\bshd\shd,\\
\n
&&\psi= \hat\psi+\frac{\D^2}{rs}\bshd\shd+\frac{\D}{s}\shd\bchd+\frac{\D}{r}\bshd\chd,\\
&&\tilde\chi= \frac{\D}{\psi}\left(e^{-2\pi i T x^4}\frac{\shd}{s}+\frac{\bshd}{r}\right),
\qquad\Mphi= \frac{\psi}{\hat\psi}\,.
\nea
\noi
We have introduced short-hand notations for hyperbolic functions:
\bea\n
\shd &\equiv &\sinh(s\v),\qquad \chd\equiv\cosh(s\v),\\
\n
\bshd &\equiv &\sinh(r\bv),\qquad \bchd\equiv\cosh(r\bv) \,.
\nea
The first term corresponds to
a constant $A_4$ component at spatial infinity ($ A_4 \approx  i\v\frac{\tau^3}{2} $)
and gives rise to the non-trivial holonomy.
One can see that $A_\mu$ is periodic in time with period $1/T$. A
useful formula for the field strength squared is
$$
\Tr\,F_{\mu\nu}F_{\mu\nu}=\partial^2\partial^2\log\psi.
$$

In the situation when the separation between dyons $\D$ is large
compared to both their core sizes $\frac{1}{\v}$ (M) and
$\frac{1}{\bv}$ (L), the caloron field can be approximated by the
sum of individual BPS dyons. To demonstrate it, we give below
the field inside the cores and far away from both cores.

\begin{figure}[h]
{\epsfig{figure=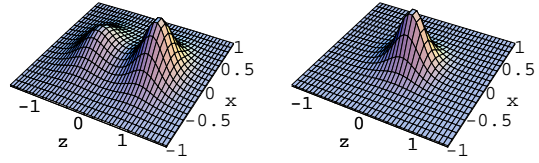,width=7.5cm}}
\caption{The action density of the KvBLL caloron as
function of $z,x$ at fixed $t=y=0$. At large separations $r_{12}$ the caloron
is a superposition of two BPS dyon solutions ({\it left}: $r_{12}=1.5/T$). At
small separations they merge ({\it right}: $r_{12}=0.6/T$).}
\end{figure}

\subsection{Inside dyon cores}

In the vicinity of the L dyon center ${\bf x}_1$ and far away from
the M dyon ($s \v \gg 1$) the field becomes that of the L dyon.
It is instructive to write it in spherical coordinates centered at ${\bf x}_1$.
In the `stringy' gauge in which the $A_4$ component is
constant and diagonal at spatial infinity, the L dyon field is
\bea\nn
A_4^L\!\!\!&=&\!\!\!\frac{\tau^3}{2}\left(\frac{1}{r}+2\pi-\bv\coth(\bv r)\right),\;\;\;A^L_r=0,\\
\nn
A^L_\theta\!\!\! &=&\!\!\! \bv\frac{-\sin(2\pi T x^4-\phi)\;\tau^1
+\cos(2\pi x^4-\phi)\tau^2}{2\sinh(\bv r)},\\
\n
A^L_\phi\!\!\! &=&\!\!\!\bv\frac{\cos(2\pi x^4-\phi)\;\tau^1+\sin(2\pi T x^4-\phi)\tau^2}{2\sinh(\bv r)}\\
\!\!\!&-&\!\!\!\tau^3\frac{\tan(\theta/2)}{2r}\;.
\nea
Here $A_\theta$, for example, is the projection of $\vec A$ onto the direction
$\vec{n}_\theta=(\cos\theta\cos\phi,\cos\theta\sin\phi,-\sin\theta)$.
The $\phi$ component has a string singularity along the $z$ axis
going in the positive direction. Notice that inside the core region ($\bv r\leq 1$)
the field is time-dependent, although the action density is static.
At large distances from the L dyon center, {\it i.e.} far outside the core
one neglects exponentially small terms ${\cal O}\left(e^{-\bv r}\right)$
and the surviving components are
\bea
\nn
A^L_4&\stackrel{r\to\infty}{\longrightarrow}&\left(\v+\frac{1}{r}\right)\,\frac{\tau^3}{2},\\
\nn
A^L_\phi &\stackrel{r\to\infty}{\longrightarrow}&
-\frac{{\rm tan}\frac{\theta}{2}}{r}\,\frac{\tau^3}{2}
\nea
corresponding to the radial electric and magnetic field components of the L dyon (see subsection 6.1):
$$
E^L_r= B^L_r\stackrel{r\to\infty}{\longrightarrow}
-\frac{1}{r^2}\,\frac{\tau^3}{2}.
$$

Similarly, in the vicinity of the M dyon and far away from the L
dyon ($r \bv \gg1$) the field becomes that of the M dyon, which we
write in spherical coordinates centered at ${\bf x}_2$:
\bea\n
A_4^M \!\!\!&=&\!\!\!\frac{\tau^3}{2}\left(\v\coth(\v s)-\frac{1}{s}\right),\quad A_r^M =0,\\
\n
A_\theta^M \!\!\!&=&\!\!\! \v\frac{\sin\!\phi\;\tau^1-\cos\!\phi\;\tau^2}{2i\sinh(\v s)},\\
\n
A_\phi^M \!\!\!&=&\!\!\! \v\frac{\cos\!\phi\;\tau^1+\sin\!\phi\;\tau^2}{2\sinh(\v s)}
+\tau^3\frac{\tan(\theta/2)}{2s},
\nea
whose asymptotics is
\bea\n
A^M_4 &\stackrel{r\to\infty}{\longrightarrow}& \left(\v-\frac{1}{s}\right)\,\frac{\tau^3}{2},\\
\nn A^M_\phi &\stackrel{r\to\infty}{\longrightarrow}& \frac{{\rm  tan}\frac{\theta}{2}}{s}\,
\frac{\tau^3}{2},\\
E^M_r=B^M_r &\stackrel{r\to\infty}{\longrightarrow}&
\frac{1}{s^2}\,\frac{\tau^3}{2}.
\nea
We see that in both cases the L,M fields become Abelian at large distances,
corresponding to (electric, magnetic) charges $(-,-)$ and $(+,+)$, respectively.
The corrections to the fields are hence of the
order of $1/\D$ arising from the presence of the other dyon.

\subsection{Far away from dyon cores}

Far away from both dyon cores ($r \bv\gg 1,\;s\v \gg 1$; note that
it does not necessarily imply large separations -- the dyons may
even be overlapping) one can neglect both types of exponentially
small terms, ${\cal O}\left(e^{-r\bv }\right)$ and ${\cal  O}\left(e^{-s\v}\right)$.
With exponential accuracy the function $\tilde\chi$ is zero,
and the KvBLL field becomes Abelian:
$$
A_\mu^{\rm as}= \frac{\tau_3}{2}\left(\delta_{\mu 4}\v
+\bar\eta^3_{\mu\nu}\partial_\nu\ln\,\Phi_{\rm as}\right),
$$
where $\Phi_{\rm as}$ is
$$
\Phi_{\rm as}= \frac{r+s+\D}{r+s-\D}= \frac{s-s_3}{r-r_3}\;\;
{\rm if}\;\;{\bf r}_{12}= \D\vec e_3.
$$
It is interesting that, despite being Abelian, the asymptotic field retains its
self-duality. This is because the $3^{\rm d}$ colour component of the electric field is
$$
E_i^3=\partial_iA_4^3=\partial_i\partial_3\ln\,\Phi_{\rm as}
$$
while the magnetic field is
$$
B_i^3=\epsilon_{ijk}\partial_jA_k^3= \partial_i\partial_3\ln\,\Phi_{\rm as}
-\delta_{i3}\partial^2\ln\,\Phi_{\rm as},
$$
where the last term is zero, except on the line connecting the dyon centers where
it is singular; however, this singularity is an artifact of the exponential
approximation used. Explicit evaluation gives the following nonzero
components of the $A_\mu$ field far away from both dyon centers:
\bea\la{KvBLL_asympt}
A_4^{\rm  as}\!\!\!&=&\!\!\! \frac{\tau_3}{2}\left(\v+\frac{1}{r}-\frac{1}{s}\right),\\
\n
A_\varphi^{\rm  as}\!\!\!&=&\!\!\! -\frac{\tau_3}{2}\left(\frac{1}{r}+\frac{1}{s}\right)
\sqrt{\frac{(\D\!-\!r\!+\!s)(\D\!+\!r\!-\!s)}{(\D\!+\!r\!+\!s)(r\!+\!s\!-\!\D)}}\;.
\nea
In particular, far away from both dyons, $A_4$ is the Coulomb field
of two opposite charges. The azimuthal $A_\varphi$ component is that of the
magnetic field created by a current going from ${\bf x}_1$ to ${\bf x}_2$.

\section{QUANTUM WEIGHT OF A NEUTRAL CLUSTER OF $N$ DYONS}

Remarkably, the small-oscillation determinant about the KvBLL instanton can be computed
exactly; this has been first done for the $SU(2)$ group in Ref.~\cite{DGPS}
and later generalized to $SU(N)$ in Ref.~\cite{S}. The quantum weight of the KvBLL instanton
can be schematically written as an integral over $3N$ coordinates of dyons (the
weight does not depend on the $U(1)$ angles $\psi_m$, hence they can be integrated out):
\bea\n
W_1\!\!\! &=&\!\!\! \int\!d{\bf x}_1...d{\bf x}_N\,\sqrt{\det g}\,\left(\frac{4\pi}{g^4}\frac{\mu^4}{T}\right)^N\,
\\
\!\!\!&\cdot &\!\!\!\exp\left(-\frac{8\pi^2}{g^2}\right)\left(\Det(-\triangle)\right)^{-1}_{\rm reg,\,norm},
\la{W10}\eea
where $g$ is the full $4N\times 4N$ metric tensor of the moduli space, defined as
the zero modes overlap matrix, and $\Det(-\triangle)$ is the functional determinant over
nonzero modes, normalized to the free one and regularized by the background Pauli--Villars
method; $\mu$ is the Pauli--Villars ultraviolet cutoff and $g^2$ is the bare coupling
constant defined at that cutoff. The Jacobian $\det g$ turns out to be a square of
the determinant of an $N\times N$ matrix $G^{(1)}$ such that $\sqrt{\det g}=\det G^{(1)}$ where
\bea\la{G1}
G_{mn}^{(1)}\!\!\!&=&\!\!\!\delta_{mn}\,\left(4\pi\nu_m\right.\\
\n
\!\!\!&+&\!\!\!\left.\frac{1}{T|{\bf x}_m-{\bf x}_{m-1}|}
+\frac{1}{T|{\bf x}_m-{\bf x}_{m+1}|}\right)\\
\n
\!\!\!&-&\!\!\!\frac{\delta_{m,n-1}}{T|{\bf x}_m-{\bf x}_{m+1}|}
-\frac{\delta_{m,n+1}}{T|{\bf x}_m-{\bf x}_{m-1}|}
\eea
is a matrix whose entries are Coulomb interactions between dyons that are nearest neighbours
in kind. The Coulomb interactions in the zero mode overlap matrix arise naturally from the
Coulomb asymptotics of the dyon field \ur{KvBLL_asympt}, so it is quite simple to check that \Eq{G1}
is correct at large separations. A nontrivial fact is that \Eq{G1} is actually {\em exact}
for all separations between dyons, including the case when they strongly overlap like in
Fig.~12, bottom. This has been first conjectured by Lee, Weinberg and Yi~\cite{LWY}
and then proved to be indeed exact at all separations by a direct calculation by
Kraan~\cite{Kraan} and later checked in Ref.~\cite{DG-05}. In the last paper it has been
also shown that in the limit of trivial holonomy ($\mu_m=0$) or vanishing temperature
the measure given by \Eqs{W10}{G1} reduces to the standard instanton measure \ur{1instw} written
in terms of the conventional ``center-size-orientation'', which is a rather nontrivial
but gratifying statement.

The functional determinant over nonzero modes $\Det^{-1}(-\triangle)$ together with
the classical action and the Pauli--Villars cutoff combine into the renormalized scale
parameter $\Lambda_{\rm PV}$, times a function of dyon separations, $\Lambda_{\rm PV}$
and T~\cite{DGPS,S}. It is a complicated function which, for the time being, we approximate
by its most essential part: a constant equal to $\exp\left(-P^{\rm pert}\right)$, where
$P^{\rm pert}$ is the perturbative gluon loop \ur{Ppert} in the background of a
constant field $A_4$ \ur{A4}. This part is necessarily present in $\Det^{-1}(-\triangle)$
as most of the $3d$ space outside the instanton's core is just a constant $A_4$ background,
and indeed the calculation~\cite{DGPS,S} exhibits this piece which is the only one
proportional to the 3-volume.

Therefore, we write the weight of the KvBLL instanton {\it i.e.} a neutral cluster
of $N$ different-kind dyons as
\bea\la{W1}
W_1 \!\!\!&\approx &\!\!\! \int\!d{\bf x}_1...d{\bf x}_N\,\det G_1\,f^N\\
\n
\!\!\!&\cdot &\!\!\!\exp\left(-P^{\rm pert}(\mu_1,\ldots,\mu_N)\right)
\eea
where the fugacity $f$ is
\beq
f=\frac{4\pi}{g^4}\,\frac{\Lambda^4}{T}={\cal O}(N^2).
\la{f}\eeq
The bare coupling constant $g^2$ in the pre-exponent is renormalized and starts to ``run''
only at the 2-loop level not considered here. Eventually, its argument will be
the largest scale in the vacuum, be it the temperature or the equilibrium density of dyons.

\section{QUANTUM WEIGHT OF MANY DYONS}

In the vacuum problem, one needs to use not one but ${\cal O}(V)$ number of KvBLL instantons
as the saddle point. Solutions with the topological charge greater than 2 will be hardly
ever known explicitly as their construction runs into the problem of resolving the nonlinear
Atiyah--Drinfeld--Hitchin--Manin--Nahm constraints. At present 2-instanton solutions
characterized by a nontrivial holonomy have been found~\cite{Nogradi} but it is insufficient.
Nevertheless, the moduli space {\it measure} of an arbitrary number $K$ of KvBLL instantons
can be constructed despite the lack of explicit solutions, at least in the approximation
which seems to be relevant for the large-volume thermodynamics, if not exactly.

When one takes a configuration of $K$ instantons each made of $N$ different-kind dyons
one encounters also same-kind dyons for which the metric \ur{G1} is inapplicable. However,
the case of identical dyons has been considered separately by Gibbons and Manton~\cite{GM2}.
The integration measure for $K$ identical dyons following from that work is, in our
notations,
\bea\la{Gident}
W^{\rm ident}\!\!\! &=&\!\!\! \frac{1}{K!}\int\!d{\bf x}_1...d{\bf x}_K\,\det G^{\rm ident}_{K\times K}\,,\\
\n
G^{\rm ident}_{ij}\!\!\! &=&\!\!\! \left\{\begin{array}{cc}4\pi\nu_m-\sum_{k\neq i}
\frac{2}{T|{\bf x}_i-{\bf x}_k|},& i=j,\\ \frac{2}{T|{\bf x}_i-{\bf x}_j|},
& i\neq j\end{array}\right.\,,
\nea
where the identity factorial is inserted to avoid counting same configurations more
than once.

As in the case of different-kind dyons, this result for the metric can be easily obtained
at large separations from considering the asymptotics of the zero modes' overlap.
The four zero modes $\phi^{(\kappa)}_\mu (\kappa=1,2,3,4)$ of individual dyons
are given by the components of the field strength: $\phi^{(\kappa)}_\mu=F_{\mu\kappa}$.
The zero modes for the $m^{\rm th}$ kind of dyon are normalized to its action,
$\int \Tr \phi^{(\kappa)}_\mu\phi^{(\lambda)}_\mu\sim\delta^{\kappa\lambda}\nu_m$
(see \Eq{dyon-action}) and hence depend on the holonomy. Since the field strengths
decay as $1/r^2$ the overlaps between zero modes are Coulomb-like, and only those that
are nearest neighbors in $m$ do interact. In fact, the diagonal components of the metric
tensor also acquire Coulomb-like corrections since the action of individual dyons
is actually normalized to its asymptotic field $A_4$ that gets Coulomb corrections from other dyons.

However, in contrast to the different-kind dyons, it is not possible to prove that this
expression is correct at all separations. Moreover, such an extension of \Eq{Gident} is
probably wrong. The metric for two same-kind dyons has been found exactly at all separations
by Atiyah and Hitchin~\cite{AH} by an undirect method: it is more complicated than what follows from \Eq{Gident}
at $K=2$ but differs from it by terms that are exponentially small at large separations~\cite{GM1}.
We shall neglect the difference and use the Gibbons--Manton metric at face value. The point
is, \Eq{Gident} imposes very strong repulsion between same-kind dyons (as does the
Atiyah--Hitchin metric), hence the range of the moduli space where the two metrics
differ is, statistically, not frequently visited by dyons. We do not have a proof that all
thermodynamic quantities will be computed correctly with this simplification: proving or
disproving it is an interesting and important problem. To remain on the safe side, one has
to admit today that the metric \ur{Gident} is applicable if the dyon ensemble is sufficiently
dilute, that is at high temperatures. Nevertheless, physical observables we compute
have a smooth limit even at $T\to 0$. Therefore, it may well prove to be a correct computation
at any temperatures, but this remains to be seen.

It is possible to combine the metric tensors for different-kind \ur{G1} and same-kind \ur{Gident}
dyons into one metric appropriate for the moduli space of $K_1$ dyons of kind 1,
$K_2$ dyons of kind 2,..., $K_N$ dyons of kind $N$. It is a matrix whose dimension is
the total number of dyons, that is a $(K_1+\ldots +K_N)\times(K_1+\ldots +K_N)$ matrix:
\bea\n
\!\!\!&&\!\!\!G_{mi,nj}=\delta_{mn}\delta_{ij}\,
\left(4\pi\nu_m-2\sum_{k\neq i}\frac{1}{T|{\bf x}_{mi}\!-\!{\bf x}_{mk}|}\right.\\
\n
\!\!\!&&\!\!\!+\left.
\sum_k\frac{1}{T|{\bf x}_{mi}\!-\!{\bf x}_{m+1,k}|}
+\sum_k\frac{1}{T|{\bf x}_{mi}\!-\!{\bf x}_{m-1,k}|} \right)\\
\n
\!\!\!&&\!\!\!-\;\frac{\delta_{m,n-1}}{T|{\bf x}_{mi}\!-\!{\bf x}_{m+1,j}|}
-\frac{\delta_{m,n+1}}{T|{\bf x}_{mi}\!-\!{\bf x}_{m-1,j}|}\\
\!\!\!&&\!\!\!+\;2\left.\frac{\delta_{mn}}{T|{\bf x}_{mi}\!-\!{\bf x}_{mj}|}\right|_{i\neq j},
\la{G}\eea
where ${\bf x}_{mi}$ is the coordinate of the $i^{\rm th}$ dyon of kind $m$.
Since the statistical weight of a configuration of dyons is large when $\det G$
is large and small when it is small, $\det G$ imposes an attraction between
dyons that are nearest neighbours in kind, and a repulsion between same-kind dyons.
The coefficients -1,2,-1 in front of the Coulomb interactions are actually the
scalar products of the Cartan generators that determine the asymptotics of the
dyons' field, see subsection 6.2.

The matrix $G$ has the following nice properties:
\begin{itemize}
\item symmetry: $G_{mi,nj}=G_{nj,mi}$
\item overall ``neutrality'': the sum of Coulomb interactions in non-diagonal entries
cancel those on the diagonal:
$\sum_{nj}G_{mi,nj}=4\pi\nu_m$
\item identity loss: dyons of the same kind are indistinguishable, meaning
mathematically that $\det G$ is symmetric under permutation of any pair of dyons
$(i\!\leftrightarrow\!j)$ of the same kind $m$. Dyons do not `know' to which
instanton they belong to
\item factorization: in the geometry when dyons fall into $K$ well separated
neutral clusters of $N$ dyons of different kinds in each, $\det G$ factorizes
into a product of exact integration measures for $K$ KvBLL instantons,
$\det G=(\det G^{(1)})^K$ where $G^{(1)}$ is given by \Eq{G1}
\item last but not least, the metric corresponding to $G$ is hyper-K\"ahler,
as it should be for the moduli space of a self-dual classical field~\cite{AH}.
In fact, it is a severe restriction on the metric.
\end{itemize}

\section{ENSEMBLE OF DYONS}

In the semiclassical approximation we thus replace the YM partition function
\ur{Zph2} by the partition function of an interacting ensemble of an arbitrary number
of dyons of $N$ kinds:
\bea\n
{\cal Z}\!&=&\!\sum_{K_1...K_N}\frac{1}{K_1!...K_N!}\prod_{m=1}^N\,\prod_{i=1}^{K_m}
\int (d{\bf x}_{mi}\,f)\\
&\!\cdot\!\! &\det G({\bf x}_{mi}),
\la{Z3}\eea
where ${\bf x}_{mi}$ is the coordinate of the $i^{\rm th}$ dyon of kind $m$,
the matrix $G$ is given by \Eq{G} and the fugacity $f$ is given by \Eq{f}.
The overall exponent of the perturbative potential energy as function of the
holonomy $\{\mu\}$ is understood, as in \Eq{W1}.

The ensemble defined by a determinant of a matrix whose dimension is the number of
particles, is not a usual one. More customary, the interaction is given by
the Boltzmann factor $\exp\left(-U_{\rm int}({\bf x}_1,\ldots)\right)$. Of course,
one can always present the determinant in that way using the identity
$\det G\!=\!\exp(\Tr\log G)\equiv \exp(-U_{\rm int})$ but the interactions will then
include many-body forces. At the same time, it is precisely
the determinant form of the interaction that makes the statistical physics of dyons
an exactly solvable problem.

\subsection{Dyons' free energy: confining holonomy preferred}

The partition function \ur{Z3} can be computed directly and exactly, just by writing
the determinant of $G$ by definition as a sum of permutations of products of the matrix
entries. The result is astonishingly simple: all Coulomb interactions cancel exactly
after integration over dyons' positions,
provided the overall neutrality condition is satisfied, {\it viz.} $K_1=K_2=\ldots=K_N
=K$; otherwise the partition function is divergent.
We shall derive this interesting result by a more general method in section 11.
Therefore, the recipe for computing the
partition function is just to impose the neutrality condition and then to throw
out all Coulomb interactions! We have thus to take the product of $(4\pi\nu_m)$'s
from the diagonal of $G$:
$$
{\cal Z}=\sum_K\frac{(4\pi f V)^{KN}}{(K!)^N}\prod_{m=1}^N\nu_m^K\,.
$$
The quantity $4\pi f V$ is dimensionless and large for large volumes $V$. The sum can be
therefore computed from the saddle point in $K$ using the Stirling asymptotics for large
factorials, and we obtain
\bea\n
{\cal Z}\!\!\!&=&\!\!\!\exp\left(4\pi f V N\left(\nu_1\nu_2\ldots\nu_N\right)^{\frac{1}{N}}\right),\\
&&\nu_1+\nu_2+\ldots + \nu_N=1.
\la{Z0}\eea
By definition, $F=-T\log {\cal Z}$ is the nonperturbative dyon-induced free energy as function
of the holonomy; for $N=2,3$ it is plotted in Fig.~9. Evidently, it has the minimum at
\beq
\nu_1=\nu_2=\ldots=\nu_N=\frac{1}{N}
\la{nuconf}\eeq
corresponding to equidistant, that is confining values of $\mu$'s \ur{muconf}!
At the minimum, the free energy is
\bea\n
F_{\rm min}&=&-T\log {\cal Z}_{\rm min}=-4\pi f V T\\
&=&-\frac{16\pi^2}{g^4}\,\Lambda^4\,V=\frac{N^2}{4\pi^2}\,\frac{\Lambda^4}{\lambda^2}\,V,
\la{freeen}\eea
and there are no Coulomb corrections to this result. In the last equation we have introduced
the $N$-independent 't Hooft coupling $\lambda\equiv\alpha_sN/2\pi$.

We note that the free energy is ${\cal O}(N^2)$ as expected on general $N$-counting
grounds and that it is temperature-independent. It corresponds to $\log {\cal Z}$ being
proportional to the 4-volume $V^{(4)}=V/T$, demonstrating the expected extensive behaviour
at low temperature.

\subsection{Cautionary remark on the numerical simulation of the dyon ensemble}

On the one hand, the determinant form for the weight of a dyon configuration is a very welcome
feature as it allows an exact computation not only of the free energy but also of various correlation
functions, see sections 11,12. On the other hand, it calls for questions that have been recently raised
in Ref.~\cite{BDIMW}. The point is that the determinant \ur{G} is not positive definite. Moreover, even
if for some dyon configuration it is positive, individual eigenvalues of $G$ need not be positive.
If one takes many dyons on random in a dense ensemble, most of the eigenvalues will be
negative~\cite{BDIMW}. It rises the question of whether one can use the weight given by $\det G$.

Although the result for the free energy \ur{Z0} {\it looks} as if the Coulomb interactions can be
altogether neglected, the ensemble governed by $\det G$ is in fact very strongly correlated. The
correlation functions computed in section 12 exhibit the Debye screening phenomenon typical for
a Coulomb-like plasma. If one takes the equidistant holonomy \ur{nuconf} minimizing the free energy
one can present $G=\frac{4\pi}{N}(E+C)$ where $E$ is a unity $KN\times KN$ matrix and $C$ is a matrix
formed by Coulomb interactions. One has then $\det G={\rm const.}\,\exp(\Tr \log (1+C))=
{\rm const.}\,\exp(\Tr C-\half\Tr C^2+\ldots)$. As a matter of fact this weight is positive definite
if the series is truncated at any finite number of terms. If one truncates it at the first term
(which would be justifiable if the density is low) it is precisely the Boltzmann weight for
a Coulomb plasma. The Coulomb plasma, however, is not an exactly solvable system: there is an infinite
series of corrections to the free energy coming from interactions. The role of the higher terms
in the expansion of $\log(1+C)$ is to cancel, order by order in density, those corrections to the free energy.
The full series leads to a stronger correlation of dyons than even in the standard Coulomb plasma.
This is clearly seen if one returns back to the $\det G$ weight: it goes to zero when two same-kind
dyons come close enough together, corresponding to an infinitely high barrier in terms of the effective potential,
whereas in the standard Coulomb plasma such configurations are only exponentially suppressed.

Therefore, the ensemble governed by the $\det G$ weight strongly favours a clustering
of dyons into neutral clusters of $N$ different-kind dyons. For such configurations, all eigenvalues
of $G$ are positive definite. In the limiting case of well-separated neutral clusters $\det G$ factorizes
into a product of the one-instanton measures~\cite{DG-05}. Therefore, if in a Monte Carlo simulation one starts
with a well-clustered dyon configuration and then thermolizes it by small enough Metropolis steps, the ensemble
will never cross the line where a negative eigenvalue appears, since, by continuity, the determinant must
first become zero which would mean a zero weight for a configuration. However, implementing it in a practical
simulation may be a difficult task.

\section{STATISTICAL PHYSICS OF DYONS AS A QUANTUM FIELD THEORY IN 3D}

We follow here Ref.~\cite{DP-07}.

Although the Coulomb interactions of dyons cancel exactly in the free energy, the
dyon ensemble defined by \Eq{Z3} is not a free gas but a highly correlated system.
To facilitate computing observables through correlation functions, we rewrite
\Eq{Z3} as an equivalent quantum field theory. As a byproduct, we shall also check
that the result for the free energy \ur{Z0} is correct.

To proceed to the quantum field theory description we use two mathematical tricks.\\

1. \underline{``Fermionization''} (Berezin~\cite{Berezin}). It is helpful to exponentiate
the Coulomb interactions rather than keeping them in $\det G$.
To that end one presents the determinant of a matrix as an integral over a finite
number of anticommuting Grassmann variables:
$$
\det (G_{AB})
=\int\!\prod_A d\psi_A^\dagger\,d\psi_A\,
\exp\left(\psi_A^\dagger\,G_{AB}\,\psi_B\right)\,.
$$

Now we have the two-body Coulomb interactions in the exponent and it is possible
to use the second trick presenting Coulomb interactions with the help of a functional
integral over an auxiliary boson field.\\

2. \underline{``Bosonization''} (Polyakov~\cite{Pol}). One can write
\bea\n
&&\exp\left(\sum_{m,n}\frac{Q_mQ_n}{|{\bf x}_m-{\bf x}_n|}\right)\\
\n
&&=\int D\phi\,\exp\left[-\int\!d{\bf x}\left(\frac{1}{16\pi}\partial_i\phi\partial_i\phi
+\rho\phi\right)\right]\\
\n
&&=\exp\left(\int \rho\frac{4\pi}{\triangle}\rho\right),\quad
\rho = \sum Q_m\,\delta({\bf x}-{\bf x}_m).
\eea

After applying the first trick the ``charges'' $Q_m$ become Grassmann variables but
after applying the second one, it becomes easy to integrate them out since the square of
a Grassmann variable is zero. In fact one needs $2N$ boson fields $v_m,w_m$ to reproduce
diagonal elements of $G$ and $2N$ anticommuting (``ghost'') fields $\chi^\dagger_m,\chi_m$
to present the non-diagonal elements. The chain of identities is accomplished in
Ref.~\cite{DP-07} and the result for the partition function \ur{Z3} is, identically,
a path integral defining a quantum field theory in 3 dimensions:
\bea\la{Z4}
{\cal Z}\!\!\!&=&\!\!\!\int\!D\chi^\dagger\,D\chi\,Dv\,Dw\,\exp\int\!d^3x\\
\n
\!\!\!&\cdot &\!\!\!\left\{\frac{T}{4\pi}\,\left(\partial_i\chi_m^\dagger\partial_i\chi_m
+\partial_iv_m\partial_iw_m\right)\right.\\
\n
\!\!\!&+&\!\!\!\left.f\!\left[\!(\!-4\pi\mu_m\!+\!v_m)\frac{\partial{\cal F}}{\partial w_m}
\!+\!\chi^\dagger_m\frac{\partial^2{\cal F}}{\partial w_m\partial w_n}\chi_n\!\right]\!
\right\},\\
\la{Todapot}
{\cal F}\!\!\!&=&\!\!\!\sum_{m=1}^N e^{w_m-w_{m+1}}\,.
\eea
The subscript $m$ is periodic: $m=N+1$ is equivalent to $m=1$, and $m=0$ is equivalent to $m=N$.

In the above path integral, the fields $v_m$ have the meaning of the asymptotic Abelian
electric potentials of dyons,
\bea\la{Av}
\left(A_4\right)_{mn}\!\!\!&=&\!\!\!\delta_{mn}\,A_{m\,4},\\
\n
A_{m\,4}({\bf x})/T\!\!\! &=&\!\!\! 2\pi \mu_m-\half v_m({\bf x}),\quad {\bf E}_m={\bf \nabla}A_{m\,4},
\eea
while $w_m$ have the meaning of the dual (or magnetic) Abelian potentials.
Note that the kinetic energy for the $v_m,w_m$ fields has only the mixing term
$\partial_iv_m\partial_iw_m$ which is nothing but the Abelian duality transformation
${\bf E}\cdot{\bf B}$. The function ${\cal F}(w)$ \ur{Todapot} is known as the periodic
(or affine) Toda lattice.

Although the Lagrangian in \Eq{Z4} describes a highly nonlinear interacting quantum field
theory, it is in fact exactly solvable! To prove it, one observes that the fields $v_m$
enter the Lagrangian only linearly, therefore one can integrate them out. It leads
to a functional $\delta$-function:
\beq
\int\!Dv_m\quad\longrightarrow\quad \delta\left(-\frac{T}{4\pi}\partial^2w_m
+f\frac{\partial{\cal F}}{\partial w_m}\right).
\la{UD}\eeq
This $\delta$-function restricts possible fields $w_m$ over which one still has
to integrate in \Eq{Z4}. Let $\bar w_m$ be a solution to the argument of the
$\delta$-function. Integrating over small fluctuations about $\bar w$ gives
the Jacobian
\beq\la{Jac}
{\rm Jac}={\rm det}^{-1}\left(-\frac{T}{4\pi}\partial^2\delta_{mn}
+f\frac{\partial^2{\cal F}(\bar w)}{\partial w_m\partial w_n}\,\right)\,.
\eeq
Remarkably, exactly the same functional determinant (but in the numerator)
arises from integrating over the ghost fields, for any background $\bar w$.
Therefore, all quantum corrections cancel {\em exactly} between the boson and
ghost fields (a characteristic feature of supersymmetry), and the ensemble of dyons
is basically governed by a classical field theory.

This remarkable cancelation is of course not accidental but can be traced to the
hyper-K\"ahler nature of the dyons' metric, which in its turn stems from the
self-duality of dyons.

To find the ground state we examine the fields' potential energy being
$-4\pi f\mu_m\partial{\cal F}/\partial w_m$ which we prefer to write restoring
$\nu_m=\mu_{m\!+\!1}-\mu_m$ and ${\cal F}$ as
\beq
{\cal P}=-4\pi f V\sum_m \nu_m\,e^{w_m-w_{m\!+\!1}}
\la{calP1}\eeq
(the volume factor arises for constant fields $w_m$). One has first to find the
stationary point in $w_m$ for a given set of $\nu_m$'s. It leads to the equations
$$
\frac{\partial {\cal P}}{\partial w_m}=0
$$
whose solution is
\bea\n
e^{w_1-w_2}&=&\frac{(\nu_1\nu_2\nu_3...\nu_N)^{\frac{1}{N}}}{\nu_1},\\
\n
e^{w_2-w_3}&=&\frac{(\nu_1\nu_2\nu_3...\nu_N)^{\frac{1}{N}}}{\nu_2},\quad
{\rm etc.}
\la{extr2}\eea
Putting it back into \Eq{calP1} we obtain
\bea\la{calP2}
{\cal P}&=&-4\pi f V N (\nu_1\nu_2...\nu_N)^{\frac{1}{N}},\\
\n
&&\nu_1+\nu_2+...+\nu_N=1,
\eea
which is exactly what one gets from a direct calculation of the partition function,
outlined in the previous section, see \Eq{Z0}. The minimum is achieved at the
equidistant, confining value of the holonomy, see \Eqs{nuconf}{muconf}. Using field-theoretic
methods, we have also proven that the result is exact, as all potential quantum corrections
cancel. It is in line with the exact cancelation of the Coulomb interactions in the
determinant.

Given this cancelation, the key finding -- that the dyon-induced free energy has the
minimum at the confining value of holonomy -- is in a sense trivial. If all Coulomb interactions cancel
after integration over dyons' positions, the weight of a many-dyon configuration
is the same as if they were infinitely dilute (although they are).
Then the weight, what concerns the holonomy, is proportional to
the product of diagonal matrix elements of $G$ in the dilute limit, that is to the
normalization integrals for dyon zero modes. These are nothing but the field strengths
$F_{\mu\nu}$ of individual dyons, hence the normalization is proportional to the product of
the dyon actions $S_m=2\pi\nu_m/\alpha_s$. The sum of all $N$ kinds of dyons' actions is
fixed and equal to the instanton action, however, it is the {\em product} of actions
that defines the weight. The product is maximal when all actions are equal, hence
the equidistant or confining $\mu$'s are statistically preferred. Thus, the average
Polyakov line is zero, $<\Tr L>=0$.

\section{HEAVY QUARK POTENTIAL}

The field-theoretic representation of the dyon ensemble enables one to compute
various YM correlation functions in the semiclassical approximation.
The key observables relevant to confinement are the correlation function
of two Polyakov lines (defining the heavy quark potential), and the average of
large Wilson loops. A detailed calculation of these quantities is performed in Ref.~\cite{DP-07};
here we only present the results and discuss the meaning.

\subsection{$N$-ality and $k$-strings}

From the viewpoint of confinement, all irreducible representations of the $SU(N)$
group fall into $N$ classes: those that appear in the direct product of any number
of adjoint representations, and those that appear in the direct product of any
number of adjoint representations with the irreducible representation being the
rank-$k$ antisymmetric tensor, $k=1,\ldots , N\!-\!1$. ``$N$-ality'' is said
to be zero in the first case and equal to $k$ in the second. $N$-ality-zero
representations transform trivially under the center of the group $Z_N$;
the rest acquire a phase $2\pi k/N$.

One expects that there is no asymptotic linear potential between static colour
sources in the adjoint representation as such sources are screened by gluons.
If a representation is found in a direct product of some number of adjoint
representations and a rank-$k$ antisymmetric representation, the adjoint ones
``cancel out'' as they can be all screened by an appropriate number of gluons.
Therefore, from the confinement viewpoint all $N$-ality $=k$ representations are
equivalent and there are only $N-1$ string tensions $\sigma_{k,N}$ being the
coefficients in the {\em asymptotic} linear potential for sources in the antisymmetric
rank-$k$ representation. They are called ``$k$-strings''. The representation
dimension is $d_{k,N}=\frac{N!}{k!(N-k)!}$ and the eigenvalue of the quadratic
Casimir operator is $C_{k,N}=\frac{N+1}{2N}\,k(N-k)$.

The value $k\!=\!1$ corresponds to the fundamental representation whereas
$k=N\!-\!1$ corresponds to the representation conjugate to the fundamental
[quarks and anti-quarks]. In general, the rank-$(N\!-\!k)$ antisymmetric
representation is conjugate to the rank-$k$ one; it has the same dimension
and the same string tension, $\sigma_{k,N}=\sigma_{N\!-\!k,N}$. Therefore, for
odd $N$ all string tensions appear in equal pairs; for even $N$, apart from
pairs, there is one privileged representation with $k=\frac{N}{2}$ which
has no pair and is real. The total number of different string tensions
is thus $\left[\frac{N}{2}\right]$.

The behaviour of $\sigma_{k,N}$ as function of $k$ and $N$ is an important issue as
it discriminates between various confinement mechanisms. On general $N$-counting
grounds one can only infer that at large $N$ and $k\ll N$, $\sigma_{k,N}/\sigma_{1,N}=
(k/N)(1+{\cal O}(1/N^2))$. Important, there should be no ${\cal O}(N^{-1})$
correction~\cite{Shifman}. A popular version called ``Casimir scaling'', according
to which the string tension is proportional to the Casimir operator for a
given representation (it stems from an idea that confinement is somehow related
to the modification of a one-gluon exchange at large distances), does not satisfy
this restriction~\footnote{There are many lattice studies claiming Casimir scaling
for various groups and representations. It refers to certain ``intermediate'' separations
between static sources. We are now discussing the string tension at asymptotic separations.}.

\subsection{Correlation function of Polyakov lines}

To find the potential energy $V_{k,N}$ of static ``quark'' and ``antiquark''
transforming according to the antisymmetric rank-$k$ representation, one has
to consider the correlation of Polyakov lines in the appropriate representation:
\bea\la{VkN}
&&\left<\Tr L_{k,N}({\bf z}_1)\;\;\Tr L^\dagger_{k,N}({\bf z}_2)\right>\\
\n
&&={\rm const.}\,\exp\left(-\frac{V_{k,N}({\bf z}_1-{\bf z}_2)}{T}\right).
\eea
Far away from dyons' cores the field is Abelian and in the field-theoretic language
of \Eq{Z4} is given by \Eq{Av}. Therefore, the Polyakov line in the fundamental
representation is
\bea\la{Lv}
\Tr L({\bf z})&=&\sum_{m=1}^N Z_m,\\
\n
Z_m&=&\exp\left(2\pi i\mu_m-\frac{i}{2}v_m({\bf z})\right).
\eea
In the general antisymmetric rank-$k$ representation
\beq
\Tr L_{k,N}({\bf z})=\sum_{m_1<m_2<...<m_k}^NZ_{m_1}Z_{m_2}...Z_{m_k}
\la{LkNv}\eeq
where cyclic summation from 1 to $N$ is assumed.

The average \ur{VkN} can be computed from the quantum field theory \ur{Z4}.
Inserting the two Polyakov lines \ur{LkNv} into \Eq{Z4} we observe that
the Abelian electric potential $v_m$ enters linearly in the exponent as before.
Therefore, it can be integrated out, leading to a $\delta$-function for the
dual field $w_m$, which is now shifted by the source (cf. \Eq{UD}):
\bea\n
&&\int\!Dv_m \longrightarrow  \prod_m\delta\left(-\frac{T}{4\pi}\partial^2w_m+f\frac{\partial{\cal F}}{\partial w_m}\right.\\
\n
&&-\left.\frac{i}{2}\,\delta({\bf x}\!-\!{\bf z_1})(\delta_{mm_1}+\ldots +\delta_{mm_k})\right.\\
&&\left.+\frac{i}{2}\,\delta({\bf x}\!-\!{\bf z_2})(\delta_{mn_1}+\ldots +\delta_{mn_k})\right).
\la{UDL}\eea
One has to find the dual field $w_m({\bf x})$ nullifying the argument of this
$\delta$-function, plug it into the action
\beq
\exp\left(\!\int\!d{\bf x}\,\frac{4\pi f}{N}{\cal F}(w)\right),
\la{action}\eeq
and sum over all sets $\{m_1<m_2<...<m_k\}$, $\{n_1<n_2<...<n_k\}$ with the weight
$\exp\left(2\pi i(m_1+\ldots +m_k-n_1-\ldots - n_k)/N\right)$. The Jacobian from resolving
the $\delta$-function again cancels exactly with the determinant arising from ghosts.
Therefore, the calculation of the correlator \ur{VkN}, sketched above, is exact.

At large separations between the sources $|{\bf z}_1\!-\!{\bf z}_2|$, the fields $w_m$
resolving the $\delta$-function are small and one can expand the Toda chain:
\bea\n
{\cal F}(w)&=&\sum_m e^{w_m-w_{m\!+\!1}}\approx N+\frac{1}{2}\, w_m\,{\cal M}_{mn}\,w_n,\\
\frac{\partial {\cal F}}{\partial w_m}&\approx &{\cal M}_{mn}\,w_n,
\la{calFsm}\eea
where
\beq
{\cal M}=\left(\begin{array}{cccccc}2&-1&0&\ldots&0 & -1\\ -1&2&-1&\ldots &0& 0\\
0&-1&2&-1&\ldots &0\\ \ldots & \ldots & \ldots & \ldots & \ldots & \ldots\\-1&0&0&\ldots
&-1&2\end{array}\right).
\la{calM}\eeq
As apparent from \Eq{calFsm}, the eigenvalues of ${\cal M}$ determine the spectrum of the
dual fields $w_m$. There is one zero eigenvalue which decouples from everywhere,
and $N\!-\!1$ nonzero eigenvalues
\beq
{\cal M}^{(k)}=\left(2\sin\frac{\pi k}{N}\right)^2,\quad k=1,...,N-1.
\la{eig2}\eeq
Certain orthogonality relation imposes the selection rule: the asymptotics of the correlation
function of two Polyakov lines in the antisymmetric rank-$k$ representation is determined
by precisely the  $k^{\rm th}$ eigenvalue. We obtain~\cite{DP-07}
\bea\n
&&\left<\Tr L_{k,N}({\bf z}_1)\;\;\Tr L^\dagger_{k,N}({\bf z}_2)\right>\\
&&\stackrel{z_{12}\to\infty}{=}{\rm const.}
\exp\left(-|{\bf z}_1-{\bf z}_2|\,M\sqrt{{\cal M}^{(k)}}\right)
\la{corrLk2}\eea
where $M$ is the `dual photon' mass,
\beq
M=\sqrt{\frac{4\pi f}{T}}=\frac{N\Lambda^2}{2\pi\lambda T}={\cal O}(N).
\la{M}\eeq

Comparing \Eq{corrLk2} with the definition of the heavy quark potential \ur{VkN} we find
that there is an asymptotically linear potential between static ``quarks'' in
any $N$-ality nonzero representation, with the $k$-string tension
\bea\la{sigma-k}
\sigma_{k,N}&=&MT\sqrt{{\cal M}^{(k)}}=2MT\,\sin\frac{\pi k}{N}\\
\n
&=&\frac{\Lambda^2}{\lambda}\,\frac{N}{\pi}\,\sin\frac{\pi k}{N}
\eea
independent of temperature and of $N$ at large $N$ and fixed $k$.
This is the so-called `sine regime': it has been found before in certain supersymmetric
theories~\cite{sine}. Lattice simulations~\cite{DelDebbio-k} support this regime,
whereas another lattice study~\cite{Teper-k} gives somewhat smaller values
but within two standard deviations from the values following from \Eq{sigma-k}.
For a general discussion of the sine regime for $k$-strings, which is favoured
from many viewpoints, see~\cite{Shifman}.

We see that at large $N$ and $k\ll N$, $\sigma_{k,N}/\sigma_{1,N}=
(k/N)(1+{\cal O}(1/N^2))$, as it should be on general grounds, and that
all $k$-string tensions have a finite limit at zero temperature.

\subsection{Debye screening and clustering of dyons}

Essentially the same calculation gives the correlation function of dyons in the
ensemble defined by \Eq{Z3}. The correlation is large at small distances and
decays exponentially at large distances, see \Eq{corrLk2}. Such behaviour is known
as Debye screening and is typical for the Coulomb plasma. Screening implies that
every dyon of any kind is on the average surrounded by $N-1$ dyons of other kinds
at a range of the order or less than the Debye radius $r_D\sim 1/M$. Therefore, the dyon
ensemble is in fact strongly correlated into neutral clusters, as discussed in subsection 10.2.

A neutral cluster of $N$ different kinds of dyons is in fact an instanton (section 7).
It is interesting to estimate its average size. The KvBLL instanton size is
related to the perimeter of the polygon formed by $N$ constituent dyons where
the nearest `neighbours in kind' are connected~\cite{KvBSUN}:
\beq\la{KvBLL_size}
\rho=\sqrt{\frac{\sum_m r_{m,m+1}}{2\pi T}}.
\eeq
Assuming for an estimate that the average separation between dyons of neighbour kind
in a cluster is of the order of the Debye radius,  $r_{m,m+1}\sim r_D$, we find that the
average instanton size is
\beq
<\rho>\,\sim \,\frac{\sqrt{\lambda}}{\Lambda}
\la{rho_aver}\eeq
being independent either of temperature or of $N$. This is the expected typical size
of the $4d$ lumps formed by correlated neutral clusters of $N$ dyons.

Let us compare the Debye radius $r_D\sim\frac{1}{M}=\frac{2\pi T\lambda}{N\Lambda^2}$,
and the size of the core of an individual dyon $r_{\rm core}\sim \frac{1}{2\pi T \nu}=\frac{N}{2\pi T}$,
see subsection 6.2. We observe that at temperatures $T<\frac{N\Lambda}{2\pi\sqrt{\lambda}}$
the Debye range lies within the cores of individual dyons. Since the deconfinement temperature
is expected to be $T_c\sim\Lambda$ which is less than the above estimate, it means that
in the confinement phase the screening occurs inside the typical core range at all temperatures!
This may seem paradoxical but one has to remember that the measure \ur{G1} as due to zero modes
is exact even if dyons overlap, and that it is this measure that leads to the Debye-like screening
and to the clustering of dyons into neutral clusters.

\section{AREA LAW FOR LARGE WILSON LOOPS}

When dealing with the ensemble of dyons, it is convenient to use a gauge
where $A_4$ is diagonal ({\it i.e.} Abelian). This necessarily implies Dirac
string singularities sticking from dyons, which are however gauge artifacts as
they do not carry any energy, see subsection 6.1. Moreover, the Dirac strings' directions
are also subject to the freedom of the gauge choice. For example, one can
choose the gauge in which $N$ dyons belonging to a neutral cluster are connected
by Dirac strings. This choice is, however, not convenient for the ensemble
as dyons have to loose their ``memory'' to what particular instanton they belong to.
The natural gauge is where all Dirac strings of all dyons are directed to infinity
along some axis, {\it e.g.} along the $z$ axis, see subsection 6.1.

In this gauge, the magnetic field of dyons beyond their cores is Abelian
and is a superposition of the Abelian fields of individual dyons. For large
Wilson loops we are interested in, it is this superposition field of a large
number of dyons that contributes most as they have a slowly decreasing
$1/|{\bf x}\!-\!{\bf x}_i|$ asymptotics, hence the use of the field outside
the cores is justified. Owing to self-duality,
\beq
\left[B_i({\bf x})\right]_{mn}=\left[\partial_iA_4({\bf x})\right]_{mn}
=-\frac{T}{2}\,\delta_{mn}\,\partial_iv_m({\bf x}),
\la{B}\eeq
cf. \Eq{Av}. Since $A_i$ is Abelian beyond the cores, one can use the Stokes
theorem for the spatial Wilson loop:
\bea\n
W&\equiv &\Tr\,{\cal P}\exp\,i\oint\!A_idx^i=\Tr\exp\,i\int\!B_i\,d^2\sigma^i\\
&=&\sum_m\exp\left(-i\frac{T}{2}\int\!d^2\sigma^i\,\partial_iv_m\right).
\la{Wi1}\eea
\Eq{Wi1} may look contradictory as we first use $B_i={\rm curl}A_i$ and then
$B_i=\partial_iA_4$. Actually there is no contradiction as the last equation
is true up to Dirac string singularities which carry away the magnetic flux.
If the Dirac string pierces the surface spanning the loop it gives a quantized
contribution $\exp(2\pi i\!\cdot\!{\rm integer})=1$; one can also use the gauge
freedom to direct Dirac strings parallel to the loop surface in which case
there is no contribution from the Dirac strings at all.

Let us take a flat Wilson loop lying in the $(xy)$ plane at $z\!=\!0$. Then
\Eq{Wi1} is continued as
\bea\n
W\!\!\!&=&\!\!\!\sum_m\exp\left(-i\frac{T}{2}\int_{x,y\in {\rm Area}}\!d^3x\,\partial_zv_m\delta(z)\right)\\
\!\!\!&=&\!\!\!\sum_m\exp\left(i\frac{T}{2}\int_{x,y\in {\rm Area}}\!d^3x\,v_m\,\partial_z\delta(z)\right).
\la{Wi2}\eea
It means that the average of the Wilson loop in the dyons ensemble is given by
the partition function \ur{Z4} with the source
$$
\sum_m\exp\left(i\frac{T}{2}\int\!d^3x\,v_m\,\frac{d\delta(z)}{dz}\,
\theta(x,y\in {\rm Area})\right)
$$
where $\theta(x,y\in {\rm Area})$ is a step function equal to unity if $x,y$ belong
to the area inside the loop and equal to zero otherwise. As in the case of the
Polyakov lines the presence of the Wilson loop shifts the argument of the
$\delta$-function arising from the integration over the $v_m$ variables, and the
average Wilson loop in the fundamental representation is given
by the equation
\bea\n
\left<W\right>&=&\sum_{m_1}\int\!Dw_m\exp\left(\!\int\!d{\bf x}\,\frac{4\pi f}{N}{\cal F}(w)\right)\\
\n
&\cdot\!&\!\!\!\!\!\det\left(\!-\frac{T}{4\pi}\partial^2\delta_{mn}+
f\frac{\partial^2{\cal F}}{\partial w_m\partial w_n}\right)\\
\n
&\cdot\!&\!\!\!\!\!\prod_m
\delta\left(\!-\frac{T}{4\pi}\partial^2w_m+f\frac{\partial{\cal F}}{\partial w_m}\right.\\
&&+\left.\frac{i T}{2}\,\delta_{mm_1}\,\frac{d\delta(z)}{dz}\,\theta(x,y\in {\rm Area})
\right).
\la{Wi3}\eea
Therefore, one has to solve the non-linear equations on $w_m$'s with
a source on the surface of the loop,
\bea\n
&&-\partial^2w_m+M^2\left(e^{w_m-w_{m+1}}-e^{w_{m-1}-w_m}\right)\\
&&=-2\pi i\,\delta_{mm_1}\,\frac{d\delta(z)}{dz}\,\theta(x,y\in {\rm Area}),
\la{UD3}\eea
for all $m_1$, plug it into the action $(4\pi f/N){\cal F}(w)$, and sum over
$m_ 1$. In order to evaluate the average of the Wilson loop in a general antisymmetric
rank-$k$ representation, one has to take the source in \Eq{UD3} as $-2\pi i\,\delta'(z)\,
\left(\delta_{mm_1}\!+\!\ldots\!+\!\delta_{mm_k}\right)$ and sum over $m_1\!<\!\ldots\!<\!m_k$
from 1 to $N$, see \Eq{LkNv}. Again, the ghost determinant cancels exactly the Jacobian from
the fluctuations of $w_m$ about the solution, therefore the classical-field calculation
is exact.

Contrary to the case of the Polyakov lines, one cannot, generally speaking,
linearize \Eq{UD3} in $w_m$ but has to solve the non-linear equations as they are.
The Toda equations \ur{UD3} with a $\delta'(z)$ source in the r.h.s. define
``pinned soliton'' solutions $w_m(z)$ that are $1d$ functions in the direction
transverse to the surface spanning the Wilson loop but do not depend on the
coordinates $x,y$ provided they are taken inside the loop. Beyond that surface $w_m=0$.
Along the perimeter of the loop, $w_m$ interpolate between the soliton and zero.
For large areas, the action \ur{action} is therefore proportional to the area of the surface
spanning the loop, which gives the famous area law for the average Wilson loop.
The coefficient in the area law, the `magnetic' string tension, is found
from integrating the action density of the solution in the $z$ direction.

The exact solutions of \Eq{UD3} for any $N$ and any representation $k$ have been
found in Ref.~\cite{DP-07}, and the resulting `magnetic' string tension turns out to be
\beq
\sigma_{k,N}=\frac{\Lambda^2}{\lambda}\,\frac{N}{\pi}\,\sin\frac{\pi k}{N}\,,
\la{sigmaM-k}\eeq
which coincides with the `electric' string tension \ur{sigma-k} found
from the correlators of the Polyakov lines, for all $k$-strings!

Several comments are in order here.
\begin{itemize}
\item The `electric' and `magnetic' string tensions should coincide only in the
limit $T\to 0$ where the Euclidean $O(4)$ symmetry is restored. Both calculations
have been in fact performed in that limit as we have ignored the temperature-dependent
perturbative potential \ur{Ppert}. If it is included, the `electric' and `magnetic'
string tensions split.
\item despite that the theory \ur{Z4} is 3-dimensional, with the temperature entering
just as a parameter in the Lagrangian, it ``knows'' about the restoration of
Euclidean $O(4)$ symmetry at $T\to 0$.
\item the `electric' and `magnetic' string tensions are technically obtained in
very different ways: the first is related to the mass of the elementary excitation
of the dual fields $w_m$, whereas the latter is related to the mass of the dual field
soliton.
\end{itemize}

\section{ADDING ANTI-DYONS}

So far we have considered the YM vacuum filled with self-dual dyon solutions only.
However, the $CP$ invariance of the vacuum requests, in the language of the
semiclassical approximation, that there is an equal number of anti-self-dual dyons
in the vacuum, up to thermodynamic fluctuations of the order of $\sqrt{K}\sim\sqrt{V}$.
A superposition of self-dual and anti-self-dual solutions is not, strictly speaking,
a solution of the YM equation of motion anymore, unless they are well separated.

In particular, at distances {\em exceeding the core sizes}, dyons and anti-dyons experience
classical Coulomb interactions (see {\it e.g.} Ref.~\cite{Unsal})
\bea\la{class_inter}
U_{\rm class}^{\rm d\bar d}\!\!\!&=&\!\!\! - \frac{4\pi}{g^2T}\sum_{i,j}\sum_m
\left(\frac{2}{|{\bf x}_{mi}\!-\!{\bf y}_{mj}|}\right.\\
\n
\!\!\!&-&\!\!\!\left.\frac{1}{|{\bf x}_{mi}\!-\!{\bf y}_{m+1,j}|}
-\frac{1}{|{\bf x}_{mi}\!-\!{\bf y}_{m-1,j}|}\right)
\eea
where ${\bf x}_{mi}$ is the position of the $i^{\rm th}$ dyon of kind $m$, and
${\bf y}_{mi}$ is the position of the $i^{\rm th}$ anti-dyon of kind $m$.
We expect, however, that at separations larger than the inverse Debye screening
mass $1/M$ which in fact is smaller than the core size (see subsection 12.3) the classical
Coulomb interaction is totally screened and therefore can be neglected.

Dyons and anti-dyons experience also quantum interactions that are of the order
of unity in the coupling constant $g^2$. These arise from ({\it i}) non-factorization
of the fluctuation determinant, ({\it ii}) non-factorization of the integration measure.
It is quite difficult to define these quantities properly as the dyon--anti-dyon
configuration is not a strict saddle point. The topic has been much discussed in
the context of building the `instanton liquid' model of the YM vacuum where a
similar problem arises with respect to instantons and anti-instantons~\cite{D-02}.
There is a way to define properly both `classical' and `quantum' interactions of instantons
and anti-instantons by means of unitarity and analyticity, however for dyons this
approach has not been developed.

Therefore, we adopt here a simple model that the leading source of the dyon--anti-dyon
interaction is ({\it ii}). By definition, the integration measure over the collective
coordinates is the determinant of a matrix built of the overlaps of the (quasi) zero modes.
It is known exactly for a neutral cluster of $N$ kinds of dyons, the KvBLL caloron, see \Eq{G1},
and approximately for many dyons, \Eq{G}. The $(K_+N\times K_+N)$ matrix $G$, where $K_+$ is the
(equal) number of dyons of any kind $m$, will now become a $((K_++K_-)N\times (K_++K_-)N)$ matrix
${\cal G}$ where the diagonal $(K_\pm N\times K_\pm N)$ blocks are built from the overlaps for
the same-duality dyons whereas the off-diagonal $(K_\pm N\times K_\mp N)$ blocks describe the interaction
of opposite-duality dyons, that is $K_+$ dyons and $K_-$ anti-dyons.

The diagonal blocks for same-duality dyons are the same as before (see \ur{G}), with the
only modification that the diagonal elements composed of the normalization integrals for zero modes,
being the dyon action and thus proportional to $A_4$, get Coulomb corrections also from opposite-duality dyons.
Therefore, the first two lines in \Eq{G} acquire an addition
\bea\n
\!\!\!&&\!\!\!\Delta G_{mi,nj}=\delta_{mn}\delta_{ij}\,
\left(-2\sum_k\frac{1}{T|{\bf x}_{mi}\!-\!{\bf y}_{mk}|} \right.\\
\n
\!\!\!&&\!\!\!+\left.
\sum_k\frac{1}{T|{\bf x}_{mi}\!-\!{\bf y}_{m+1,k}|}
+\sum_k\frac{1}{T|{\bf x}_{mi}\!-\!{\bf y}_{m-1,k}|}
\right)
\eea
The off-diagonal blocks for opposite-duality dyons are composed of Coulomb bonds as
in the second two lines of \Eq{G}, with the replacement $|{\bf x}_{mi}-{\bf x}_{nj}|\to
|{\bf x}_{mi}-{\bf y}_{nj}|$. Note that the signs of all Coulomb bonds, both at the diagonal
and off-diagonal, are the same as in the case of same-duality overlaps. The resulting
$((K_++K_-)N\times (K_++K_-)N)$ matrix ${\cal G}$ whose determinant defines the dyon--anti-dyon
ensemble retains the important property that all Coulomb bonds cancel in the sum of the
elements of ${\cal G}$ along any row or any column.

With this natural modification of the measure over the moduli space to incorporate anti-dyons,
one can again rewrite the partition function of the ensemble, now composed of dyons and anti-dyons,
as a path integral for a $3d$ quantum field theory. It is exactly the same theory as given
by \Eq{Z4} but with a double fugacity, $f\to 2f$. The derivation of the `electric' and `magnetic'
string tensions goes as before (see sections 12 and 13) with the simple replacement $\sigma\to\sqrt{2}\sigma$.

We note that had we altogether neglected the interactions between dyons and anti-dyons treating
them as two independent `liquids' we would have to introduce two sets of auxiliary fields $v,w,\chi^\dagger,\chi$
to write down the ensemble as a quantum field theory. The free energy of the system would then also double
but the correlation functions would be different. Meanwhile, the boson fields $v_m$ have the meaning
of the Abelian potentials \ur{Av} and $w_m$'s have the meaning of the dual Abelian potentials,
and there should be no `double sets' of those fields in the effective action. This is achieved naturally
if one introduces the dyon--anti-dyon interaction through the overlaps of the quasi-zero modes as described above.

The proposed model for accounting anti-dyons is still a guesswork inviting for a better understanding.
Nevertheless, a quantity that is sensitive to the dyon--anti-dyon interactions -- the so-called topological
susceptibility -- turns out to be quite reasonable.

\subsection{Topological susceptibility and the gluon condensate}

An important characterization of the vacuum is the topological susceptibility
defined as the zero-momentum correlation function of topological charge densities:
$$
<\!Q_T^2\!>=\int\!d^4x\,\left<\!\frac{\Tr F\tilde F(x)}{16\pi^2}\,
\frac{\Tr F\tilde F(0)}{16\pi^2}\!\right>.
$$
If one adds the $\theta$ term to the action,
$$
\frac{i\theta}{16\pi^2}\int\!d^4x \Tr F_{\mu\nu}\tilde F_{\mu\nu},
$$
and computes the free energy $F$ of the system, the topological susceptibility can be obtained as
$$
<\!Q_T^2\!>
=\frac{1}{V}\left.\frac{\partial^2F}{\partial\theta^2}\right|_{\theta=0}.
$$

The $m^{\rm th}$ dyon carries the topological charge $\nu_m$ whereas the
$m^{\rm th}$ anti-dyon carries the topological charge $-\nu_m$, see \Eq{dyon-action}.
Therefore from the point of view of dyons, the introduction of the $\theta$ term changes
the fugacity of the $m^{\rm th}$ dyon $f\to f\,e^{i\theta\nu_m}$ and the fugacity of the
$m^{\rm th}$ anti-dyon $f\to f\,e^{-i\theta\nu_m}$. The potential energy of the ensemble \ur{calP1}
changes therefore to
$$
{\cal P}=-4\pi f V\sum_m \nu_m\,\left(e^{i\theta\nu_m}+e^{-i\theta\nu_m}\right)\,e^{w_m-w_{m\!+\!1}}.
$$
Minimizing it in $w_m$ and $\nu_m$ we find again that the confining holonomy $\nu_m=\frac{1}{N}$ gives
the minimum, and the free energy at the minimum is (cf. \Eq{freeen})
$$
F=-8\pi f\, V T\,\cos\frac{\theta}{N}=-\frac{N^2}{2\pi^2}\,\frac{\Lambda^4}{\lambda^2}\,V\,\cos\frac{\theta}{N}\,.
$$
Differentiating it twice with respect to $\theta$ we find the topological susceptibility
$$
<\!Q_T^2\!>=\frac{1}{2\pi^2}\frac{\Lambda^4}{\lambda^2}\,.
$$
We see that the topological susceptibility is stable at large $N$ as it is
expected from the $N$-counting rules.

Another important quantity characterizing the vacuum is the gluon condensate also related,
via the trace anomaly, to the free energy (see {\it e.g.} Ref.~\cite{D-02}):
\bea\n
\frac{<\!\Tr F_{\mu\nu}^2\!>}{16\pi^2}\!\!\!&=&\!\!\!-\frac{1}{V}\,\frac{12}{11N}\,F\\
\n
\!\!\!&=&\!\!\!N\,\frac{12}{11}\,\frac{1}{2\pi^2}\,\frac{\Lambda^4}{\lambda^2}
=N\,\frac{12}{11}\,<\!Q_T^2\!>\,.
\eea
The $N$-dependence of the condensate is also as expected.

Let us calculate the average density of dyons (and anti-dyons) in the ensemble.
To that end, we artificially introduce separate fugacities $f_m$ for dyons
of the $m^{\rm th}$ kind into the partition function \ur{Z3}; then the average number of dyons
is found from the obvious relation
$$
<\!K_m\!>=\left.\frac{\partial \log{\cal Z}}{\partial \log f_m}\right|_{f_m=f}.
$$
With separate fugacities, the result \ur{freeen} is modified by replacing
$f\to(f_1f_2...f_N)^{1/N}$, hence
\bea\n
<\!K_m\!>\!\!\!&=&\!\!\!-\frac{1}{T}\left.f_m\frac{\partial F}{\partial f_m}\right|_{f_m=f}\\
\n\\
\n
\!\!\!&=&\!\!\!\frac{1}{N}\,4\pi f\,V=\frac{N}{4\pi^2}\,\frac{\Lambda^4}{\lambda^2}\,V^{(4)},
\la{avn}\eea
{\it i.e.} a finite and equal density of each kind of dyons and anti-dyons in the 4-volume
$V^{(4)}$, meaning also the finite density of the KvBLL instantons and anti-instantons.
From the 3-dimensional point of view, the $3d$ density of dyons (and KvBLL instantons) is increasing
as the temperature goes down: there are more and more instantons sitting on top
of each other in $3d$ but spread over the (compactified) time direction.

Although we have not expected that our theory is valid at small temperatures (where the
approximate measure we use for same-kind and opposite-duality dyons is probably incomplete
since dyons are dense) the zero-temperature limit is obtained in a smooth way. We have seen
it already examining the `electric' and `magnetic' string tensions, and now we see that
the topological susceptibility and the gluon condensate have also a reasonable zero-temperature
limit, while the logarithm of the partition function is extensive in the $4d$ volume as it should be.
It is interesting that these results are found from a theory that is essentially 3-dimensional.
A popular word used in such situation is ``holography'': the $3d$ theory apparently `knows' about
and adequately describes the $4d$ world.

Finally, let us make a numerical estimate of the calculated quantities.
For a numerical estimate at $N\!=\!3$, we take $\lambda=1/4$ compatible with
the commonly assumed freezing of $\alpha_s$ at the value of 0.5, and
$\Lambda=200\,{\rm MeV}$ in the Pauli--Villars scheme.
We then obtain the topological susceptibility,
the gluon condensate and the string tension $(190\,{\rm MeV})^4,\;
(255\,{\rm MeV})^4$ and $(433\,{\rm MeV})^2$, respectively, being in good
agreement with the phenomenological and lattice values.\\

\section{CANCELATION OF GLUONS IN THE CONFINEMENT PHASE}

To prove confinement, it is insufficient to demonstrate the area law for large
Wilson loops and the zero average for the Polyakov line: it must be shown that
there are no massless gluons left in the spectrum. We give an argument
that this indeed happens in the dyon vacuum.

A manifestation of massless gluons in perturbation theory is the Stefan--Boltzmann
law for the free energy:
\bea\n
&&\frac{F_{\rm SB}}{V}=-\frac{T}{V}\log {\cal Z}_{\rm SB}\\
\n\\
\n
&&=\!2\,({\rm number\;of\;gluons})\!\int\!\!\frac{d^3{\bf p}}{(2\pi)^3}\log\left(1\!-\!e^{-\frac{|{\bf p}|}{T}}\right)\\
\n\\
&&=-\frac{\pi^2}{45}\,T^4\,(N^2-1).
\la{SB}\eea
It is proportional to the number of gluons $N^2\!-\!1$ and has the $T^4$ behaviour
characteristic of massless particles. In the confinement phase, neither is permissible:
If only glueballs are left in the spectrum the free energy must be ${\cal O}(N^0)$
and the temperature dependence must be very weak until $T\approx T_c$ where it abruptly
rises owing to the excitation of many glueballs.

As explained in Section 8, the ensemble of dyons has a nonperturbative free energy
\beq
\frac{F_{\rm dyon}}{V}=-\frac{N^2}{2\pi^2}\,\frac{\Lambda^4}{\lambda^2}.
\la{Fdyon}\eeq
It is ${\cal O}(N^2)$ but temperature-independent.
Dyons force the system to have the `most nontrivial' holonomy \ur{muconf}.
For that holonomy, the perturbative potential energy \ur{Ppert} is at its maximum equal to
\beq
\frac{F_{\rm pert,\, max}}{V}=\frac{\pi^2}{45}\,T^4\,\left(N^2-\frac{1}{N^2}\right).
\la{Fmax}\eeq
The full free energy is the sum of the three terms above:
\bea\n
\frac{F}{V}\!\!\!&=&\!\!\!-\frac{N^2}{2\pi^2}\,\frac{\Lambda^4}{\lambda^2}+\frac{\pi^2}{45}\,T^4\,\left(N^2-\frac{1}{N^2}\right)\\
\!\!\!&-&\!\!\!\frac{\pi^2}{45}\,T^4\,(N^2-1).
\la{full_freeen}\eea

We see that the leading ${\cal O}(N^2)$ term in the Stefan--Boltzmann law is {\em canceled}
by the potential energy. It happens only at precisely the `most nontrivial', confining value
of the holonomy and nowhere else! In fact it seems to be the only way how ${\cal O}(N^2)$ massless gluons can be
canceled out of the free energy, and the main question shifts to why does the
system prefer the `confining' holonomy. Dyons seem to be in a position to answer the question.

There is a ${\cal O}(T^4N^0)$ term left in \Eq{full_freeen}, which is also unacceptable
in the confinement phase and has to be canceled, too. However at the ${\cal O}(N^0)$ level
there appear also quantum corrections to the free energy not considered here. Although it
has not been checked explicitly there is a good chance that all ${\cal O}(T^4)$ terms
in the free energy will be absent since there are no massless degrees of freedom left
in the theory. Actually the spectrum is determined by string excitations but they are not
studied yet.

\section{DECONFINEMENT PHASE TRANSITION}

As the temperature rises, the perturbative free energy grows as $T^4$ and eventually it
overcomes the negative nonperturbative free energy \ur{Fdyon}. At this point, the trivial
holonomy for which both the perturbative and nonperturbative free energy are zero,
becomes favourable. Therefore an estimate of the critical deconfinement temperature
comes from equating the sum of \Eq{Fdyon} and \Eq{Fmax} to zero, which gives
\beq
T_c^4=\frac{45}{2\pi^4}\,\frac{N^4}{N^4-1}\,
\frac{\Lambda^4}{\lambda^2}\,.
\la{Tc}\eeq
As expected, it is stable in $N$. A more robust quantity, both from the theoretical
and lattice viewpoints, is the ratio $T_c/\sqrt{\sigma}$ since in this ratio
the poorly known parameters $\Lambda$ and $\lambda$ cancel out:
\bea\la{ratio}
\frac{T_c}{\sqrt{\sigma}}\!\!\!&=&\!\!\!\left(\frac{45}{4\pi^4}\,\frac{\pi^2 N^2}{(N^4-1)\sin^2\frac{\pi}{N}}
\right)^{\frac{1}{4}}\\
\n
\!\!\!&\stackrel{N\to\infty}{\longrightarrow}&\!\!\!
\frac{1}{\pi}\left(\frac{45}{4}\right)^{\frac{1}{4}}+{\cal O}\left(\frac{1}{N^2}\right).
\eea

\begin{table}[h]
\begin{tabular}{|c|c|c|c|c|}
\hline
&$\!\!\!SU(3)\!\!\!$ & $\!\!\!SU(4)\!\!\!$ & $\!\!\!SU(6)\!\!\!$ & $\!\!\!SU(8)\!\!\!$ \\
\hline
&&&&\\
$\!\!\!\begin{array}{c}T_c/\sqrt{\sigma}\\ {\rm theory}\end{array}\!\!\!$& $\!\!\!0.6430\!\!\!$ &
$\!\!\!0.6150\!\!\!$ & $\!\!\!0.5967\!\!\!$ & $\!\!\!0.5906\!\!\!$ \\
&&&&\\
\hline
&&&&\\
$\!\!\!\begin{array}{c}T_c/\sqrt{\sigma}\\ {\rm lattice}\end{array}\!\!\!$& $\!\!\!\begin{array}{c}0.6462\\(30)\end{array}\!\!\!$ &
$\!\!\!\begin{array}{c}0.6344\\(81)\end{array}\!\!\!$ & $\!\!\!\begin{array}{c}0.6101\\(51)\end{array}\!\!\!$ &
$\!\!\!\begin{array}{c}0.5928\\(107)\end{array}\!\!\!$ \\
&&&&\\
\hline
\end{tabular}
\end{table}

In the Table, we compare the values from \Eq{ratio} to those measured in lattice simulations
of the pure $SU(N)$ gauge theories~\cite{Teper}; there is a surprisingly good agreement.

For the $SU(2)$ gauge group, the phase transition is of the 2nd order, as seen from Fig.~14; for higher-rank
groups it is of the 1st order.

\begin{figure}[h]
\hspace{1.5cm}{\epsfig{figure=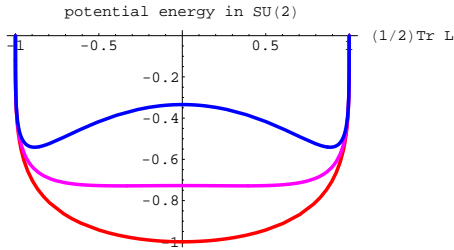,width=6cm}}
\caption{Potential energy of the pure $SU(2)$ YM theory as function of $\half \Tr L$ at zero ({\it lower curve}),
critical ({\it middle curve}) and high ({\it upper curve}) temperatures. At the critical temperature the curve
is flat exhibiting the second order phase transition.}
\end{figure}

To illustrate the confinement-deconfinement phase transition, we sum the nonperturbative potential
energy following from \ur{Z0} and the perturbative potential energy \ur{Ppert}, and plot it for the $SU(3)$
gauge group as function of the holonomy in Fig.~15. The contour plot has the symmetry of the extended root
diagram for the group and is double periodic in $A_4^3$ and $A_4^8$, if one takes the diagonal components
of $A_4$ as a basis. One can concentrate on the fundamental domain corresponding to the Weyl camera in
the root diagram; all the rest are obtained from this one by shifts and reflections. At zero temperature
the perturbative potential energy is zero while the nonperturbative one plotted in the lower part of Fig.~9
has the minimum at the `confining holonomy' corresponding to $\Tr L=0$. It is the confining phase.
\begin{figure}[h]
{\epsfig{figure=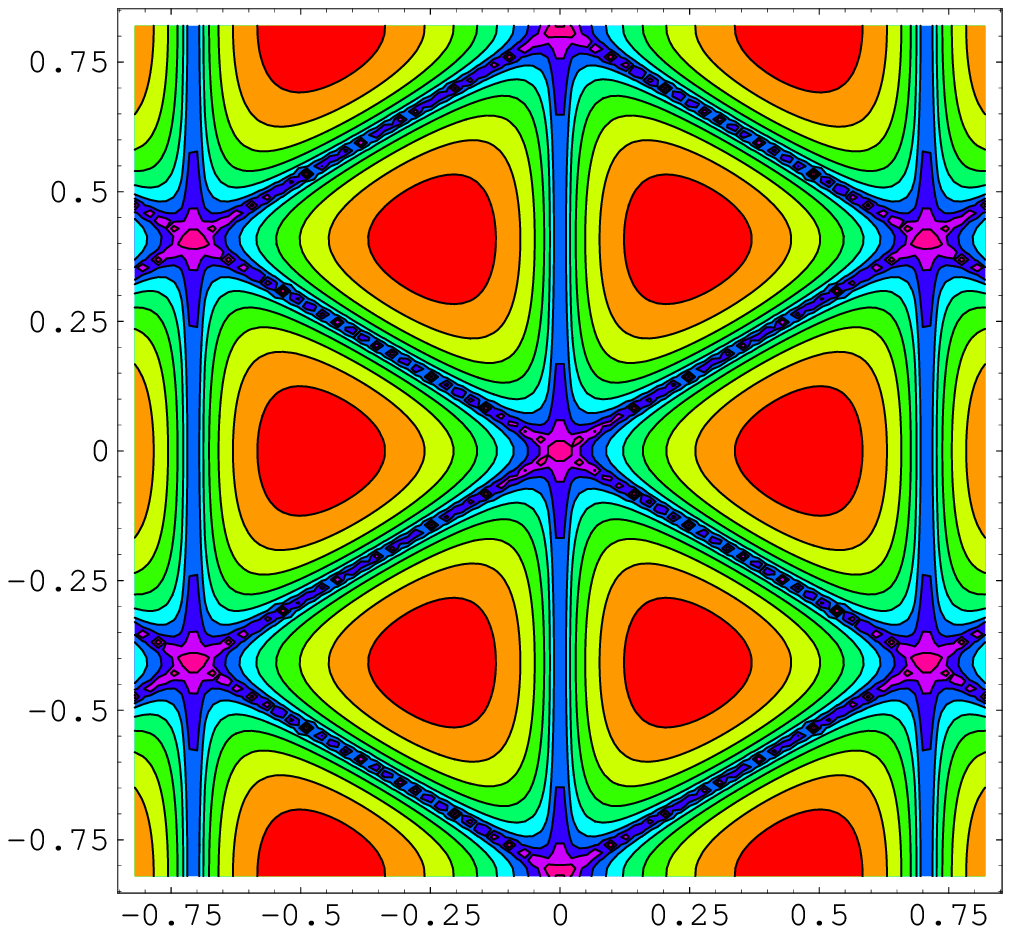,width=3.5cm}}
{\epsfig{figure=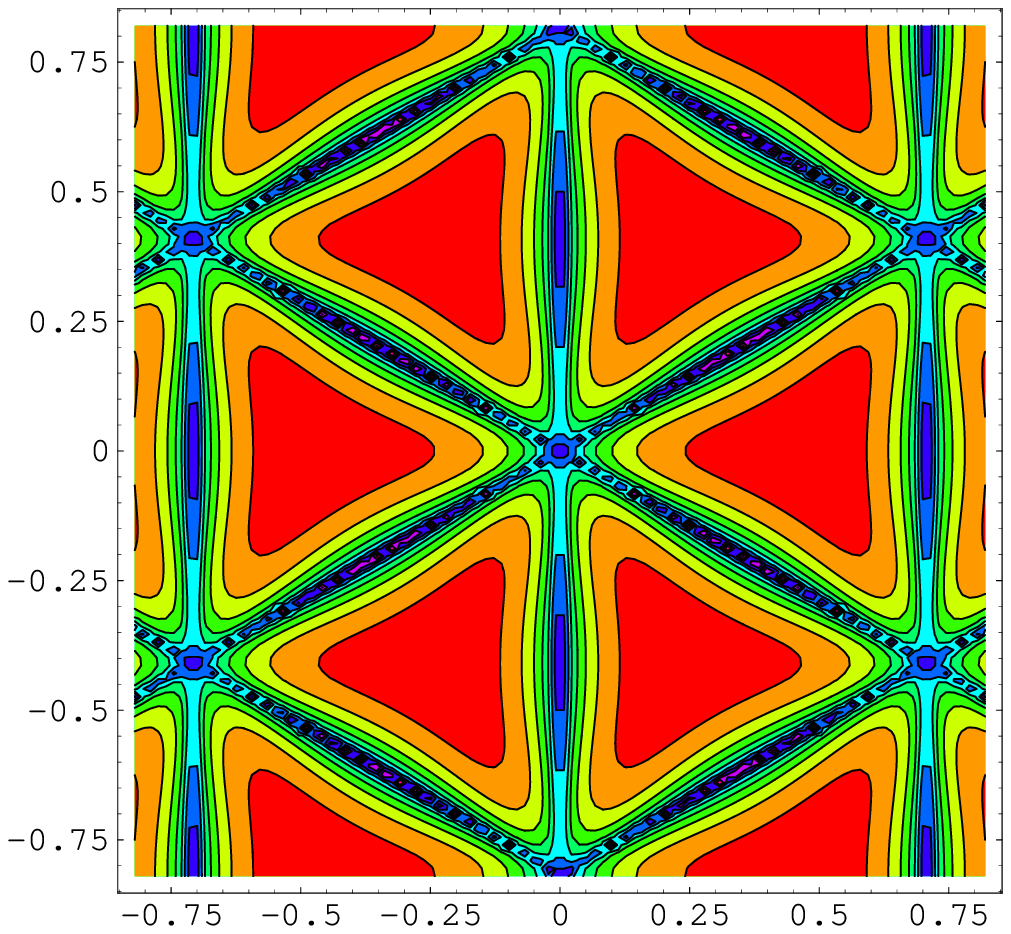,width=3.5cm}}
{\epsfig{figure=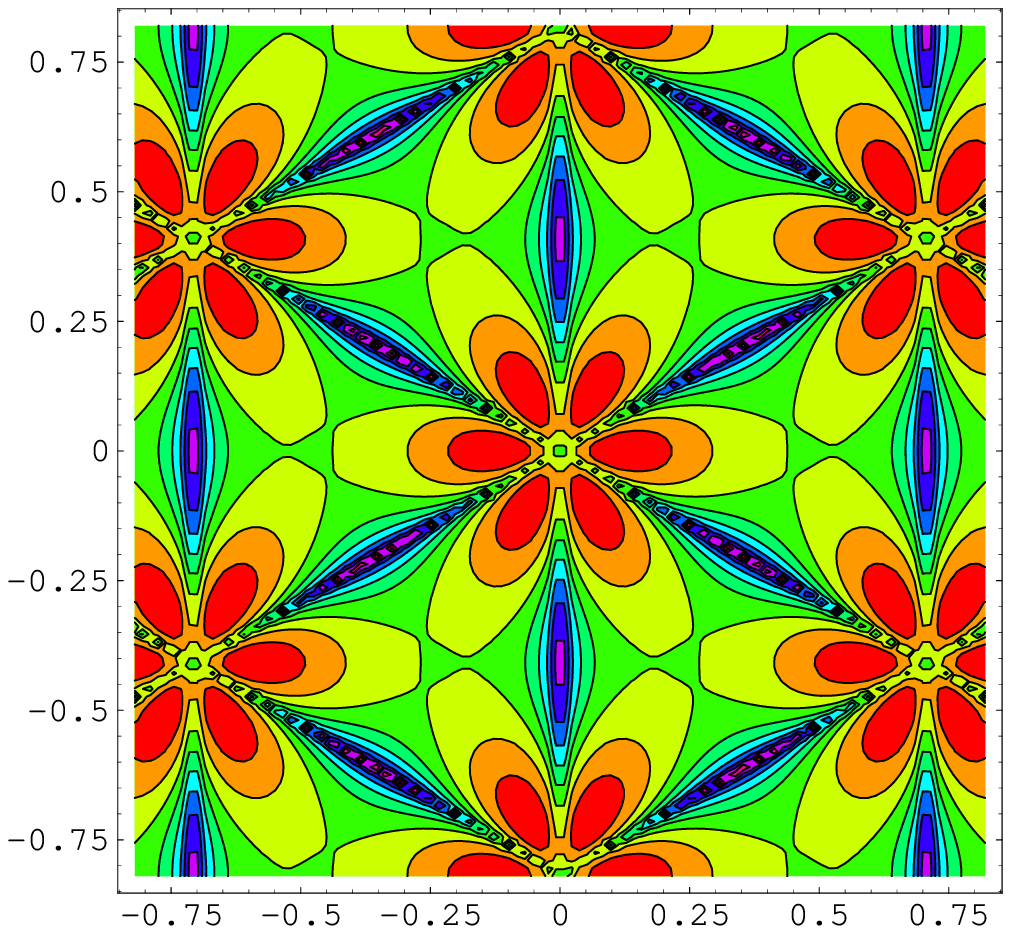,width=3.5cm}}
\hspace{0.6cm}\vspace{-0.3cm}
{\epsfig{figure=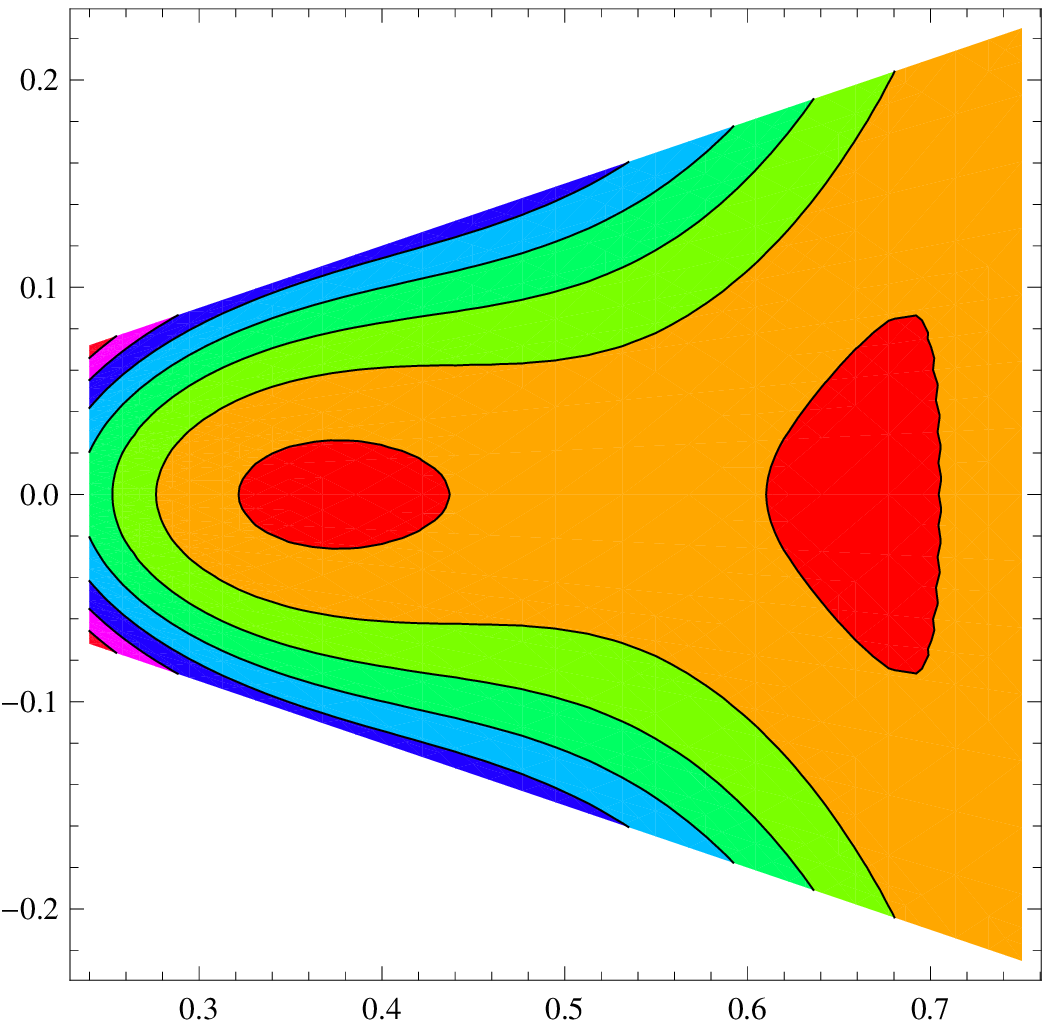,width=3.3cm}}
\caption{Contour plots for the potential energy in the $(A_4^3,A_4^8)$ plane at zero ({\it upper left}),
critical ({\it upper right}) and high ({\it lower left}) temperatures for the $SU(3)$ gauge group. A zoom
at critical temperature ({\it lower right}) clearly shows the coexistence of two minima.}
\end{figure}

As the temperature rises, the perturbative potential energy starts to play an increasingly important role.
At high temperature it is dominant, and asymptotically the minimum is at the trivial field $A_4^3=A_4^8=0$ (at the center of the plots),
and at its copies by periodicity. At a temperature close to that estimated in \ur{Tc} there are two minima:
one is at the `confining' holonomy, the other is at small values of $A_4^3,\,A_4^8$. Such a situation is
referred to as the phase transition of the 1st order: when one of the two minima becomes more deep than the
other the system jumps to the deeper one; the other is then meta-stable. The first derivative of the free
energy {\it e.g.} with respect to the temperature is discontinuous as the system jumps.

\section{CONFINEMENT AND DECONFINEMENT IN THE G(2) GROUP}

It is interesting to check the dyon mechanism of confinement and of deconfinement in the $G(2)$ gauge group.
$G(2)$ is an exceptional group of rank 2 with 14 generators, hence 14 gluons, out of which 8 are the same
as in $SU(3)$ and the other 6 have the colour quantum numbers of 3 quarks and 3 anti-quarks.

$G(2)$ can be considered as a subgroup of $SO(7)$ whose group elements $\Omega$ are represented by $7\times 7$
orthogonal matrices ($\Omega\Omega^T={\bf 1},\;\det\Omega=1$) with an additional constraint
$$
T_{abc}\Omega_{aa^\prime}\Omega_{bb^\prime}\Omega_{cc^\prime}=T_{a^\prime b^\prime c^\prime}
$$
where $T_{abc}$ is an antisymmetric matrix with the following nonzero elements:
\bea\n
\!\!\!&&T_{127}=T_{154}=T_{163}=T_{235}\\
\n
\!\!\!&&=T_{264}=T_{374}=T_{576}=1,\quad T_{acd}T_{bcd}=6\delta_{ab}.
\eea
The lowest-dimensional representation of $G(2)$ is ${\bf 7}$ (we shall call it fundamental),
the next are ${\bf 14}$ (the adjoint representation), ${\bf 27}$, {\it etc.}

The main reason why $G(2)$ gauge theory is interesting is that the $G(2)$ group has a trivial center:
only the unity matrix commutes with all the other group elements. In contrast, in the $SU(N)$ group
the center is nontrivial since $N$ group elements of the form $\exp(2\pi i k/N){\bf 1}_N,\;k=1,...N,$
commute with other elements of the group; they form a discrete $Z_N$ subgroup of $SU(N)$. Therefore,
the $G(2)$ gauge theory serves as an excellent laboratory to check various ideas about the mechanism
of confinement. In particular, it is almost universally accepted that confinement in the $SU(N)$ gauge theory
is related to the center group $Z_N$, and that deconfinement is directly related to the spontaneous
breaking of the symmetry with respect to $Z_N$. If the mechanism of confinement is universal for all
Lie groups and if there is confinement in the centerless $G(2)$ group, this popular view has to be reconsidered.

The gauge theory based on the $G(2)$ group has been extensively studied on the lattice in the last few
years by several groups~\cite{G2-Greensite}-\cite{G2-Gattringer}. The conclusion is unanimous: There
is confinement at low temperature, in the sense that $<\!\Tr L\!>=0$ where $L$ is the Polyakov line
in the fundamental representation ${\bf 7}$, and there is a 1st-order deconfinement transition to
a phase where $<\!\Tr L\!>\neq 0$, -- just as in the $SU(3)$ gauge theory. It means that the presence
or the absence of a nontrivial center of the gauge group is irrelevant to the mechanism of confinement,
and that the deconfinement transition is unrelated to the breaking of center symmetry, as there is
no such symmetry in $G(2)$. As a matter of fact this has been long advocated by Smilga~\cite{Smilga}
on rather general grounds.

The question is whether dyons are capable to explain confinement and deconfinement in the
$G(2)$ gauge theory, as they manage to do it for $SU(N)$. The answer is ``yes'', and
it serves as a very nontrivial check of the philosophy and of the formalism. The material below
is based on the work in collaboration with Victor Petrov, in preparation.

\subsection{Dyons in the $G(2)$ gauge group}

There are 14 generators in $G(2)$; in the lowest 7-dimensional representation eight of them can be
chosen in the form
$$
\Lambda_a = \frac{1}{2}\left(\begin{array}{ccc}
0&0&0\\
0&\lambda_a& 0\\
0& 0 & -\lambda_a
\end{array}
\right)\,,
\quad
\Tr\left(\Lambda_a\Lambda_a\right)=1,
$$
where $\lambda_a$ are the standard $SU(3)$ Gell-Mann matrices.
We choose the two Cartan generators $H_{3,8}=\Lambda_{3,8}$ in similarity with $SU(3)$.
There are 12 ``shift'' generators $E^\pm_i$ satisfying the commutation relations
$$
[H_3,E^\pm_i] = \alpha_{3i} E^\pm_i, \quad [H_8,E^\pm_i] = \alpha_{8i} E^\pm_i\,.
$$
The set of 12 two-dimensional vectors $\mbox{\boldmath$\alpha$}_i=(\alpha_{3i},\alpha_{8i})$ forms a root system of $G_2$.
We choose $\mbox{\boldmath$\alpha$}_1=\left(-\frac{1}{2},-\frac{\sqrt{3}}{2}\right),\;\mbox{\boldmath$\alpha$}_1^2=1$, and
$\mbox{\boldmath$\alpha$}_2=\left(\frac{1}{2},\frac{1}{2\sqrt{3}}\right),\;\mbox{\boldmath$\alpha$}_2^2=\frac{1}{3},$ as the simple roots.
All the rest roots are linear combinations of these. We also introduce the third root
\mbox{\boldmath$\alpha_0$}$=\left(-\frac{1}{2},\frac{\sqrt{3}}{2}\right),\;\mbox{\boldmath$\alpha$}_0^2=1$, which is not simple but
has the largest negative coefficients when expanded in simple roots. Correspondingly, there
are three types of fundamental dyons in $G(2)$, whose magnetic charges are determined by the dual roots,
${\bf m}_i=$\mbox{\boldmath$\frac{\alpha_i}{|\alpha_i|^2}$}:
\bea\la{magnetic-c}
{\bf m}_0\!\!\!&=&\!\!\!\left(-\frac{1}{2},\frac{\sqrt{3}}{2}\right),\\
\n
{\bf m}_1\!\!\!&=&\!\!\!\left(-\frac{1}{2},-\frac{\sqrt{3}}{2}\right),\qquad
{\bf m}_2=\left(\frac{3}{2},\frac{\sqrt{3}}{2}\right).
\nea
It implies that the asymptotic electric and magnetic fields of those three
dyons are
\beq
{\bf E}_{0,1,2}={\bf B}_{0,1,2}=\frac{{\bf x}}{|{\bf x}|^3}\,({\bf m}_{0,1,2}\cdot {\bf H})
\la{G2-EB}\eeq
where ${\bf H}=(H_3,H_8)$ is a diagonal matrix. The dual roots appear here because the dyon is
an $SU(2)$ object: When constructing it along the lines of section 6 one picks up an $SU(2)$
subgroup whose generators commuting as the three Pauli matrices are
\bea\n
\tau^1\!\!\!&=&\!\!\!\sqrt{2}\frac{E^+_i\!+\!E^-_i}{|\mbox{\boldmath$\alpha_i$}|}, \quad
\tau^2= \sqrt{2}\frac{E^+_i\!-\!E^-_i }{i|\mbox{\boldmath$\alpha_i$}|}, \\
\n
\tau^3\!\!\!&=&\!\!\! 2\frac{(\mbox{\boldmath$\alpha_i$}\cdot \vec{H})}{|\mbox{\boldmath$\alpha_i$}|^2},
\qquad [\tau^i\tau^j]=2i\epsilon^{ijk}\tau^k.
\nea

As usually, dyon solutions are characterized by the holonomy or the eigenvalues of the Polyakov line
(here: in the 7-dimensional representation) at spatial infinity. In the gauge where $A_4$ is static and
diagonal we label the holonomy by two numbers ${\bf \mu}=(\mu_3,\mu_8)$ such that asymptotically
\beq
A_4(\infty) = 2\pi T\,(\mbox{\boldmath$\mu$}\cdot {\bf H})=2\pi T(\mu_3H_3+\mu_8 H_8).
\la{G2-A4}\eeq
Correspondingly, the Polyakov line's eigenvalues are
\beq
L=\exp\left(2\pi i (\mbox{\boldmath$\mu$}\cdot {\bf H})\right).
\la{G2-Pol-line}\eeq
The actions of the three individual dyons are
\bea
S_{1,2}\!\!\!&=&\frac{4\pi}{\alpha s}(\mbox{\boldmath$\mu$}\cdot {\bf m}_{1,2}),\\
\n
S_0\!\!\!&=&\frac{4\pi}{\alpha s}(1+\mbox{\boldmath$\mu$}\cdot {\bf m}_0).
\la{G2-actions}\eea

The above construction \ur{G2-EB}-\ur{G2-actions} can be generalized to an arbitrary gauge group~\cite{DHK}:
One takes $r$ dyon solutions associated with the simple roots $\mbox{\boldmath$\alpha$}_i,\;i=1,...,r$,
where $r$ is the rank, and supplements them by a dyon associated with a non-simple lowest root $\mbox{\boldmath$\alpha$}_0$
that has the most negative coefficients in the expansion over the simple roots. In the gauge where
$A_4$ is time independent the first $r$ dyons are static while the last or zeroth dyon is obtained
by a time-dependent gauge transformation; in the notations of subsection 6.1 it is the ``$L$'' dyon.
The magnetic charges of all $r+1$ dyons are given by the dual roots: ${\bf m}_i=$\mbox{\boldmath$\frac{\alpha_i}{|\alpha_i|^2}$},
$i=0,1,...,r$.

Since ${\bf m}_0$ is not linearly independent as an $r$-dimensional vector it can be expanded in the other
dual roots, ${\bf m}_0=-\sum_{i=1}^rk_i^*{\bf m}_i$, or
\beq
\sum_{i=0}^rk_i^*{\bf m}_i=0,
\la{identity}\eeq
where the integers $k_i^*$ are called dual Kac labels, or co-marks; by construction $k_0^*=1$. In the case of $G(2)$
we have $k_0^*=1,\;k_1^*=2,\;k_2^*=1$. The sum
\beq
c_2=\sum_{i=0}^rk_i^*
\la{dual_Coxeter}\eeq
(equal to 4 for $G(2)$ and $N$ for $SU(N)$) is called the dual Coxeter number. \Eq{identity} tells us that in order to build an
electric- and magnetic-neutral object (the KvBLL instanton with a nontrivial holonomy) one needs $c_2$
fundamental dyons, some of which may have a multiplicity other than unity. For example, in the $G(2)$ gauge
theory one has to take the dyon of the 1st kind twice since ${\bf m}_0+2{\bf m}_1+{\bf m}_2=0$, meaning
that in this case the KvBLL instanton is made of four dyons but only of three kinds. As a curiosity,
the KvBLL instanton of the exceptional $E(8)$ gauge theory is built out of 30 dyons of 9 kinds (since the rank is
eight)~\cite{DHK}.

Returning to the $G(2)$ group, let us introduce the numbers $\nu_{1,2}$ characterizing the holonomy \ur{G2-Pol-line}:
\bea\la{nu012}
\frac{\alpha_s}{2\pi}S_1\!\!\!\!&=&\!\!\!\nu_1=-\mu_3-\sqrt{3}\mu_8,\\
\n
\frac{\alpha_s}{2\pi}S_2\!\!\!\!&=&\!\!\!\nu_2=3\mu_3+\sqrt{3}\mu_8,\\
\n
\frac{\alpha_s}{2\pi}S_0\!\!\!\!&=&\!\!\! (2-2\nu_1-\nu_2).
\nea
We imply that $S_{0,1,2}>0$; if $\mu_{3,8}$ are such that this condition is violated, one can always
perform a time-dependent gauge transformation which shifts $\mu_{3,8}$ to the `fundamental' domain where
the right hand sides of the above equations are non-negative.

We now build the semiclassical vacuum as an ensemble of 3 kinds of dyons (and anti-dyons) satisfying the
neutrality condition $K_1=2K_2=2K_0$. This is done in the same way and in the same approximation as
for the $SU(N)$ gauge theory, see sections 8-11. The main conclusion there is that the free energy of the
ensemble is proportional to the geometrical mean of the individual dyon actions. In the $G(2)$ case the
action of the dyon of the 1st kind needs to be squared as it enters twice more often than the other two,
to satisfy the neutrality condition. It results in the nonperturbative free energy induced by dyons
({\it cf.} \Eq{Z0})
\beq
F=-4\pi f V T\left(\nu_1^2\nu_2(2-2\nu_1-\nu_2)\right)^{\frac{1}{4}}.
\la{G2-F}\eeq
The minimum of this expression in $\nu_{1,2}$ or, equivalently, in $\mu_{3,8}$ is achieved when
the actions for all three kinds of dyons are equal, $S_0=S_1=S_2$, or $\nu_1=\nu_2=2-2\nu_1-\nu_2=\frac{1}{2}$,
or
\beq
\mu_3=\frac{1}{2},\qquad \mu_8=-\frac{1}{\sqrt{3}}.
\la{G2-mu}\eeq
This is the value of the holonomy that is dynamically preferred by the ensemble of dyons.

The final step is to substitute \Eq{G2-mu} into the Polyakov line \ur{G2-Pol-line}, and we
find
$$
\!\!\!\Tr L\! =\! 1\!+\!2\cos\frac{\pi}{6}\!+\!2\cos(\frac{4\pi}{6})\!+\!2\cos(\frac{5\pi}{6})\! =\! 0\quad(!)
$$
We conclude that the confining (zero) value of the trace of the Polyakov line follows from the
minimization of the free energy induced by dyons, although {\em there are no a priori symmetry reasons}
why this holonomy is privileged.

\subsection{Deconfinement phase transition in $G(2)$}

As temperature increases, the perturbative free energy, also a function of the holonomy, starts to play
an increasingly important role. The perturbative potential for any gauge group is given by the
equation
\beq
{\cal P}^{\rm pert}=\frac{2\pi^2}{3}T^3\sum_n\!\left(\!\frac{\lambda_n}{2\pi T}\!\right)^2
\left(\!1\!-\!\left|\frac{\lambda_n}{2\pi T}\right|\!\right)^2
\la{G2-Ppert}\eeq
where $\lambda_n$ are the `charged' gluon masses in the background of a constant $A_4$. More precisely,
$\lambda_n$ are the eigenvalues of $A_4$ in the adjoint representation. In the $G(2)$ case there are
12 nonzero eigenvalues:
\bea\n
\frac{\lambda_n}{2\pi T}\!\!\!&=&\!\!\!\pm\mu_3,\quad\pm\frac{\mu_8}{\sqrt{3}},\quad
\pm\frac{1}{2}\left(\mu_3\pm\frac{\mu_8}{\sqrt{3}}\right),\\
\n
\!\!\!&\pm &\!\!\!\frac{1}{2}\left(\mu_3\pm\mu_8\sqrt{3}\right)
\nea
which should be plugged into \Eq{G2-Ppert}.

The sum of the perturbative \ur{G2-Ppert} and nonperturbative \ur{G2-F} parts of the free energy
as function of the holonomy $\mu_{3,8}$ is plotted in Fig.~16. The contour plots are very similar
to those of Fig.~15 depicting the free energy at different temperatures for the $SU(3)$ gauge theory.
The 12 symmetrical sectors in Fig.~16 correspond to 12 different domains for $\nu_1,\,\nu_2$. They
can be related to each other by time-dependent gauge transformations. Going from one
sector to another corresponds to changing the roots used to define the dyon solutions; physically all
of them are equivalent.

\begin{figure}[h]
{\epsfig{figure=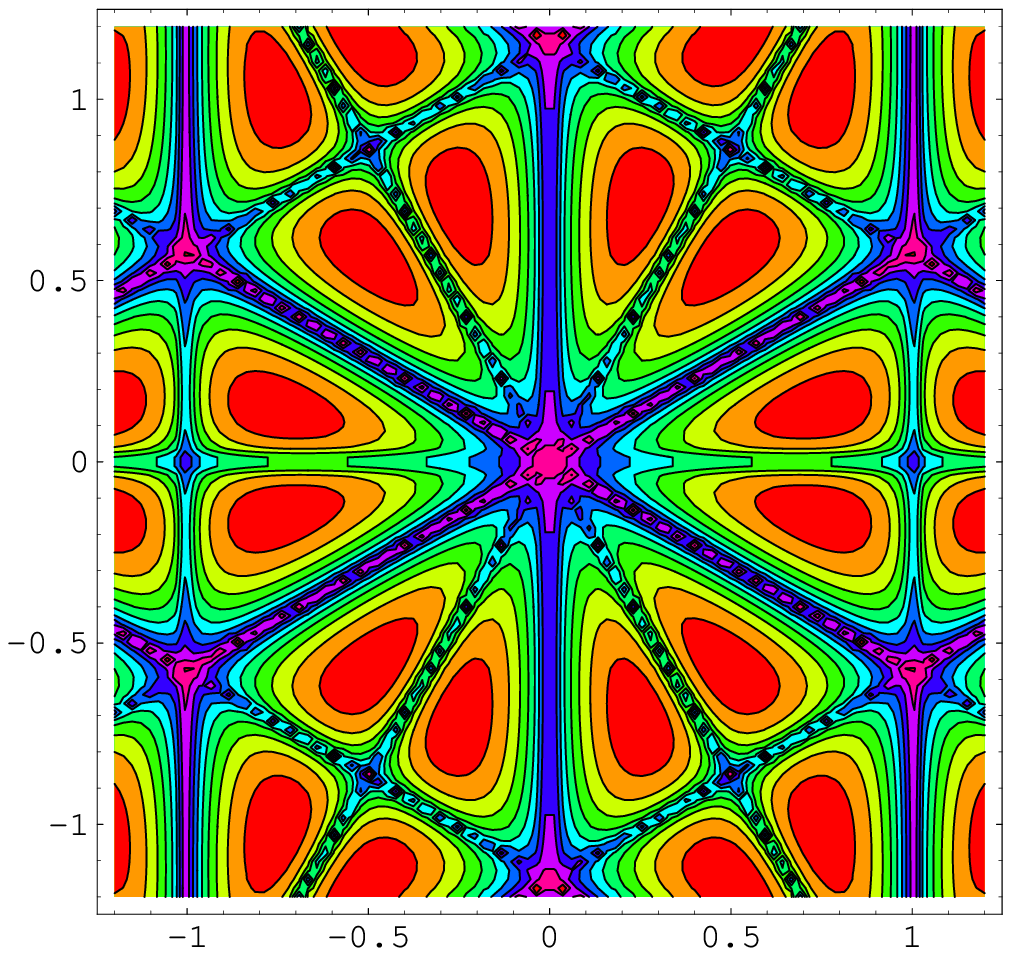,width=3.5cm}}
{\epsfig{figure=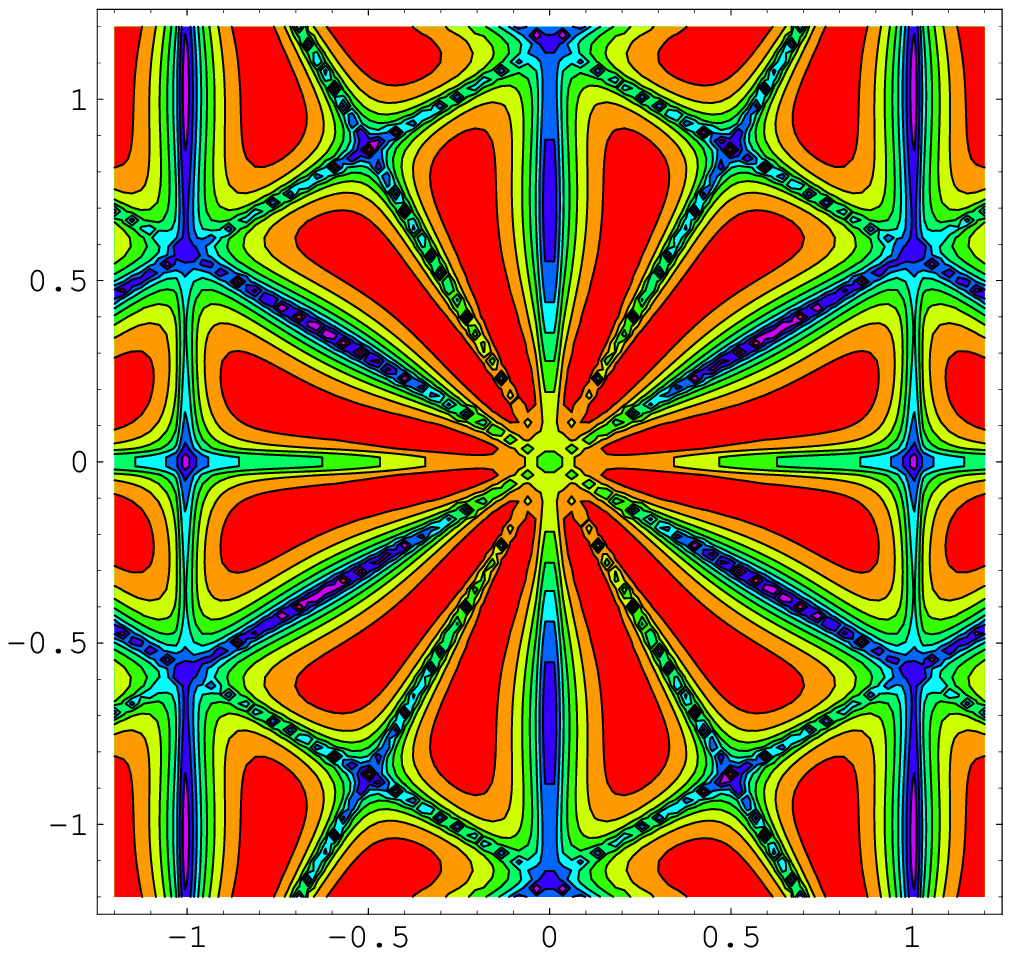,width=3.5cm}}
{\epsfig{figure=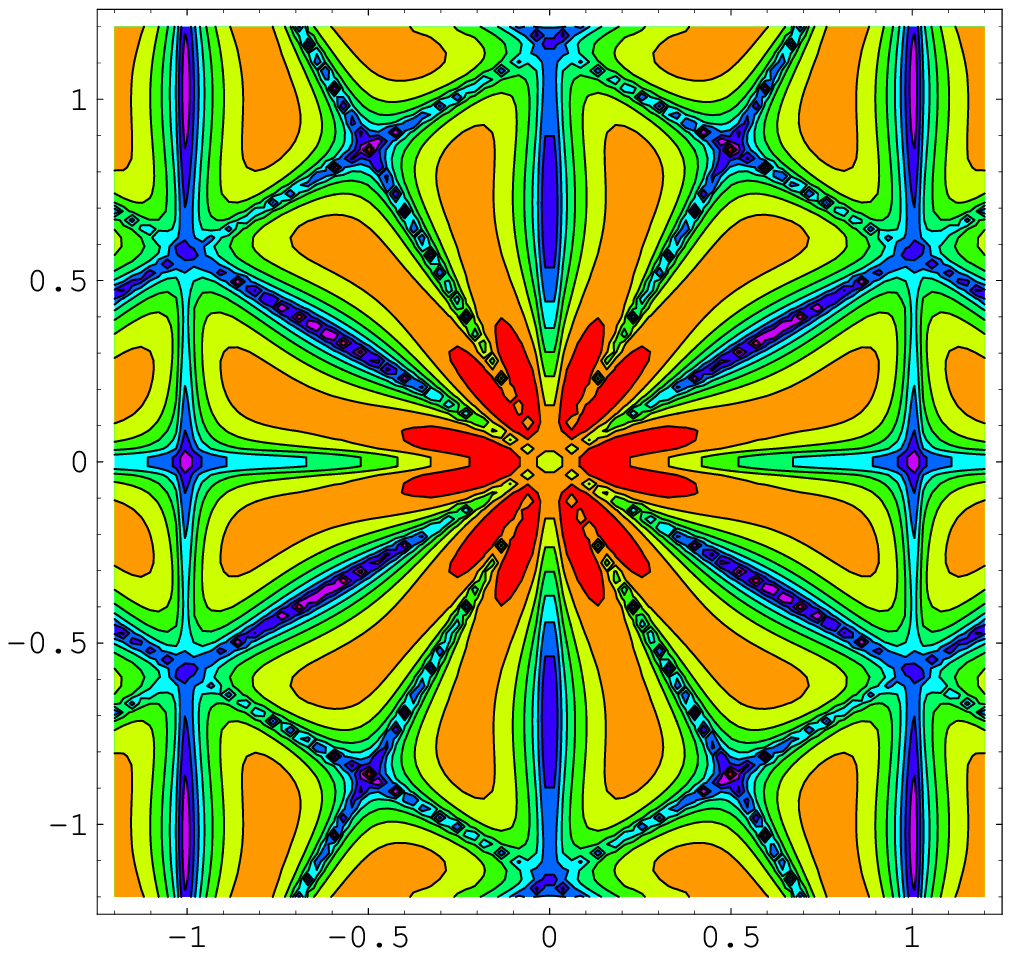,width=3.5cm}}
\hspace{0.4cm}
{\epsfig{figure=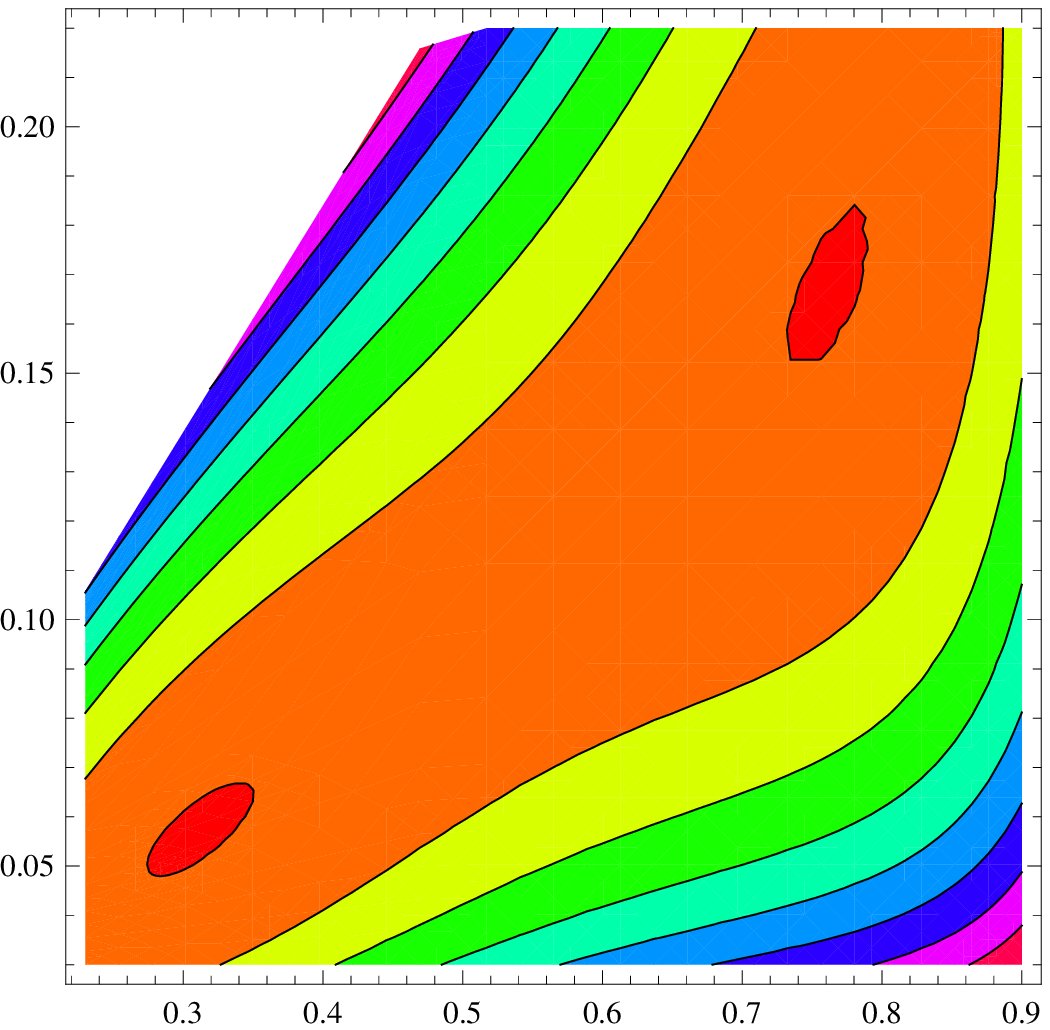,width=3.3cm}}
\vskip -0.3true cm

\caption{Contour plots for the potential energy as function of the holonomy $\mu_3$ and $\mu_8$
in the $(\mu_3,\mu_8)$ plane at zero ({\it upper left}),
critical ({\it upper right}) and high ({\it lower left}) temperatures for the $G(2)$ gauge group. A zoom
at critical temperature ({\it lower right}) clearly shows the coexistence of two minima, typical for the 1st
order phase transition.}
\vskip -0.16true cm

\end{figure}

At temperatures below critical the minimum is at a point in the $(\mu_3,\mu_8)$ plane that
has no obvious privilege from the point of view of symmetry. Nevertheless, it is precisely
the point where the Polyakov loop vanishes, see the previous subsection.

At certain temperature which can be estimated through the $\Lambda$ scale of the theory
a second minimum starts to develop. At a critical temperature it becomes exactly as deep
as the first minimum. It is the 1st order transition point. Above the phase transition
the system stays in the second minimum which also has no obvious symmetry but it now
leads to $\Tr L\neq 0$. Eventually at very high temperatures the system will move to the
trivial holonomy at the origin of the contour plots in Fig.~16. This picture is in accordance
with what is observed in recent lattice studies~\cite{G2-Greensite}-\cite{G2-Gattringer}.

Qualitatively, exactly the same story happens in the $SU(3)$ gauge theory, see section 16
and Fig.~15, with the only unprinciple difference that in the $SU(3)$ case the trajectory of the
minimum in the holonomy plane is along a definite ray, and it appears more symmetric to the eye.\\

We conclude that the center $Z_N$ symmetry of the $SU(N)$ gauge theory is not of fundamental
importance in the confinement mechanism, as a very similar picture of confinement and deconfinement
emerges in the centerless $G(2)$ gauge theory. Dyons, however, are in a position to explain
both gauge theories as in both cases they dynamically drive the system to the confining value
of the holonomy.

\section{CONCLUSIONS}

In these lectures, we have reviewed some essential tools used to study Yang--Mills theories
at zero, as well as at nonzero temperatures, most of which are related to the underlying topology
of gauge theories. In particular, we have briefly outlined
\begin{itemize}
\item instantons in Quantum Mechanics and Yang--Mills (YM) theory
\item the Hamiltonian or Schr\"odinger formulation of the YM theory
\item periodic generalization of instantons to nonzero temperatures
\item topological classification of YM classical solutions
\item Bogomolny--Prasad--Sommerfield monopoles, also called dyons in the pure YM theory
\item the construction of dyons for arbitrary gauge groups
\item instantons with a nontrivial holonomy (the KvBLL calorons).
\end{itemize}

We have then moved to an uncharted territory in an attempt to build a semiclassical picture
based on the ensemble of dyons, that would give an explanation of the puzzling phenomena
associated with the confinement of quarks. In particular, we have discussed
\begin{itemize}
\item integration measure over the collective coordinates of many dyons
\item reduction of the dyons' partition function to that of an exactly solvable $3d$ quantum field theory
\item why the `confining' holonomy is dynamically preferred by the ensemble of dyons
\item the calculation of the `electric' string tensions for any nonzero $N$-ality representation
from the correlation function of the Polyakov lines
\item the 'magnetic' string tension (equal to the `electric' one despite the lack of apparent Euclidean symmetry
in the $3d$ formalism used) from the area law for large Wilson loops
\item the cancelation of perturbative gluons in the Stefan--Boltzmann law in the confining phase
\item an estimate of the critical deconfinement temperature in units of the fundamental string tension
\item confinement and deconfinement in the centerless $G(2)$ gauge theory.
\end{itemize}

There are still many points in the presented dyon mechanism of confinement, that need better understanding if not justification.
Therefore at the moment it is rather a nonperturbative model of strong interactions, than a
rigorous theory. Nevertheless, the overall picture with its good accordance with phenomenological and lattice
results, even at a quantitative level, looks rather satisfactory and hopefully is a step in the right direction.

\section{ACKNOWLEDGEMENTS}

I am grateful to Victor Petrov for numerous helpful discussions and a collaboration. Many thanks to Mikhail
Vysotsky and Leonid Glozman and their colleagues in Moscow and Graz for kind hospitality during the ITEP and
Schladming Winter Schools, respectively, where these lectures have been delivered. This work is supported
in part by the Russian Foundation for Basic Research grant RFBR-09-02-01198.

\end{document}